\documentclass[%
 reprint,
 amsmath,amssymb,
 aps,
 prl
]{revtex4-2}

\usepackage{graphicx}
\usepackage{dcolumn}
\usepackage{bm}
\usepackage{hyperref}
\hypersetup{colorlinks=true,
linkcolor=cyan,
urlcolor=blue,
citecolor=red}

\begin{document}

\title{Transgression Completion for Chern-Simons Black Hole Thermodynamics}


\author{\textbf{Rong-Gen Cai}$^ {2,1,3}$}%
\email{cairg@itp.ac.cn}
\author{\textbf{Li Li}$^{1,3,4}$}%
  \email{liliphy@itp.ac.cn}
\author{\textbf{Jun-Kun Zhao}$^{1}$}
  \email{junkunzhao@itp.ac.cn}

\affiliation{${}^{1}$Institute of Theoretical Physics, Chinese Academy of Sciences, Beijing 100190, China}
\affiliation{${}^{2}$Institute of Fundamental Physics and Quantum  Technology, Ningbo University, Ningbo 315211, China}
\affiliation{${}^{3}$School of Fundamental Physics and Mathematical Sciences, Hangzhou Institute for Advanced Study, UCAS, Hangzhou 310024, China}
\affiliation{${}^{4}$School of Physical Sciences, University of Chinese Academy of Sciences, No.19A Yuquan Road, Beijing 100049, China}

\begin{abstract}
The quantum statistical relation, which identifies the thermodynamic free energy with the on-shell Euclidean action, is a cornerstone of black hole thermodynamics. In this Letter, we demonstrate that it fails for charged magnetic black holes in five-dimensional Einstein-Maxwell-Chern-Simons theory unless the action is formulated in a gauge-invariant manner. In particular, the free energy computed via the quantum statistical relation violates both the first law of thermodynamics and hydrodynamics. We identify the missing finite contribution---a nonlocal radial integral---whose inclusion restores exact thermodynamic consistency. Remarkably, this term is not an ad hoc correction: combined with the local Chern-Simons representative, it forms the transgression form, yielding a strictly gauge-invariant action. The correct action is therefore not the local Chern-Simons term alone but its transgression completion---a physical requirement rather than a mere mathematical refinement. The same mechanism persists for helical, rotating, scalarized, higher-derivative, and higher-dimensional black holes. Our work establishes that topological interactions can fundamentally alter the relation between the Euclidean action and the thermodynamic potential and identifies the gauge-invariant transgression form as the correct framework for black hole thermodynamics in Chern-Simons-coupled theories.
\end{abstract}

\maketitle
\textbf{Introduction.--}Black hole (BH) thermodynamics provides one of the most profound connections between gravitation, quantum mechanics, and statistical physics~\cite{Wald:1984rg}. A widely used framework for studying BH thermodynamics is the Euclidean gravitational path integral, pioneered by Gibbons and Hawking~\cite{Gibbons:1976ue}. In the semiclassical approximation, the gravitational partition function is dominated by the on-shell Euclidean action $I_E$, and the quantum statistical relation (QSR)~\cite{Gibbons:1976ue,Gibbons:2004ai} then conventionally identifies the thermodynamic potential, or free energy, as
\begin{equation}\label{relation}
W_{QSR}= T I_E \,.
\end{equation}
Its validity has been remarkably robust, extending from the Kerr-Newman family to asymptotically AdS spacetimes, and even to black rings with intricate dipole charges~\cite{Astefanesei:2005ad,Copsey:2005se}. This success conveys an implicit expectation: the QSR, when properly applied, captures the full thermodynamic content of any BH solution.

\begin{figure}[ht!]
\includegraphics[width=0.45\textwidth]{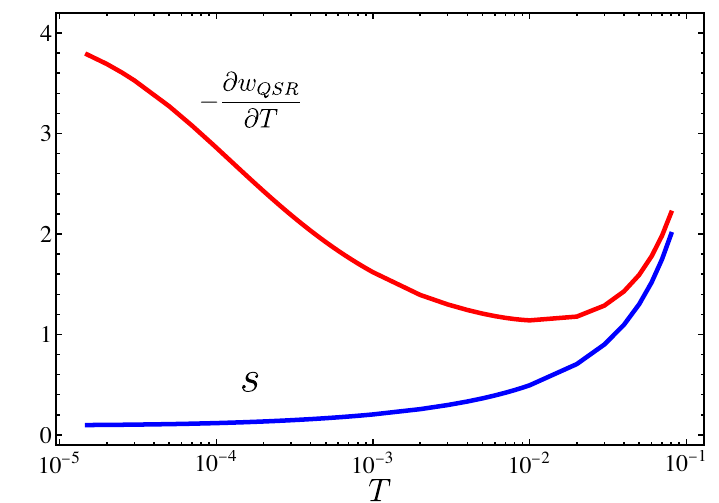}
\includegraphics[width=0.45\textwidth]{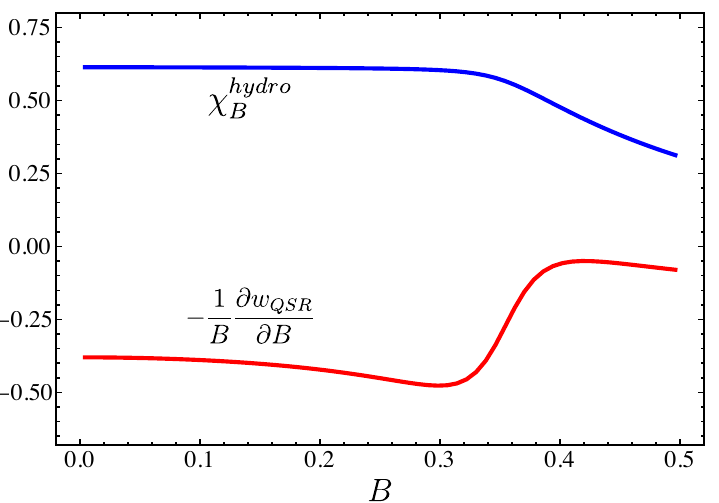}
\caption{Numerical tests of the first law using the free energy $w_{QSR}$ from the standard QSR~\eqref{relation}. \textbf{Top:} Comparison of the Bekenstein-Hawking entropy $s$ and $-(\partial w_{QSR}/\partial T)|_{\mu, B}$ at $B=0.33$. \textbf{Bottom:} Comparison of magnetic susceptibility from hydrodynamics $\chi_B^{hydro}$ and $-\frac{1}{B}(\partial w_{QSR}/\partial B)|_{\mu, T}$ at $T=0.005$. The systematic deviations indicate that $w_{QSR}$ is incomplete. The plots are for minimal supergravity with $k=k_{\text{susy}}=2/\sqrt{3}$ and $\mu=1$.}\label{fig:check0}
\end{figure}

\begin{figure}[ht!]
\includegraphics[width=0.45\textwidth]{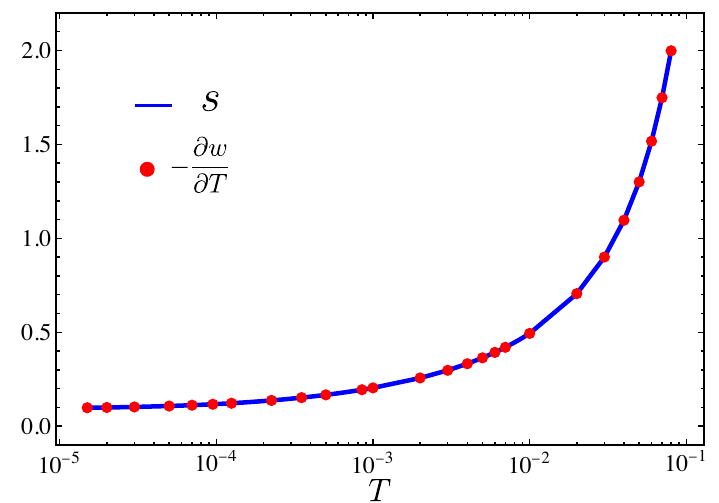}
\includegraphics[width=0.45\textwidth]{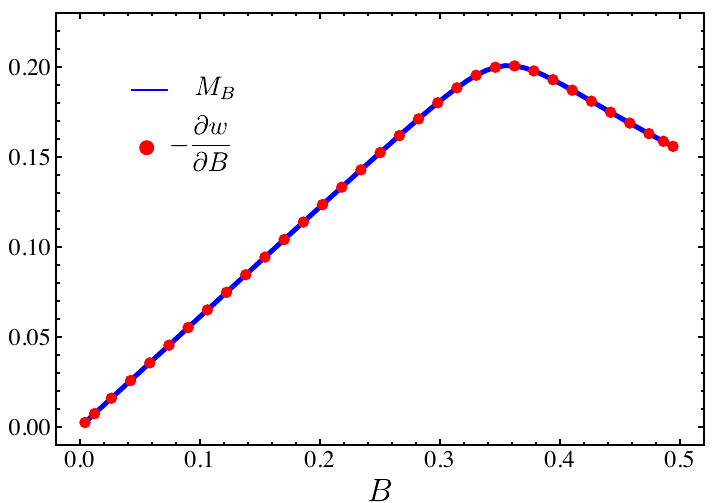}
\caption{Numerical verification of the corrected thermodynamics after including $k Q_{cs}$ in the free energy. \textbf{Top:} Perfect agreement between the Bekenstein–Hawking entropy $s$ of~\eqref{eqTS} and $-(\partial w/\partial T)|_{\mu, B}$ at $B=0.33$. \textbf{Bottom:} Exact match between the magnetization $M_B$ from hydrodynamics and $-(\partial w/\partial B)|_{\mu, T}$ at $T=0.005$. We consider $k=k_{\text{susy}}= 2/\sqrt{3}$ and $\mu=1$.}\label{fig:check1}
\end{figure}

In this Letter, we demonstrate a fundamental limitation of this expectation for BHs with Chern-Simons (CS) interactions. The CS terms enter action principles across a broad range of physical theories, reflecting a deep interplay between geometry and physics~\cite{Dunne:1998qy,Deser:1998mk}. In gravitational settings, they arise naturally in supergravity and string theory constructions, with important implications for BH dynamics and thermodynamics~\cite{Cremmer:1978km,Chamblin:1999tk,Witten:1998qj}. Here we show that, for charged magnetic BHs in five-dimensional Einstein-Maxwell-Chern-Simons (EMCS) theory, the free energy computed via the standard QSR~\eqref{relation} produces thermodynamics that systematically violates both the first law and the hydrodynamic relation (Fig.~\ref{fig:check0}). No local counterterms at the asymptotically AdS boundary can cure the inconsistency. The failure is traced to a missing nonlocal contribution $Q_{cs}$---an integral over the entire bulk radial coordinate---that the conventional Euclidean action fails to capture. Including this single term restores exact agreement with the first law and the hydrodynamic prediction (Fig.~\ref{fig:check1}). Beyond the numerical evidence, the above conclusion is independently supported by two analytical approaches: the variation of the Euclidean action~\cite{Gauntlett:2009bh,Donos:2012wi} and the Iyer–Wald Noether charge formalism~\cite{Wald:1993nt,Iyer:1994ys,Li:2020spf,Xiao:2023lap}.

More importantly, we find that the origin of $Q_{cs}$ has a natural geometric explanation. It is not an independent correction to the thermodynamic potential; together with the CS term $A\wedge F\wedge F$, it forms the transgression form~\cite{Nakahara:2003nw,Mora:2004kb,Mora:2006ka,Pais:2023zqz,Mora:2010gr,Mora:2014fba}. Unlike the bare CS term, the transgression form is strictly invariant under $U(1)$ gauge transformations. This is not a mathematical nicety—it is a physical imperative. The local CS term, as conventionally written, is not a valid global action in the presence of magnetic flux; its gauge variance is precisely what produced the thermodynamic violation. The transgression form removes this variance and, in doing so, automatically generates the previously missing $Q_{cs}$ as its finite global completion.

This revelation elevates our finding from a thermodynamic patch to a fundamental paradigm shift: for a broad class of gravitational theories with CS-type interactions, the correct action—and hence the starting point for the Euclidean path integral—is not the local CS term, but rather its gauge-invariant transgression completion. The universality of this statement is striking: we verify the same transgression mechanism across helical, rotating, scalarized, higher-derivative, and higher-dimensional BHs, with the correction taking the same structural form in every case. This universality also explains why this correction has remained unnoticed in the extensive literature on CS-coupled BHs: previous analyses, while numerous, have focused on simpler configurations---often with vanishing magnetic fields or restricted gauge-field configurations---where the transgression completion happens to vanish. Our work thus establishes that the QSR is not a priori guaranteed in theories with topological interactions, and identifies the transgression form as the correct geometric framework for BH thermodynamics and quantum gravity path integrals in such systems. 

\textbf{Model and Incompleteness of the QSR.}--We begin with the five‑dimensional EMCS theory
\begin{equation}
\begin{split} \label{action}
S=\frac{1}{16\pi G} \int& d^5x \sqrt{-g} \Big( R+\frac{12}{L^2}-\frac{1}{4}F_{ab}F^{ab}  \\
&+\frac{k}{24} \epsilon^{abcde}A_a F_{bc} F_{de} \Big) \,,
\end{split}
\end{equation}
where $F=dA$ is the field strength of the $U(1)$ gauge field $A_a$, $G$ is Newton's constant, $L$ is the AdS radius, and $k$ is the CS coupling. At the supersymmetric value $k=k_{\text{susy}}=2/\sqrt{3}$, this is the bosonic sector of minimal supergravity~\cite{Buchel:2006gb,Gauntlett:2006ai}. Our analysis is independent of $k$. We set $16\pi G=L=1$. The dyonic BH solutions~\cite{DHoker:2010zpp,DHoker:2010onp} take the form
\begin{equation}\label{BHs}
\begin{split}
ds^2=&\frac{1}{r^2}\Big[-\big(f e^{-\chi}-h^2p^2\big)dt^2+2ph^2 dtdz  \\
 &+dx^2+dy^2+h^2 dz^2+\frac{dr^2}{f}\Big]\,,  \\
A=& A_t dt+\frac{B}{2}(x dy-ydx)-A_z dz \,,
\end{split}
\end{equation}
where $f,\, \chi,\, h,\, p,\, A_t$ and $A_z$ are functions of the radial coordinate $r$. The constant $B$ is the background magnetic field perpendicular to the $x$-$y$ plane. The AdS boundary is located at $r=0$, while the BH horizon is at $r=r_h$ where regularity conditions $f(r_h)=A_t(r_h)=p(r_h)=0$ are imposed. The system admits the AdS Reissner-Nordstr\"{o}m solution when $B=0$. However, for $B\neq0$, one must solve the system numerically. 

We now compute all thermodynamic quantities using standard methods in the literature. For example, the temperature and entropy density are given by 
\begin{equation}\label{eqTS}
T=-\frac{1}{4\pi} f'(r_h) e^{-\chi(r_h)/2}   \,, \quad s=\frac{4\pi h(r_h)}{r_h^3}  \,.
\end{equation}
The latter comes from the Bekenstein-Hawking area law. Recalling that in the presence of external magnetic flux, the standard laws of thermodynamics take the form~\cite{Landau:1995}
\begin{equation}\label{eq:1st}
\begin{split}
\delta w&= -s\delta T -\rho \delta\mu -M_B \delta B \,,\\
\delta \epsilon&= T\delta s+\mu\delta \rho -M_B \delta B \,,
\end{split}
\end{equation}
where $w=W/V$ is the free energy density in a fixed spatial volume $V$, and the energy density $\epsilon=w+Ts+\mu\rho$~\cite{dfenergy}. Here $T, \mu$ and $B$ denote temperature, chemical potential, and magnetic field, respectively, while their conjugate quantities are entropy density $s$, charge density $\rho$, and magnetization $M_B$. For the planar ansatz~\eqref{BHs}, these are the only independent thermodynamic conjugate pairs; additional pairs, if present in more general configurations, can be included without affecting the analysis below.

We then compute the free energy density $w_{QSR}$ from the on‑shell Euclidean action--which includes appropriate counterterms--via the QSR~\eqref{relation}. Surprisingly, for the BHs~\eqref{BHs}, a direct calculation of the entropy and magnetic susceptibility $\chi_B\equiv M_B/B$ yields that
\begin{equation}
-\left(\frac{\partial w_{QSR}}{\partial T }\right)_{B,\mu} \neq s,\,\, -\frac{1}{B}\left(\frac{\partial w_{QSR}}{\partial B }\right)_{T,\mu} \neq \chi_B^{hydro} ,
\end{equation}
with $s$ the Bekenstein-Hawking entropy and $\chi_B^{hydro}$ the magnetic susceptibility from hydrodynamics \cite{Kovtun:2016lfw,Jensen:2013kka,Ammon:2017ded}. All technical details can be found in the Supplemental Material~\cite{ourSMs}. As shown in Fig.~\ref{fig:check0}, the deviations are systematic and significant across the entire parameter space.

The correlated deviations point to a missing contribution to the thermodynamic potential: $w_{QSR}$ fails to reproduce both entropy and magnetic susceptibility, two independent thermodynamic derivatives. Their simultaneous failure is unlikely to arise from a numerical artifact or a quantity-specific error and instead signals an omitted free-energy contribution. The standard Euclidean action underlying $w_{QSR}$ already contains all allowed local bulk terms and the standard boundary counterterms at the AdS boundary, $(r=0)$. The discrepancy therefore points to a contribution nonlocal in the boundary data---a radial bulk integral. The CS interaction, being topological, is a natural candidate for generating such a nonlocal contribution in the presence of magnetic flux. Below, we identify this term, derive it independently in two ways, and show that it restores both the first law and the hydrodynamic relation.

\textbf{Identifying the Missing Contribution.}--The preceding analysis leads to a clear conclusion: the standard QSR~\eqref{relation} for the CS-coupled theory systematically misses an intrinsic contribution to the free energy. What is this contribution, and where does it originate? To answer these questions, we turn to two independent first-principles methods: direct variation of the Euclidean action with respect to boundary data~\cite{Gauntlett:2009bh,Donos:2012wi} and the Iyer–Wald Noether charge formalism~\cite{Wald:1993nt,Iyer:1994ys}. Remarkably, both methods yield the same correction, which we now present. The technical details are provided in the Supplemental Material~\cite{ourSMs}.

The missing term is indeed a nonlocal radial integral,
\begin{equation}\label{eqQcs}
Q_{cs}=\frac{B}{6}\int_0^{r_h} (A_z' A_t- A_z A_t') dr\,.
\end{equation}
This object is precisely what the conventional Euclidean action fails to capture. Including it in the free energy,
\begin{equation} \label{eq:free}
w=w_{QSR}+k Q_{cs} \,,
\end{equation}
we obtain the complete thermodynamic potential. The transformation is not ad hoc: Eq.~\eqref{eq:free} follows rigorously from both the Euclidean variation and the Iyer–Wald procedure. With this inclusion, we exactly restore the standard thermodynamic relations given in textbook~\cite{Landau:1995}:
\begin{equation} 
\begin{split}
w =& \epsilon-Ts-\mu \rho \,, \\  
\delta w =&-s\delta T -\rho \delta\mu -M_B \delta B\,,
\end{split}
\end{equation}
as well as the hydrodynamic susceptibility~\cite{Kovtun:2016lfw,Jensen:2013kka,Ammon:2017ded}. The numerical verification is shown in Fig.~\ref{fig:check1}. The entropy from $-\partial w/\partial T$ now agrees precisely with the Bekenstein–Hawking area law (top panel), and the magnetization from $-\partial w/\partial B$ exactly matches the hydrodynamic prediction (bottom panel).

Two features of $Q_{cs}$ deserve emphasis. First, it is non-local in the radial direction: it is an integral over the entire bulk from the horizon to the boundary. This non-locality distinguishes it from conventional thermodynamic quantities, which are often determined by local data at the horizon or at infinity. Second, its derivation from two entirely independent methods provides strong evidence that it is a genuine component of the on‑shell action, not an artifact of a particular computational scheme.

\textbf{Gauge Invariance and the Transgression Completion.}--The identification of $Q_{cs}$ by two independent methods raises a fundamental question: why does the conventional Euclidean action fail to capture this contribution? The key observation is that, under a $U(1)$ gauge transformation, the CS form $\mathcal{C}_5(A)=A\wedge F\wedge F$ varies by a total derivative whose integral need not vanish in the presence of magnetic flux. The action should therefore be reformulated in terms of a genuinely gauge-invariant object.

Interestingly, the relevant construction is well known in mathematics but, as we show here, is essential for BH thermodynamics: the local CS representative must be promoted to a transgression form~\cite{Nakahara:2003nw,Mora:2006ka}
\begin{equation}\label{T5}
\mathcal T_{5} =\mathcal C_{5}(A) -\mathcal C_{5}(\bar A) -d\mathcal B_{4}(A,\bar A)\,,
\end{equation}
where $\bar A$ is a reference gauge potential carrying the same magnetic flux as $A$, with $\bar F=d\bar A$, and $\mathcal B_4(A,\bar A)=A\wedge\bar A\wedge(F+\bar F)$. By construction, $\mathcal T_{5}$ is strictly invariant under simultaneous gauge transformations of $A$ and $\bar{A}$. Accordingly, replacing the CS term $S_{CS}=\frac{k}{6}\int\mathcal C_5(A)$ in the action~\eqref{action} with $S_{\mathcal T_5}=\frac{k}{6}\int\mathcal T_5$ and evaluating the corresponding free energy density $w_{\mathcal T_5}$, we obtain
\begin{equation}\label{wT5}
 w_{\mathcal T_5}=w_{QSR}+k Q_{cs}=w \,,
\end{equation}
where $\bar{A}=\frac{B}{2}(x dy-ydx)$. Thus, $Q_{cs}$ is not an ad hoc completion; it is precisely the finite global contribution required to complete the local CS representative into the gauge-invariant transgression form. The transgression form~\eqref{T5} automatically generates the previously missing contribution, thereby establishing the correct action principle for the thermodynamics of CS-coupled theories.

More generally, the mechanism described above extends naturally to CS interactions of the form  $``A\wedge F^n"$ in $D=2n+1$ ($n\geq 2$) spacetime dimensions, for which the transgression form reads~\cite{Nakahara:2003nw,Mora:2006ka}
\begin{equation}\label{Tform}
\mathcal T_{2n+1}(A,\bar{A})=\mathcal C_{2n+1}(A) -\mathcal C_{2n+1}(\bar A) -d\mathcal B_{2n}(A,\bar A)\,.
\end{equation}
Here $\mathcal C_{2n+1}(A)\equiv A\wedge F^{\,n}$, and $\mathcal B_{2n}(A,\bar A)=A \wedge \bar A\wedge \sum_{j=0}^{n-1} F^{\,n-1-j}\wedge\bar F^{\,j}$. We next test whether the transgression completion extends beyond the simple planar dyonic solution~\eqref{BHs}.

\textbf{Universality.--}Having established that the transgression completion is necessary for thermodynamic consistency in the planar dyonic case, we now demonstrate that this mechanism extends far beyond this specific background. To this end, we analyze five additional classes of BH backgrounds spanning distinct symmetries, horizon topologies, matter sectors, higher-curvature corrections, and spacetime dimensions: (i) 5D helical BHs with Bianchi VII$_0$ symmetry; (ii) 5D rotating BHs with squashed $S^3$ horizons; (iii) scalarized BHs with arbitrary scalar couplings; (iv) Gauss–Bonnet-corrected BHs; and (v) BHs in arbitrary odd spacetime dimensions, $D=2n+1$. Across all backgrounds, thermodynamic consistency requires the same transgression completion, establishing it as a generic feature of CS-type interactions. Details of these cases are given in the End Matter.

More specifically, the standard QSR yields a free energy that violates both the first law and the hydrodynamic relation, whereas the transgression form~\eqref{Tform} restores both exactly. Across these backgrounds, $Q_{cs}$ retains the same radial-integral structure, up to an overall prefactor. For instance, in the 7D background with two independent magnetic fields $B_1$ and $B_2$, the correction reads
\begin{equation}
Q_{cs} = 12 B_1 B_2 \int_0^{r_h} \big( A_z' A_t - A_z A_t' \big) dr \,.
\end{equation}
This expression differs from its 5D counterpart~\eqref{eqQcs} only in the overall prefactor.

This universality demonstrates that the transgression completion is not an artifact of a particular solution but a generic feature of CS-type interactions. Whenever a CS term of the form $A\wedge F^n$ appears in an odd-dimensional spacetime with $D=2n+1$, strict gauge invariance demands the transgression form as the correct action. The QSR~\eqref{relation}, though valid in principle, must be applied to this gauge-invariant action rather than the gauge-variant representative.

It is instructive to place our result in the broader context of transgression forms in gravitational physics. In pure CS AdS gravity, the transgression form renders the Euclidean BH action finite and provides a well-defined action principle, resolving both problems ``in one stroke just by requiring strict gauge invariance"~\cite{Mora:2006ka}. Subsequent work showed that transgression forms automatically supply the boundary terms required for finite conserved charges and correct thermodynamics, with implications for Lovelock gravity~\cite{Mora:2010gr,Mora:2014aca}. In these earlier investigations, the thermodynamic results obtained with and without the transgression form were identical; the transgression form served as a regularization tool rather than a correction to the thermodynamic potential itself. Our setting is qualitatively different. In EMCS theory with a dynamical $U(1)$ gauge field, the renormalized Euclidean action is already finite, yet its local CS representative yields, through the QSR, a thermodynamically inconsistent free energy. The transgression completion instead supplies the finite global contribution---$Q_{cs}$---required to restore consistency. Thus, the same mathematical construction addresses a distinct physical problem: not a divergence of the Euclidean action, but a finite mismatch in the thermodynamic potential. To our knowledge, this is the first example in which the transgression completion modifies a finite BH thermodynamic potential and is required for thermodynamic consistency.

\textbf{Conclusion.--}We have uncovered a universal nonlocal completion required for thermodynamic consistency in charged magnetic BHs in EMCS theory. The standard QSR~\eqref{relation}, applied to theories with CS interaction $A\wedge F\wedge F$, fails systematically: the resulting free energy violates both the first law of thermodynamics and the hydrodynamic relation. Two independent first-principles methods---direct variation of the Euclidean action and the Iyer–Wald formalism---identify the missing contribution $Q_{cs}$, a nonlocal radial integral whose inclusion restores exact thermodynamic consistency.

The significance of $Q_{cs}$, however, extends far beyond its role as a correction to the free energy. It is the finite global term that promotes the gauge-variant local CS term into the transgression form~\eqref{Tform}—a strictly gauge-invariant action. This reformulation is not merely a mathematical refinement; it is a physical requirement. In the present setting, the gauge dependence of the local CS action prevents it from defining a well-posed path integral, while the thermodynamic inconsistency shown in Fig.~\ref{fig:check0} is a direct physical manifestation of this obstruction. The correct starting point for the Euclidean path integral in CS-coupled theories is therefore the transgression action, not the local CS representative. The QSR itself remains valid, provided that it is evaluated using this gauge-invariant action.

The universality of this transgression completion across helical, rotating, scalarized, higher-derivative, and higher-dimensional BHs shows that this is not a solution-specific artifact, but a generic feature of CS-type interactions in odd-dimensional spacetimes. In four-dimensional theories with a $\theta F\wedge F$ term, no such completion is needed, as the term is itself gauge-invariant, underscoring that the phenomenon is tied to the nature of the CS form in odd dimensions. While our explicit demonstration has focused on the $U(1)$ gauge field, we anticipate that the same conclusion holds for non-Abelian CS terms: gauge invariance, as a fundamental organizing principle, likewise points to the transgression form as the appropriate action.

More broadly, our work establishes a general principle: BH thermodynamics in any theory with CS-type interactions requires the corresponding transgression completion. 
In particular, via the AdS/CFT correspondence, this transgression completion directly affects holographic predictions for systems exhibiting chiral anomalies, including their magnetic susceptibility and equation of state in external fields---quantities of central interest in quantum critical materials~\cite{DHoker:2012rlj,Zhao:2026boj,2003PhRvL91f6404Z,garst2005sign,wolf2014cooling,xiang2024giant} and strongly coupled QCD matter~\cite{Fukushima:2008xe,Son:2009tf,Kharzeev:2013ffa,Fukushima:2013zga,Ooguri:2010xs,Bali:2014kia}. Taken together, our results show that CS-type topological interactions can fundamentally alter the relation between the Euclidean action and the thermodynamic potential and establish the transgression form as the appropriate geometric framework for BH thermodynamics in such theories.

\vspace{.2 cm}
\textbf{Acknowledgments}--We thank Mirjam Cvetic, Elias Kiritsis, Hong Lu, Peng-Zhang He, Liang Ma, Yong Xiao, Amos Yarom and Hua Zhan for useful discussions. This work was partly supported by the National Natural Science Foundation of China Grants No.\,12525503, No.\,12247156, No.\,12505087, No.\,12588101 and No.\,12447101. We acknowledge the use of the High Performance Cluster at the Institute of Theoretical Physics, Chinese Academy of Sciences.

\section*{End Matter}

\textbf{Appendix A: Transgression form for $U(1)$ gauge field.}--Let $A$ be a $U(1)$ gauge potential and $\bar A$ a reference gauge potential, with $F=dA$ and $\bar F=d\bar A$. Defining $\mathfrak a=A-\bar A$, $A_s=\bar A+s\mathfrak a$, and $F_s=dA_s$, the transgression form $\mathcal{T}_{2n+1}(A,\bar A)$ in $D=2n+1$ dimensions is~\cite{Nakahara:2003nw,Mora:2006ka}
\begin{equation}  \label{eq:app-transgression}
\mathcal T_{2n+1} =(n+1)\int_0^1 ds\,\mathfrak a\wedge F_s^n \,.
\end{equation}
It satisfies the Chern-Weil relation $d\mathcal T_{2n+1}=F^{n+1}-\bar F^{n+1}$ and can equivalently be written as
\begin{equation}  \label{eq:app-transgression-cs}
 \mathcal T_{2n+1}=\mathcal C_{2n+1}(A)-\mathcal C_{2n+1}(\bar A) -d\mathcal B_{2n}(A,\bar A) \,,
\end{equation}
where $\mathcal C_{2n+1}(A)\equiv A\wedge F^{\,n}$, and $\mathcal B_{2n}(A,\bar A)= \mathfrak{a} \wedge\bar A\wedge \sum_{j=0}^{n-1} F^{\,n-1-j}\wedge\bar F^{\,j}=A \wedge\bar A\wedge \sum_{j=0}^{n-1} F^{\,n-1-j}\wedge\bar F^{\,j}$.

Under the simultaneous $U(1)$ transformations $A\to A+d\Lambda$ and $\bar A\to\bar A+d\Lambda$, $\mathfrak a$, $F$, and $\bar F$ remain unchanged; hence $\mathcal T_{2n+1}$ is strictly gauge invariant,~\emph{i.e.}, $\delta_\Lambda\mathcal T_{2n+1}=0$. By contrast, the local CS form is invariant only up to an exact differential, $\delta_\Lambda\mathcal C_{2n+1}=d(\Lambda F^n)$. The formal choice $\bar A=0$ recovers the local CS form $\mathcal{C}_{2n+1}$ whenever a global zero reference is admissible. In a background with nonzero magnetic flux, however, $\bar A$ should instead be chosen to carry the same magnetic flux as $A$. Accordingly, $\bar A$ is varied together with the magnetic fields, so that $\delta\bar A=\sum_i(\partial_{B_i}\bar A)\delta B_i$. As in the planar background~\eqref{BHs} discussed in the main text, natural choices for $\bar A$, for example, in the helical~\eqref{BH-helical}, rotating~\eqref{BH-spherical}, and seven-dimensional~\eqref{BH-7D} backgrounds presented below are $\bar A_{  hel}=\frac{B}{2}(x dy-y dx)$, $\bar A_{  rot}=\frac{B}{4}\cos\theta\,d\varphi$, and $\bar A_{7{  D}}=\frac{B_1}{2}(x_1\,dx_2-x_2\,dx_1)+\frac{B_2}{2}(x_3\,dx_4-x_4\,dx_3)$, respectively.

\textbf{Appendix B: Additional BH backgrounds.}--Here we summarize five additional classes of BH backgrounds used to establish the universality of $Q_{cs}$. These backgrounds span distinct symmetries, horizon topologies, matter sectors, higher-curvature corrections, and spacetime dimensions. In each case, $Q_{cs}$ retains the same radial-integral structure, up to a background-dependent overall prefactor, and restores thermodynamic consistency. Complete thermodynamic analysis is provided in the Supplemental Material~\cite{ourSMs}.

\textit{(i) 5D Helical BH:} This solution possesses Bianchi VII$_0$ symmetry, with the invariant one-forms $\omega_i$ defined by $\omega_1+i\omega_2 =e^{i\lambda_L z}(dx+idy)$ and $\omega_3=dz$, where $\lambda_L$ is the pitch of the helical lattice. Such inhomogeneous BHs of~\eqref{action} read~\cite{Ammon:2016szz}
\begin{eqnarray}\label{BH-helical}
\begin{split}
ds^2&=\frac{1}{r^2}\Big[- \big(f e^{-\chi}-h^2p^2\big)dt^2+2ph^2 dtdz +h^2 dz^2 \\
&\quad +\frac{\omega_2^2}{u^2} +u^{2}\left( \omega_1 +\alpha dz+\beta dt\right)^2+\frac{dr^2}{f}\Big]\,,  \\
A&=A_t dt  +\frac{B}{2}(x dy-ydx) +b\, \omega_1-A_z dz \,,
\end{split}
\end{eqnarray}
where $f,\, \chi,\, h,\, p,\, u,\, \alpha,\, \beta,\, A_t,\, b$ and $A_z$ are functions of $r$. The free energy is $w=\epsilon-Ts-\mu \rho =w_{QSR}+k Q_{cs}$ with $Q_{cs}$ exactly the same as~\eqref{eqQcs}, although the helical BH~\eqref{BH-helical} is much more complicated than the prototypical one~\eqref{BHs}. Moreover, the first law of thermodynamics takes
\begin{equation} \label{BH-helical-1st} 
\begin{split}
\delta w=-s\delta T -\rho \delta\mu - \langle J^{b}\rangle \delta\mu_b -M_B\delta B \,.
\end{split}
\end{equation}
Here, $\mu_b$ and $\langle J^b\rangle \equiv\sqrt{\langle J^x\rangle^2+\langle J^y\rangle^2}$ are the source and expectation value of the helical current, respectively. In this more intricate case, the CS term enters explicitly into the expressions for $\rho$ and $\langle J^b\rangle$, thereby producing a significant effect on the BH thermodynamics.

\textit{(ii) 5D rotating BH:} This BH is a solution of~\eqref{action} with equal angular momenta ($\mathcal{J}_1 = \mathcal{J}_2 = \mathcal{J}$), thereby exhibiting a local $SU(2) \times U(1)$ symmetry. The corresponding ansatz is
\begin{eqnarray} \begin{split} \label{BH-spherical}
ds^2 &=-F_0 dt^2 +\frac{dr^2}{F_1} +\frac{F_2}{4} \left(\sigma_1^2+\sigma_2^2\right) +\frac{F_3}{4} \left( \sigma_3 +2p dt \right)^2 \,,  \\
A &= A_t dt + \frac{B\cos{\theta}}{4} d\phi  - \frac{A_z}{2} \sigma_3 \,,
\end{split}\end{eqnarray}
where $F_i (i=0,\cdots,3),\, p,\, A_t$ and $A_z$ depend on $r$. The $\sigma_1= \cos{\psi} d\theta+ \sin\psi \sin\theta d\phi, \sigma_2= -\sin{\psi} d\theta+ \cos\psi \sin\theta d\phi, \sigma_3= d \psi + \cos \theta \, d \phi$ denote left-invariant 1-forms on $SU(2)$ expressed in terms of Euler angles $(\theta,\phi,\psi)$. The rotating BH ansatz considered here is inspired by the configurations studied in~\cite{Chong:2005hr,Blazquez-Salcedo:2016rkj}, but is distinguished from them by the inclusion of an independent magnetic field. These BHs have an asymptotically AdS boundary with a squashed $S^3$ geometry parameterized by $v$. Since $L$ sets the physical scale of the spherical boundary, we retain it explicitly throughout the analysis. We obtain the free energy $w=\epsilon-Ts-\mu \rho -2\Omega_H \mathcal{J} =w_{QSR}+ \frac{1}{L^3v} k Q_{cs}$, where $Q_{cs}$ takes the same form as~\eqref{eqQcs} and $\Omega_H$ is the angular velocity. One also has the first law of thermodynamics
\begin{equation} \label{BH-spherical-1st}
\delta w = -s \delta T -\rho \delta\mu -2\mathcal{J}\delta\Omega_H -M_B \delta B \,,
\end{equation}
where the factor of 2 accounts for two independent rotational directions. Notably, the planar BH~\eqref{BHs} emerges precisely as a planar limit of the rotating BH~\eqref{BH-spherical}.

\textit{(iii) BH with scalar hair:} To demonstrate that $Q_{cs}$ is insensitive to the details of the theory, we extend the original action~\eqref{action} by adding a scalar field term $\mathcal{L}_\phi=-\frac{(\partial\phi)^2}{2} -V(\phi)$ and a scalar‑electromagnetic coupling $-\frac{1}{4}Z(\phi)F_{ab}F^{ab}$. As expected, the expression for $Q_{cs}$~\eqref{eqQcs} remains unchanged, even though the scalar hair significantly alters the detailed properties of the BH. Accordingly, the first law receives an additional contribution from the scalar field
\begin{equation} \label{BH-hair-1st}
\begin{split}
\delta w &=-s\delta T -\rho \delta\mu -\langle O\rangle \delta \phi_s -M_B \delta B \,,
\end{split}
\end{equation}
where $\phi_s$ and $\langle O\rangle$ are the source and expectation value of the scalar operator $O$ dual to $\phi$.

\textit{(iv) BH with higher-derivative terms:} We examine the impact of the Gauss–Bonnet term~\cite{Cai:2001dz,Brihaye:2008xu} on the thermodynamics of planar~\eqref{BHs} and spherical~\eqref{BH-spherical} BHs. It should be emphasized that, in this case, the entropy is now given by the Wald entropy $s_{\rm Wald}$~\cite{Iyer:1994ys}, and other thermodynamic quantities are also modified by this higher‑derivative term. Once again, when these modified thermodynamic quantities are combined, we obtain the same form of the thermodynamic law:
\begin{equation} \label{BH-GB-1st}
\begin{split}
\delta w &=-s_{\rm Wald}\delta T -\rho\delta\mu -M_B \delta B \,,
\end{split}
\end{equation}
in which the free energy must include the aforementioned $Q_{cs}$. We emphasize that similar higher-derivative corrections can also be incorporated into the other BHs discussed here, such as the helical and hairy cases, although the analysis is technically more involved.

\textit{(v) BHs in other dimensions:} All the BHs discussed above admit a natural generalization to Einstein–Maxwell theory with an ``$A\wedge F^n$" CS term in any odd dimension $D=2n+1$ ($n\geq2$). As a concrete illustration, we present the 7D ($n=3$) case
\begin{equation}
S=\int d^7x \sqrt{-g}\Big( R+\frac{30}{L^2} -\frac{F^2}{4} +\frac{k}{4} \epsilon^{abcdemn}A_a F_{bc} F_{de} F_{mn} \Big) \,,
\end{equation}
where magnetic fields can be turned on along distinct spatial directions. The background takes
\begin{eqnarray} \label{BH-7D}
\begin{split}
ds^2&=\frac{1}{r^2}\Big[- \big(f e^{-\chi}-h^2p^2\big)dt^2+2ph^2 dtdz +dx_1^2  \\
&\quad +dx_2^2 +u(dx_3^2+dx_4^2) +h^2 dz^2+\frac{dr^2}{f}\Big] \,,\\
A &= A_t dt +\frac{ B_1}{2} (x_1 dx_2 -x_2dx_1) \\
&\qquad +\frac{ B_2}{2}(x_3 dx_4- x_4dx_3) - A_z dz \,,
\end{split}
\end{eqnarray}
where $f,\, \chi,\, h,\, p,\, u,\, A_t$ and $A_z$ are functions of $r$. Following the same procedure, we obtain $w=\epsilon-Ts-\mu \rho =w_{QSR}+ k Q_{cs}$ where
\begin{equation}
Q_{cs} = 12 B_1 B_2 \int_0^{r_h} \big( A_z' A_t - A_z A_t' \big)\, dr\,.
\end{equation}
This expression retains the same functional form as~\eqref{eqQcs}, differing only by an overall factor. Moreover, we obtain the first law:
\begin{equation} \label{BH-7D-1st}
\begin{split}
\delta w =-s \delta T -\rho \delta\mu -M_{B_1} \delta B_1 -M_{B_2} \delta B_2 \,,
\end{split}
\end{equation}
where $M_{B_1}$ and $M_{B_2}$ denote the magnetizations associated with each magnetic field.

\bibliography{main}%

\begin{thebibliography}{55}%
\makeatletter
\providecommand \@ifxundefined [1]{%
 \@ifx{#1\undefined}
}%
\providecommand \@ifnum [1]{%
 \ifnum #1\expandafter \@firstoftwo
 \else \expandafter \@secondoftwo
 \fi
}%
\providecommand \@ifx [1]{%
 \ifx #1\expandafter \@firstoftwo
 \else \expandafter \@secondoftwo
 \fi
}%
\providecommand \natexlab [1]{#1}%
\providecommand \enquote  [1]{``#1''}%
\providecommand \bibnamefont  [1]{#1}%
\providecommand \bibfnamefont [1]{#1}%
\providecommand \citenamefont [1]{#1}%
\providecommand \href@noop [0]{\@secondoftwo}%
\providecommand \href [0]{\begingroup \@sanitize@url \@href}%
\providecommand \@href[1]{\@@startlink{#1}\@@href}%
\providecommand \@@href[1]{\endgroup#1\@@endlink}%
\providecommand \@sanitize@url [0]{\catcode `\\12\catcode `\$12\catcode `\&12\catcode `\#12\catcode `\^12\catcode `\_12\catcode `\%12\relax}%
\providecommand \@@startlink[1]{}%
\providecommand \@@endlink[0]{}%
\providecommand \url  [0]{\begingroup\@sanitize@url \@url }%
\providecommand \@url [1]{\endgroup\@href {#1}{\urlprefix }}%
\providecommand \urlprefix  [0]{URL }%
\providecommand \Eprint [0]{\href }%
\providecommand \doibase [0]{https://doi.org/}%
\providecommand \selectlanguage [0]{\@gobble}%
\providecommand \bibinfo  [0]{\@secondoftwo}%
\providecommand \bibfield  [0]{\@secondoftwo}%
\providecommand \translation [1]{[#1]}%
\providecommand \BibitemOpen [0]{}%
\providecommand \bibitemStop [0]{}%
\providecommand \bibitemNoStop [0]{.\EOS\space}%
\providecommand \EOS [0]{\spacefactor3000\relax}%
\providecommand \BibitemShut  [1]{\csname bibitem#1\endcsname}%
\let\auto@bib@innerbib\@empty
\bibitem [{\citenamefont {Wald}(1984)}]{Wald:1984rg}%
  \BibitemOpen
  \bibfield  {author} {\bibinfo {author} {\bibfnamefont {R.~M.}\ \bibnamefont {Wald}},\ }\href {https://doi.org/10.7208/chicago/9780226870373.001.0001} {\emph {\bibinfo {title} {{General Relativity}}}}\ (\bibinfo  {publisher} {Chicago Univ. Pr.},\ \bibinfo {address} {Chicago, USA},\ \bibinfo {year} {1984})\BibitemShut {NoStop}%
\bibitem [{\citenamefont {Gibbons}\ and\ \citenamefont {Hawking}(1977)}]{Gibbons:1976ue}%
  \BibitemOpen
  \bibfield  {author} {\bibinfo {author} {\bibfnamefont {G.~W.}\ \bibnamefont {Gibbons}}\ and\ \bibinfo {author} {\bibfnamefont {S.~W.}\ \bibnamefont {Hawking}},\ }\bibfield  {title} {\bibinfo {title} {{Action Integrals and Partition Functions in Quantum Gravity}},\ }\href {https://doi.org/10.1103/PhysRevD.15.2752} {\bibfield  {journal} {\bibinfo  {journal} {Phys. Rev. D}\ }\textbf {\bibinfo {volume} {15}},\ \bibinfo {pages} {2752} (\bibinfo {year} {1977})}\BibitemShut {NoStop}%
\bibitem [{\citenamefont {Gibbons}\ \emph {et~al.}(2005)\citenamefont {Gibbons}, \citenamefont {Perry},\ and\ \citenamefont {Pope}}]{Gibbons:2004ai}%
  \BibitemOpen
  \bibfield  {author} {\bibinfo {author} {\bibfnamefont {G.~W.}\ \bibnamefont {Gibbons}}, \bibinfo {author} {\bibfnamefont {M.~J.}\ \bibnamefont {Perry}},\ and\ \bibinfo {author} {\bibfnamefont {C.~N.}\ \bibnamefont {Pope}},\ }\bibfield  {title} {\bibinfo {title} {{The First law of thermodynamics for Kerr-anti-de Sitter black holes}},\ }\href {https://doi.org/10.1088/0264-9381/22/9/002} {\bibfield  {journal} {\bibinfo  {journal} {Class. Quant. Grav.}\ }\textbf {\bibinfo {volume} {22}},\ \bibinfo {pages} {1503} (\bibinfo {year} {2005})},\ \Eprint {https://arxiv.org/abs/hep-th/0408217} {arXiv:hep-th/0408217} \BibitemShut {NoStop}%
\bibitem [{\citenamefont {Astefanesei}\ and\ \citenamefont {Radu}(2006)}]{Astefanesei:2005ad}%
  \BibitemOpen
  \bibfield  {author} {\bibinfo {author} {\bibfnamefont {D.}~\bibnamefont {Astefanesei}}\ and\ \bibinfo {author} {\bibfnamefont {E.}~\bibnamefont {Radu}},\ }\bibfield  {title} {\bibinfo {title} {{Quasilocal formalism and black ring thermodynamics}},\ }\href {https://doi.org/10.1103/PhysRevD.73.044014} {\bibfield  {journal} {\bibinfo  {journal} {Phys. Rev. D}\ }\textbf {\bibinfo {volume} {73}},\ \bibinfo {pages} {044014} (\bibinfo {year} {2006})},\ \Eprint {https://arxiv.org/abs/hep-th/0509144} {arXiv:hep-th/0509144} \BibitemShut {NoStop}%
\bibitem [{\citenamefont {Copsey}\ and\ \citenamefont {Horowitz}(2006)}]{Copsey:2005se}%
  \BibitemOpen
  \bibfield  {author} {\bibinfo {author} {\bibfnamefont {K.}~\bibnamefont {Copsey}}\ and\ \bibinfo {author} {\bibfnamefont {G.~T.}\ \bibnamefont {Horowitz}},\ }\bibfield  {title} {\bibinfo {title} {{The Role of dipole charges in black hole thermodynamics}},\ }\href {https://doi.org/10.1103/PhysRevD.73.024015} {\bibfield  {journal} {\bibinfo  {journal} {Phys. Rev. D}\ }\textbf {\bibinfo {volume} {73}},\ \bibinfo {pages} {024015} (\bibinfo {year} {2006})},\ \Eprint {https://arxiv.org/abs/hep-th/0505278} {arXiv:hep-th/0505278} \BibitemShut {NoStop}%
\bibitem [{\citenamefont {Dunne}(1998)}]{Dunne:1998qy}%
  \BibitemOpen
  \bibfield  {author} {\bibinfo {author} {\bibfnamefont {G.~V.}\ \bibnamefont {Dunne}},\ }\bibfield  {title} {\bibinfo {title} {{Aspects of Chern-Simons theory}},\ }in\ \href@noop {} {\emph {\bibinfo {booktitle} {{Les Houches Summer School in Theoretical Physics, Session 69: Topological Aspects of Low-dimensional Systems}}}}\ (\bibinfo {year} {1998})\ \Eprint {https://arxiv.org/abs/hep-th/9902115} {arXiv:hep-th/9902115} \BibitemShut {NoStop}%
\bibitem [{\citenamefont {Deser}(1998)}]{Deser:1998mk}%
  \BibitemOpen
  \bibfield  {author} {\bibinfo {author} {\bibfnamefont {S.}~\bibnamefont {Deser}},\ }\bibfield  {title} {\bibinfo {title} {{Chern-Simons terms as an example of the relations between mathematics and physics}},\ }\href@noop {} {\bibfield  {journal} {\bibinfo  {journal} {unpublished}\ } (\bibinfo {year} {1998})},\ \Eprint {https://arxiv.org/abs/math-ph/9805020} {arXiv:math-ph/9805020} \BibitemShut {NoStop}%
\bibitem [{\citenamefont {Cremmer}\ \emph {et~al.}(1978)\citenamefont {Cremmer}, \citenamefont {Julia},\ and\ \citenamefont {Scherk}}]{Cremmer:1978km}%
  \BibitemOpen
  \bibfield  {author} {\bibinfo {author} {\bibfnamefont {E.}~\bibnamefont {Cremmer}}, \bibinfo {author} {\bibfnamefont {B.}~\bibnamefont {Julia}},\ and\ \bibinfo {author} {\bibfnamefont {J.}~\bibnamefont {Scherk}},\ }\bibfield  {title} {\bibinfo {title} {{Supergravity Theory in 11 Dimensions}},\ }\href {https://doi.org/10.1016/0370-2693(78)90894-8} {\bibfield  {journal} {\bibinfo  {journal} {Phys. Lett. B}\ }\textbf {\bibinfo {volume} {76}},\ \bibinfo {pages} {409} (\bibinfo {year} {1978})}\BibitemShut {NoStop}%
\bibitem [{\citenamefont {Chamblin}\ \emph {et~al.}(1999)\citenamefont {Chamblin}, \citenamefont {Emparan}, \citenamefont {Johnson},\ and\ \citenamefont {Myers}}]{Chamblin:1999tk}%
  \BibitemOpen
  \bibfield  {author} {\bibinfo {author} {\bibfnamefont {A.}~\bibnamefont {Chamblin}}, \bibinfo {author} {\bibfnamefont {R.}~\bibnamefont {Emparan}}, \bibinfo {author} {\bibfnamefont {C.~V.}\ \bibnamefont {Johnson}},\ and\ \bibinfo {author} {\bibfnamefont {R.~C.}\ \bibnamefont {Myers}},\ }\bibfield  {title} {\bibinfo {title} {{Charged AdS black holes and catastrophic holography}},\ }\href {https://doi.org/10.1103/PhysRevD.60.064018} {\bibfield  {journal} {\bibinfo  {journal} {Phys. Rev. D}\ }\textbf {\bibinfo {volume} {60}},\ \bibinfo {pages} {064018} (\bibinfo {year} {1999})},\ \Eprint {https://arxiv.org/abs/hep-th/9902170} {arXiv:hep-th/9902170} \BibitemShut {NoStop}%
\bibitem [{\citenamefont {Witten}(1998)}]{Witten:1998qj}%
  \BibitemOpen
  \bibfield  {author} {\bibinfo {author} {\bibfnamefont {E.}~\bibnamefont {Witten}},\ }\bibfield  {title} {\bibinfo {title} {{Anti de Sitter space and holography}},\ }\href {https://doi.org/10.4310/ATMP.1998.v2.n2.a2} {\bibfield  {journal} {\bibinfo  {journal} {Adv. Theor. Math. Phys.}\ }\textbf {\bibinfo {volume} {2}},\ \bibinfo {pages} {253} (\bibinfo {year} {1998})},\ \Eprint {https://arxiv.org/abs/hep-th/9802150} {arXiv:hep-th/9802150} \BibitemShut {NoStop}%
\bibitem [{\citenamefont {Gauntlett}\ \emph {et~al.}(2010)\citenamefont {Gauntlett}, \citenamefont {Sonner},\ and\ \citenamefont {Wiseman}}]{Gauntlett:2009bh}%
  \BibitemOpen
  \bibfield  {author} {\bibinfo {author} {\bibfnamefont {J.~P.}\ \bibnamefont {Gauntlett}}, \bibinfo {author} {\bibfnamefont {J.}~\bibnamefont {Sonner}},\ and\ \bibinfo {author} {\bibfnamefont {T.}~\bibnamefont {Wiseman}},\ }\bibfield  {title} {\bibinfo {title} {{Quantum Criticality and Holographic Superconductors in M-theory}},\ }\href {https://doi.org/10.1007/JHEP02(2010)060} {\bibfield  {journal} {\bibinfo  {journal} {JHEP}\ }\textbf {\bibinfo {volume} {02}},\ \bibinfo {pages} {060}},\ \Eprint {https://arxiv.org/abs/0912.0512} {arXiv:0912.0512 [hep-th]} \BibitemShut {NoStop}%
\bibitem [{\citenamefont {Donos}\ and\ \citenamefont {Gauntlett}(2012)}]{Donos:2012wi}%
  \BibitemOpen
  \bibfield  {author} {\bibinfo {author} {\bibfnamefont {A.}~\bibnamefont {Donos}}\ and\ \bibinfo {author} {\bibfnamefont {J.~P.}\ \bibnamefont {Gauntlett}},\ }\bibfield  {title} {\bibinfo {title} {{Black holes dual to helical current phases}},\ }\href {https://doi.org/10.1103/PhysRevD.86.064010} {\bibfield  {journal} {\bibinfo  {journal} {Phys. Rev. D}\ }\textbf {\bibinfo {volume} {86}},\ \bibinfo {pages} {064010} (\bibinfo {year} {2012})},\ \Eprint {https://arxiv.org/abs/1204.1734} {arXiv:1204.1734 [hep-th]} \BibitemShut {NoStop}%
\bibitem [{\citenamefont {Wald}(1993)}]{Wald:1993nt}%
  \BibitemOpen
  \bibfield  {author} {\bibinfo {author} {\bibfnamefont {R.~M.}\ \bibnamefont {Wald}},\ }\bibfield  {title} {\bibinfo {title} {{Black hole entropy is the Noether charge}},\ }\href {https://doi.org/10.1103/PhysRevD.48.R3427} {\bibfield  {journal} {\bibinfo  {journal} {Phys. Rev. D}\ }\textbf {\bibinfo {volume} {48}},\ \bibinfo {pages} {R3427} (\bibinfo {year} {1993})},\ \Eprint {https://arxiv.org/abs/gr-qc/9307038} {arXiv:gr-qc/9307038} \BibitemShut {NoStop}%
\bibitem [{\citenamefont {Iyer}\ and\ \citenamefont {Wald}(1994)}]{Iyer:1994ys}%
  \BibitemOpen
  \bibfield  {author} {\bibinfo {author} {\bibfnamefont {V.}~\bibnamefont {Iyer}}\ and\ \bibinfo {author} {\bibfnamefont {R.~M.}\ \bibnamefont {Wald}},\ }\bibfield  {title} {\bibinfo {title} {{Some properties of Noether charge and a proposal for dynamical black hole entropy}},\ }\href {https://doi.org/10.1103/PhysRevD.50.846} {\bibfield  {journal} {\bibinfo  {journal} {Phys. Rev. D}\ }\textbf {\bibinfo {volume} {50}},\ \bibinfo {pages} {846} (\bibinfo {year} {1994})},\ \Eprint {https://arxiv.org/abs/gr-qc/9403028} {arXiv:gr-qc/9403028} \BibitemShut {NoStop}%
\bibitem [{\citenamefont {Li}(2021)}]{Li:2020spf}%
  \BibitemOpen
  \bibfield  {author} {\bibinfo {author} {\bibfnamefont {L.}~\bibnamefont {Li}},\ }\bibfield  {title} {\bibinfo {title} {{On Thermodynamics of AdS Black Holes with Scalar Hair}},\ }\href {https://doi.org/10.1016/j.physletb.2021.136123} {\bibfield  {journal} {\bibinfo  {journal} {Phys. Lett. B}\ }\textbf {\bibinfo {volume} {815}},\ \bibinfo {pages} {136123} (\bibinfo {year} {2021})},\ \Eprint {https://arxiv.org/abs/2008.05597} {arXiv:2008.05597 [gr-qc]} \BibitemShut {NoStop}%
\bibitem [{\citenamefont {Xiao}\ \emph {et~al.}(2024)\citenamefont {Xiao}, \citenamefont {Tian},\ and\ \citenamefont {Liu}}]{Xiao:2023lap}%
  \BibitemOpen
  \bibfield  {author} {\bibinfo {author} {\bibfnamefont {Y.}~\bibnamefont {Xiao}}, \bibinfo {author} {\bibfnamefont {Y.}~\bibnamefont {Tian}},\ and\ \bibinfo {author} {\bibfnamefont {Y.-X.}\ \bibnamefont {Liu}},\ }\bibfield  {title} {\bibinfo {title} {{Extended Black Hole Thermodynamics from Extended Iyer-Wald Formalism}},\ }\href {https://doi.org/10.1103/PhysRevLett.132.021401} {\bibfield  {journal} {\bibinfo  {journal} {Phys. Rev. Lett.}\ }\textbf {\bibinfo {volume} {132}},\ \bibinfo {pages} {021401} (\bibinfo {year} {2024})},\ \Eprint {https://arxiv.org/abs/2308.12630} {arXiv:2308.12630 [gr-qc]} \BibitemShut {NoStop}%
\bibitem [{\citenamefont {Nakahara}(2003)}]{Nakahara:2003nw}%
  \BibitemOpen
  \bibfield  {author} {\bibinfo {author} {\bibfnamefont {M.}~\bibnamefont {Nakahara}},\ }\href@noop {} {\emph {\bibinfo {title} {{Geometry, topology and physics}}}}\ (\bibinfo {year} {2003})\BibitemShut {NoStop}%
\bibitem [{\citenamefont {Mora}\ \emph {et~al.}(2004)\citenamefont {Mora}, \citenamefont {Olea}, \citenamefont {Troncoso},\ and\ \citenamefont {Zanelli}}]{Mora:2004kb}%
  \BibitemOpen
  \bibfield  {author} {\bibinfo {author} {\bibfnamefont {P.}~\bibnamefont {Mora}}, \bibinfo {author} {\bibfnamefont {R.}~\bibnamefont {Olea}}, \bibinfo {author} {\bibfnamefont {R.}~\bibnamefont {Troncoso}},\ and\ \bibinfo {author} {\bibfnamefont {J.}~\bibnamefont {Zanelli}},\ }\bibfield  {title} {\bibinfo {title} {{Finite action principle for Chern-Simons AdS gravity}},\ }\href {https://doi.org/10.1088/1126-6708/2004/06/036} {\bibfield  {journal} {\bibinfo  {journal} {JHEP}\ }\textbf {\bibinfo {volume} {06}},\ \bibinfo {pages} {036}},\ \Eprint {https://arxiv.org/abs/hep-th/0405267} {arXiv:hep-th/0405267} \BibitemShut {NoStop}%
\bibitem [{\citenamefont {Mora}\ \emph {et~al.}(2006)\citenamefont {Mora}, \citenamefont {Olea}, \citenamefont {Troncoso},\ and\ \citenamefont {Zanelli}}]{Mora:2006ka}%
  \BibitemOpen
  \bibfield  {author} {\bibinfo {author} {\bibfnamefont {P.}~\bibnamefont {Mora}}, \bibinfo {author} {\bibfnamefont {R.}~\bibnamefont {Olea}}, \bibinfo {author} {\bibfnamefont {R.}~\bibnamefont {Troncoso}},\ and\ \bibinfo {author} {\bibfnamefont {J.}~\bibnamefont {Zanelli}},\ }\bibfield  {title} {\bibinfo {title} {{Transgression forms and extensions of Chern-Simons gauge theories}},\ }\href {https://doi.org/10.1088/1126-6708/2006/02/067} {\bibfield  {journal} {\bibinfo  {journal} {JHEP}\ }\textbf {\bibinfo {volume} {02}},\ \bibinfo {pages} {067}},\ \Eprint {https://arxiv.org/abs/hep-th/0601081} {arXiv:hep-th/0601081} \BibitemShut {NoStop}%
\bibitem [{\citenamefont {Pais}\ \emph {et~al.}(2023)\citenamefont {Pais}, \citenamefont {Salgado-Rebolledo},\ and\ \citenamefont {Vera}}]{Pais:2023zqz}%
  \BibitemOpen
  \bibfield  {author} {\bibinfo {author} {\bibfnamefont {P.}~\bibnamefont {Pais}}, \bibinfo {author} {\bibfnamefont {P.}~\bibnamefont {Salgado-Rebolledo}},\ and\ \bibinfo {author} {\bibfnamefont {A.}~\bibnamefont {Vera}},\ }\bibfield  {title} {\bibinfo {title} {{A note on the Hamiltonian structure of transgression forms}},\ }\href {https://doi.org/10.1007/JHEP12(2023)190} {\bibfield  {journal} {\bibinfo  {journal} {JHEP}\ }\textbf {\bibinfo {volume} {12}},\ \bibinfo {pages} {190}},\ \Eprint {https://arxiv.org/abs/2309.16760} {arXiv:2309.16760 [hep-th]} \BibitemShut {NoStop}%
\bibitem [{\citenamefont {Mora}(2010)}]{Mora:2010gr}%
  \BibitemOpen
  \bibfield  {author} {\bibinfo {author} {\bibfnamefont {P.}~\bibnamefont {Mora}},\ }\bibfield  {title} {\bibinfo {title} {{Transgressions and Holographic Conformal Anomalies for Chern-Simons Gravities}},\ }\href@noop {} {\bibfield  {journal} {\bibinfo  {journal} {arXiv preprint}\ } (\bibinfo {year} {2010})},\ \Eprint {https://arxiv.org/abs/1010.5110} {arXiv:1010.5110 [hep-th]} \BibitemShut {NoStop}%
\bibitem [{\citenamefont {Mora}(2015)}]{Mora:2014fba}%
  \BibitemOpen
  \bibfield  {author} {\bibinfo {author} {\bibfnamefont {P.}~\bibnamefont {Mora}},\ }\bibfield  {title} {\bibinfo {title} {{Gauge Symmetries and Holographic Anomalies of Chern-Simons and Transgression AdS Gravity}},\ }\href {https://doi.org/10.1007/JHEP04(2015)090} {\bibfield  {journal} {\bibinfo  {journal} {JHEP}\ }\textbf {\bibinfo {volume} {04}},\ \bibinfo {pages} {090}},\ \Eprint {https://arxiv.org/abs/1408.1436} {arXiv:1408.1436 [hep-th]} \BibitemShut {NoStop}%
\bibitem [{\citenamefont {Buchel}\ and\ \citenamefont {Liu}(2007)}]{Buchel:2006gb}%
  \BibitemOpen
  \bibfield  {author} {\bibinfo {author} {\bibfnamefont {A.}~\bibnamefont {Buchel}}\ and\ \bibinfo {author} {\bibfnamefont {J.~T.}\ \bibnamefont {Liu}},\ }\bibfield  {title} {\bibinfo {title} {{Gauged supergravity from type IIB string theory on Y**p,q manifolds}},\ }\href {https://doi.org/10.1016/j.nuclphysb.2007.03.001} {\bibfield  {journal} {\bibinfo  {journal} {Nucl. Phys. B}\ }\textbf {\bibinfo {volume} {771}},\ \bibinfo {pages} {93} (\bibinfo {year} {2007})},\ \Eprint {https://arxiv.org/abs/hep-th/0608002} {arXiv:hep-th/0608002} \BibitemShut {NoStop}%
\bibitem [{\citenamefont {Gauntlett}\ \emph {et~al.}(2007)\citenamefont {Gauntlett}, \citenamefont {O~Colgain},\ and\ \citenamefont {Varela}}]{Gauntlett:2006ai}%
  \BibitemOpen
  \bibfield  {author} {\bibinfo {author} {\bibfnamefont {J.~P.}\ \bibnamefont {Gauntlett}}, \bibinfo {author} {\bibfnamefont {E.}~\bibnamefont {O~Colgain}},\ and\ \bibinfo {author} {\bibfnamefont {O.}~\bibnamefont {Varela}},\ }\bibfield  {title} {\bibinfo {title} {{Properties of some conformal field theories with M-theory duals}},\ }\href {https://doi.org/10.1088/1126-6708/2007/02/049} {\bibfield  {journal} {\bibinfo  {journal} {JHEP}\ }\textbf {\bibinfo {volume} {02}},\ \bibinfo {pages} {049}},\ \Eprint {https://arxiv.org/abs/hep-th/0611219} {arXiv:hep-th/0611219} \BibitemShut {NoStop}%
\bibitem [{\citenamefont {D'Hoker}\ and\ \citenamefont {Kraus}(2010{\natexlab{a}})}]{DHoker:2010zpp}%
  \BibitemOpen
  \bibfield  {author} {\bibinfo {author} {\bibfnamefont {E.}~\bibnamefont {D'Hoker}}\ and\ \bibinfo {author} {\bibfnamefont {P.}~\bibnamefont {Kraus}},\ }\bibfield  {title} {\bibinfo {title} {{Holographic Metamagnetism, Quantum Criticality, and Crossover Behavior}},\ }\href {https://doi.org/10.1007/JHEP05(2010)083} {\bibfield  {journal} {\bibinfo  {journal} {JHEP}\ }\textbf {\bibinfo {volume} {05}},\ \bibinfo {pages} {083}},\ \Eprint {https://arxiv.org/abs/1003.1302} {arXiv:1003.1302 [hep-th]} \BibitemShut {NoStop}%
\bibitem [{\citenamefont {D'Hoker}\ and\ \citenamefont {Kraus}(2010{\natexlab{b}})}]{DHoker:2010onp}%
  \BibitemOpen
  \bibfield  {author} {\bibinfo {author} {\bibfnamefont {E.}~\bibnamefont {D'Hoker}}\ and\ \bibinfo {author} {\bibfnamefont {P.}~\bibnamefont {Kraus}},\ }\bibfield  {title} {\bibinfo {title} {{Magnetic Field Induced Quantum Criticality via new Asymptotically AdS$_{5}$ Solutions}},\ }\href {https://doi.org/10.1088/0264-9381/27/21/215022} {\bibfield  {journal} {\bibinfo  {journal} {Class. Quant. Grav.}\ }\textbf {\bibinfo {volume} {27}},\ \bibinfo {pages} {215022} (\bibinfo {year} {2010}{\natexlab{b}})},\ \Eprint {https://arxiv.org/abs/1006.2573} {arXiv:1006.2573 [hep-th]} \BibitemShut {NoStop}%
\bibitem [{\citenamefont {Landau}\ \emph {et~al.}(1995)\citenamefont {Landau}, \citenamefont {Lifshitz},\ and\ \citenamefont {Pitaevskii}}]{Landau:1995}%
  \BibitemOpen
  \bibfield  {author} {\bibinfo {author} {\bibfnamefont {L.~D.}\ \bibnamefont {Landau}}, \bibinfo {author} {\bibfnamefont {E.}~\bibnamefont {Lifshitz}},\ and\ \bibinfo {author} {\bibfnamefont {L.~P.}\ \bibnamefont {Pitaevskii}},\ }\href@noop {} {\emph {\bibinfo {title} {{Electrodynamics of Continuous Media,}}}}\ (\bibinfo  {publisher} {Butterworth-Heinemann},\ \bibinfo {year} {1995})\BibitemShut {NoStop}%
\bibitem [{dfe()}]{dfenergy}%
  \BibitemOpen
  \href@noop {} {}\bibinfo {note} {Alternatively, we can define a total energy $\epsilon^{tot}=\epsilon+M_BB$. Then, the first law of thermodynamics is given by $\delta \epsilon^{tot}=T\delta s +\mu\delta \rho +B\delta M_B+\cdots$.}\BibitemShut {Stop}%
\bibitem [{\citenamefont {Kovtun}(2016)}]{Kovtun:2016lfw}%
  \BibitemOpen
  \bibfield  {author} {\bibinfo {author} {\bibfnamefont {P.}~\bibnamefont {Kovtun}},\ }\bibfield  {title} {\bibinfo {title} {{Thermodynamics of polarized relativistic matter}},\ }\href {https://doi.org/10.1007/JHEP07(2016)028} {\bibfield  {journal} {\bibinfo  {journal} {JHEP}\ }\textbf {\bibinfo {volume} {07}},\ \bibinfo {pages} {028}},\ \Eprint {https://arxiv.org/abs/1606.01226} {arXiv:1606.01226 [hep-th]} \BibitemShut {NoStop}%
\bibitem [{\citenamefont {Jensen}\ \emph {et~al.}(2014)\citenamefont {Jensen}, \citenamefont {Loganayagam},\ and\ \citenamefont {Yarom}}]{Jensen:2013kka}%
  \BibitemOpen
  \bibfield  {author} {\bibinfo {author} {\bibfnamefont {K.}~\bibnamefont {Jensen}}, \bibinfo {author} {\bibfnamefont {R.}~\bibnamefont {Loganayagam}},\ and\ \bibinfo {author} {\bibfnamefont {A.}~\bibnamefont {Yarom}},\ }\bibfield  {title} {\bibinfo {title} {{Anomaly inflow and thermal equilibrium}},\ }\href {https://doi.org/10.1007/JHEP05(2014)134} {\bibfield  {journal} {\bibinfo  {journal} {JHEP}\ }\textbf {\bibinfo {volume} {05}},\ \bibinfo {pages} {134}},\ \Eprint {https://arxiv.org/abs/1310.7024} {arXiv:1310.7024 [hep-th]} \BibitemShut {NoStop}%
\bibitem [{\citenamefont {Ammon}\ \emph {et~al.}(2017)\citenamefont {Ammon}, \citenamefont {Kaminski}, \citenamefont {Koirala}, \citenamefont {Leiber},\ and\ \citenamefont {Wu}}]{Ammon:2017ded}%
  \BibitemOpen
  \bibfield  {author} {\bibinfo {author} {\bibfnamefont {M.}~\bibnamefont {Ammon}}, \bibinfo {author} {\bibfnamefont {M.}~\bibnamefont {Kaminski}}, \bibinfo {author} {\bibfnamefont {R.}~\bibnamefont {Koirala}}, \bibinfo {author} {\bibfnamefont {J.}~\bibnamefont {Leiber}},\ and\ \bibinfo {author} {\bibfnamefont {J.}~\bibnamefont {Wu}},\ }\bibfield  {title} {\bibinfo {title} {{Quasinormal modes of charged magnetic black branes \& chiral magnetic transport}},\ }\href {https://doi.org/10.1007/JHEP04(2017)067} {\bibfield  {journal} {\bibinfo  {journal} {JHEP}\ }\textbf {\bibinfo {volume} {04}},\ \bibinfo {pages} {067}},\ \Eprint {https://arxiv.org/abs/1701.05565} {arXiv:1701.05565 [hep-th]} \BibitemShut {NoStop}%
\bibitem [{our()}]{ourSMs}%
  \BibitemOpen
  \href@noop {} {}\bibinfo {note} {See Supplemental Material for more technical details, including the definitions of the thermodynamic quantities, the variation of the Euclidean action, the analysis using the Iyer-Wald formalism, the transgression-form origin of $Q_{\rm cs}$ and the associated gauge-invariant completion of the Chern-Simons action, an overview of the magnetic susceptibility from hydrodynamics, and generalizations to various types of black holes exhibiting the same phenomenon. The Supplemental Material also includes Refs.~\cite{Balasubramanian:1999re,Donos:2013cka,Wald:1990mme,Blazquez-Salcedo:2017ghg,Astefanesei:2010dk}}\BibitemShut {NoStop}%
\bibitem [{\citenamefont {Mora}(2014)}]{Mora:2014aca}%
  \BibitemOpen
  \bibfield  {author} {\bibinfo {author} {\bibfnamefont {P.}~\bibnamefont {Mora}},\ }\bibfield  {title} {\bibinfo {title} {{Action Principles for Transgression and Chern-Simons AdS Gravities}},\ }\href {https://doi.org/10.1007/JHEP11(2014)128} {\bibfield  {journal} {\bibinfo  {journal} {JHEP}\ }\textbf {\bibinfo {volume} {11}},\ \bibinfo {pages} {128}},\ \Eprint {https://arxiv.org/abs/1407.6032} {arXiv:1407.6032 [hep-th]} \BibitemShut {NoStop}%
\bibitem [{\citenamefont {D'Hoker}\ and\ \citenamefont {Kraus}(2013)}]{DHoker:2012rlj}%
  \BibitemOpen
  \bibfield  {author} {\bibinfo {author} {\bibfnamefont {E.}~\bibnamefont {D'Hoker}}\ and\ \bibinfo {author} {\bibfnamefont {P.}~\bibnamefont {Kraus}},\ }\bibfield  {title} {\bibinfo {title} {{Quantum Criticality via Magnetic Branes}},\ }\href {https://doi.org/10.1007/978-3-642-37305-3_18} {\bibfield  {journal} {\bibinfo  {journal} {Lect. Notes Phys.}\ }\textbf {\bibinfo {volume} {871}},\ \bibinfo {pages} {469} (\bibinfo {year} {2013})},\ \Eprint {https://arxiv.org/abs/1208.1925} {arXiv:1208.1925 [hep-th]} \BibitemShut {NoStop}%
\bibitem [{\citenamefont {Zhao}\ \emph {et~al.}(2026)\citenamefont {Zhao}, \citenamefont {Lv}, \citenamefont {Li},\ and\ \citenamefont {Li}}]{Zhao:2026boj}%
  \BibitemOpen
  \bibfield  {author} {\bibinfo {author} {\bibfnamefont {J.-K.}\ \bibnamefont {Zhao}}, \bibinfo {author} {\bibfnamefont {E.}~\bibnamefont {Lv}}, \bibinfo {author} {\bibfnamefont {W.}~\bibnamefont {Li}},\ and\ \bibinfo {author} {\bibfnamefont {L.}~\bibnamefont {Li}},\ }\href@noop {} {\bibinfo {title} {Universally diverging gr{\"u}neisen ratio of holographic quantum criticality}} (\bibinfo {year} {2026}),\ \Eprint {https://arxiv.org/abs/2603.21320} {arXiv:2603.21320 [cond-mat.str-el]} \BibitemShut {NoStop}%
\bibitem [{\citenamefont {{Zhu}}\ \emph {et~al.}(2003)\citenamefont {{Zhu}}, \citenamefont {{Garst}}, \citenamefont {{Rosch}},\ and\ \citenamefont {{Si}}}]{2003PhRvL91f6404Z}%
  \BibitemOpen
  \bibfield  {author} {\bibinfo {author} {\bibfnamefont {L.}~\bibnamefont {{Zhu}}}, \bibinfo {author} {\bibfnamefont {M.}~\bibnamefont {{Garst}}}, \bibinfo {author} {\bibfnamefont {A.}~\bibnamefont {{Rosch}}},\ and\ \bibinfo {author} {\bibfnamefont {Q.}~\bibnamefont {{Si}}},\ }\bibfield  {title} {\bibinfo {title} {{Universally Diverging Gr{\"u}neisen Parameter and the Magnetocaloric Effect Close to Quantum Critical Points}},\ }\href {https://doi.org/10.1103/PhysRevLett.91.066404} {\bibfield  {journal} {\bibinfo  {journal} {\prl}\ }\textbf {\bibinfo {volume} {91}},\ \bibinfo {eid} {066404} (\bibinfo {year} {2003})},\ \Eprint {https://arxiv.org/abs/cond-mat/0212335} {arXiv:cond-mat/0212335 [cond-mat.str-el]} \BibitemShut {NoStop}%
\bibitem [{\citenamefont {Garst}\ and\ \citenamefont {Rosch}(2005)}]{garst2005sign}%
  \BibitemOpen
  \bibfield  {author} {\bibinfo {author} {\bibfnamefont {M.}~\bibnamefont {Garst}}\ and\ \bibinfo {author} {\bibfnamefont {A.}~\bibnamefont {Rosch}},\ }\bibfield  {title} {\bibinfo {title} {Sign change of the gr{\"u}neisen parameter and magnetocaloric effect near quantum critical points},\ }\href@noop {} {\bibfield  {journal} {\bibinfo  {journal} {Physical Review B—Condensed Matter and Materials Physics}\ }\textbf {\bibinfo {volume} {72}},\ \bibinfo {pages} {205129} (\bibinfo {year} {2005})}\BibitemShut {NoStop}%
\bibitem [{\citenamefont {Wolf}\ \emph {et~al.}(2014)\citenamefont {Wolf}, \citenamefont {Honecker}, \citenamefont {Hofstetter}, \citenamefont {Tutsch},\ and\ \citenamefont {Lang}}]{wolf2014cooling}%
  \BibitemOpen
  \bibfield  {author} {\bibinfo {author} {\bibfnamefont {B.}~\bibnamefont {Wolf}}, \bibinfo {author} {\bibfnamefont {A.}~\bibnamefont {Honecker}}, \bibinfo {author} {\bibfnamefont {W.}~\bibnamefont {Hofstetter}}, \bibinfo {author} {\bibfnamefont {U.}~\bibnamefont {Tutsch}},\ and\ \bibinfo {author} {\bibfnamefont {M.}~\bibnamefont {Lang}},\ }\bibfield  {title} {\bibinfo {title} {Cooling through quantum criticality and many-body effects in condensed matter and cold gases},\ }\href@noop {} {\bibfield  {journal} {\bibinfo  {journal} {International Journal of Modern Physics B}\ }\textbf {\bibinfo {volume} {28}},\ \bibinfo {pages} {1430017} (\bibinfo {year} {2014})}\BibitemShut {NoStop}%
\bibitem [{\citenamefont {Xiang}\ \emph {et~al.}(2024)\citenamefont {Xiang}, \citenamefont {Zhang}, \citenamefont {Gao}, \citenamefont {Schmidt}, \citenamefont {Schmalzl}, \citenamefont {Wang}, \citenamefont {Li}, \citenamefont {Xi}, \citenamefont {Liu}, \citenamefont {Jin} \emph {et~al.}}]{xiang2024giant}%
  \BibitemOpen
  \bibfield  {author} {\bibinfo {author} {\bibfnamefont {J.}~\bibnamefont {Xiang}}, \bibinfo {author} {\bibfnamefont {C.}~\bibnamefont {Zhang}}, \bibinfo {author} {\bibfnamefont {Y.}~\bibnamefont {Gao}}, \bibinfo {author} {\bibfnamefont {W.}~\bibnamefont {Schmidt}}, \bibinfo {author} {\bibfnamefont {K.}~\bibnamefont {Schmalzl}}, \bibinfo {author} {\bibfnamefont {C.-W.}\ \bibnamefont {Wang}}, \bibinfo {author} {\bibfnamefont {B.}~\bibnamefont {Li}}, \bibinfo {author} {\bibfnamefont {N.}~\bibnamefont {Xi}}, \bibinfo {author} {\bibfnamefont {X.-Y.}\ \bibnamefont {Liu}}, \bibinfo {author} {\bibfnamefont {H.}~\bibnamefont {Jin}}, \emph {et~al.},\ }\bibfield  {title} {\bibinfo {title} {Giant magnetocaloric effect in spin supersolid candidate na2baco (po4) 2},\ }\href@noop {} {\bibfield  {journal} {\bibinfo  {journal} {Nature}\ }\textbf {\bibinfo {volume} {625}},\ \bibinfo {pages} {270} (\bibinfo {year} {2024})}\BibitemShut {NoStop}%
\bibitem [{\citenamefont {Fukushima}\ \emph {et~al.}(2008)\citenamefont {Fukushima}, \citenamefont {Kharzeev},\ and\ \citenamefont {Warringa}}]{Fukushima:2008xe}%
  \BibitemOpen
  \bibfield  {author} {\bibinfo {author} {\bibfnamefont {K.}~\bibnamefont {Fukushima}}, \bibinfo {author} {\bibfnamefont {D.~E.}\ \bibnamefont {Kharzeev}},\ and\ \bibinfo {author} {\bibfnamefont {H.~J.}\ \bibnamefont {Warringa}},\ }\bibfield  {title} {\bibinfo {title} {{The Chiral Magnetic Effect}},\ }\href {https://doi.org/10.1103/PhysRevD.78.074033} {\bibfield  {journal} {\bibinfo  {journal} {Phys. Rev. D}\ }\textbf {\bibinfo {volume} {78}},\ \bibinfo {pages} {074033} (\bibinfo {year} {2008})},\ \Eprint {https://arxiv.org/abs/0808.3382} {arXiv:0808.3382 [hep-ph]} \BibitemShut {NoStop}%
\bibitem [{\citenamefont {Son}\ and\ \citenamefont {Surowka}(2009)}]{Son:2009tf}%
  \BibitemOpen
  \bibfield  {author} {\bibinfo {author} {\bibfnamefont {D.~T.}\ \bibnamefont {Son}}\ and\ \bibinfo {author} {\bibfnamefont {P.}~\bibnamefont {Surowka}},\ }\bibfield  {title} {\bibinfo {title} {{Hydrodynamics with Triangle Anomalies}},\ }\href {https://doi.org/10.1103/PhysRevLett.103.191601} {\bibfield  {journal} {\bibinfo  {journal} {Phys. Rev. Lett.}\ }\textbf {\bibinfo {volume} {103}},\ \bibinfo {pages} {191601} (\bibinfo {year} {2009})},\ \Eprint {https://arxiv.org/abs/0906.5044} {arXiv:0906.5044 [hep-th]} \BibitemShut {NoStop}%
\bibitem [{\citenamefont {Kharzeev}(2014)}]{Kharzeev:2013ffa}%
  \BibitemOpen
  \bibfield  {author} {\bibinfo {author} {\bibfnamefont {D.~E.}\ \bibnamefont {Kharzeev}},\ }\bibfield  {title} {\bibinfo {title} {{The Chiral Magnetic Effect and Anomaly-Induced Transport}},\ }\href {https://doi.org/10.1016/j.ppnp.2014.01.002} {\bibfield  {journal} {\bibinfo  {journal} {Prog. Part. Nucl. Phys.}\ }\textbf {\bibinfo {volume} {75}},\ \bibinfo {pages} {133} (\bibinfo {year} {2014})},\ \Eprint {https://arxiv.org/abs/1312.3348} {arXiv:1312.3348 [hep-ph]} \BibitemShut {NoStop}%
\bibitem [{\citenamefont {Fukushima}\ and\ \citenamefont {Morales}(2013)}]{Fukushima:2013zga}%
  \BibitemOpen
  \bibfield  {author} {\bibinfo {author} {\bibfnamefont {K.}~\bibnamefont {Fukushima}}\ and\ \bibinfo {author} {\bibfnamefont {P.}~\bibnamefont {Morales}},\ }\bibfield  {title} {\bibinfo {title} {{Spatial modulation and topological current in holographic QCD matter}},\ }\href {https://doi.org/10.1103/PhysRevLett.111.051601} {\bibfield  {journal} {\bibinfo  {journal} {Phys. Rev. Lett.}\ }\textbf {\bibinfo {volume} {111}},\ \bibinfo {pages} {051601} (\bibinfo {year} {2013})},\ \Eprint {https://arxiv.org/abs/1305.4115} {arXiv:1305.4115 [hep-ph]} \BibitemShut {NoStop}%
\bibitem [{\citenamefont {Ooguri}\ and\ \citenamefont {Park}(2011)}]{Ooguri:2010xs}%
  \BibitemOpen
  \bibfield  {author} {\bibinfo {author} {\bibfnamefont {H.}~\bibnamefont {Ooguri}}\ and\ \bibinfo {author} {\bibfnamefont {C.-S.}\ \bibnamefont {Park}},\ }\bibfield  {title} {\bibinfo {title} {{Spatially Modulated Phase in Holographic Quark-Gluon Plasma}},\ }\href {https://doi.org/10.1103/PhysRevLett.106.061601} {\bibfield  {journal} {\bibinfo  {journal} {Phys. Rev. Lett.}\ }\textbf {\bibinfo {volume} {106}},\ \bibinfo {pages} {061601} (\bibinfo {year} {2011})},\ \Eprint {https://arxiv.org/abs/1011.4144} {arXiv:1011.4144 [hep-th]} \BibitemShut {NoStop}%
\bibitem [{\citenamefont {Bali}\ \emph {et~al.}(2014)\citenamefont {Bali}, \citenamefont {Bruckmann}, \citenamefont {Endr{\"o}di}, \citenamefont {Katz},\ and\ \citenamefont {Sch{\"a}fer}}]{Bali:2014kia}%
  \BibitemOpen
  \bibfield  {author} {\bibinfo {author} {\bibfnamefont {G.~S.}\ \bibnamefont {Bali}}, \bibinfo {author} {\bibfnamefont {F.}~\bibnamefont {Bruckmann}}, \bibinfo {author} {\bibfnamefont {G.}~\bibnamefont {Endr{\"o}di}}, \bibinfo {author} {\bibfnamefont {S.~D.}\ \bibnamefont {Katz}},\ and\ \bibinfo {author} {\bibfnamefont {A.}~\bibnamefont {Sch{\"a}fer}},\ }\bibfield  {title} {\bibinfo {title} {{The QCD equation of state in background magnetic fields}},\ }\href {https://doi.org/10.1007/JHEP08(2014)177} {\bibfield  {journal} {\bibinfo  {journal} {JHEP}\ }\textbf {\bibinfo {volume} {08}},\ \bibinfo {pages} {177}},\ \Eprint {https://arxiv.org/abs/1406.0269} {arXiv:1406.0269 [hep-lat]} \BibitemShut {NoStop}%
\bibitem [{\citenamefont {Ammon}\ \emph {et~al.}(2016)\citenamefont {Ammon}, \citenamefont {Leiber},\ and\ \citenamefont {Macedo}}]{Ammon:2016szz}%
  \BibitemOpen
  \bibfield  {author} {\bibinfo {author} {\bibfnamefont {M.}~\bibnamefont {Ammon}}, \bibinfo {author} {\bibfnamefont {J.}~\bibnamefont {Leiber}},\ and\ \bibinfo {author} {\bibfnamefont {R.~P.}\ \bibnamefont {Macedo}},\ }\bibfield  {title} {\bibinfo {title} {{Phase diagram of 4D field theories with chiral anomaly from holography}},\ }\href {https://doi.org/10.1007/JHEP03(2016)164} {\bibfield  {journal} {\bibinfo  {journal} {JHEP}\ }\textbf {\bibinfo {volume} {03}},\ \bibinfo {pages} {164}},\ \Eprint {https://arxiv.org/abs/1601.02125} {arXiv:1601.02125 [hep-th]} \BibitemShut {NoStop}%
\bibitem [{\citenamefont {Chong}\ \emph {et~al.}(2005)\citenamefont {Chong}, \citenamefont {Cvetic}, \citenamefont {Lu},\ and\ \citenamefont {Pope}}]{Chong:2005hr}%
  \BibitemOpen
  \bibfield  {author} {\bibinfo {author} {\bibfnamefont {Z.~W.}\ \bibnamefont {Chong}}, \bibinfo {author} {\bibfnamefont {M.}~\bibnamefont {Cvetic}}, \bibinfo {author} {\bibfnamefont {H.}~\bibnamefont {Lu}},\ and\ \bibinfo {author} {\bibfnamefont {C.~N.}\ \bibnamefont {Pope}},\ }\bibfield  {title} {\bibinfo {title} {{General non-extremal rotating black holes in minimal five-dimensional gauged supergravity}},\ }\href {https://doi.org/10.1103/PhysRevLett.95.161301} {\bibfield  {journal} {\bibinfo  {journal} {Phys. Rev. Lett.}\ }\textbf {\bibinfo {volume} {95}},\ \bibinfo {pages} {161301} (\bibinfo {year} {2005})},\ \Eprint {https://arxiv.org/abs/hep-th/0506029} {arXiv:hep-th/0506029} \BibitemShut {NoStop}%
\bibitem [{\citenamefont {Bl{\'a}zquez-Salcedo}\ \emph {et~al.}(2017)\citenamefont {Bl{\'a}zquez-Salcedo}, \citenamefont {Kunz}, \citenamefont {Navarro-L{\'e}rida},\ and\ \citenamefont {Radu}}]{Blazquez-Salcedo:2016rkj}%
  \BibitemOpen
  \bibfield  {author} {\bibinfo {author} {\bibfnamefont {J.~L.}\ \bibnamefont {Bl{\'a}zquez-Salcedo}}, \bibinfo {author} {\bibfnamefont {J.}~\bibnamefont {Kunz}}, \bibinfo {author} {\bibfnamefont {F.}~\bibnamefont {Navarro-L{\'e}rida}},\ and\ \bibinfo {author} {\bibfnamefont {E.}~\bibnamefont {Radu}},\ }\bibfield  {title} {\bibinfo {title} {{Charged rotating black holes in Einstein-Maxwell-Chern-Simons theory with a negative cosmological constant}},\ }\href {https://doi.org/10.1103/PhysRevD.95.064018} {\bibfield  {journal} {\bibinfo  {journal} {Phys. Rev. D}\ }\textbf {\bibinfo {volume} {95}},\ \bibinfo {pages} {064018} (\bibinfo {year} {2017})},\ \Eprint {https://arxiv.org/abs/1610.05282} {arXiv:1610.05282 [gr-qc]} \BibitemShut {NoStop}%
\bibitem [{\citenamefont {Cai}(2002)}]{Cai:2001dz}%
  \BibitemOpen
  \bibfield  {author} {\bibinfo {author} {\bibfnamefont {R.-G.}\ \bibnamefont {Cai}},\ }\bibfield  {title} {\bibinfo {title} {{Gauss-Bonnet black holes in AdS spaces}},\ }\href {https://doi.org/10.1103/PhysRevD.65.084014} {\bibfield  {journal} {\bibinfo  {journal} {Phys. Rev. D}\ }\textbf {\bibinfo {volume} {65}},\ \bibinfo {pages} {084014} (\bibinfo {year} {2002})},\ \Eprint {https://arxiv.org/abs/hep-th/0109133} {arXiv:hep-th/0109133} \BibitemShut {NoStop}%
\bibitem [{\citenamefont {Brihaye}\ and\ \citenamefont {Radu}(2008)}]{Brihaye:2008xu}%
  \BibitemOpen
  \bibfield  {author} {\bibinfo {author} {\bibfnamefont {Y.}~\bibnamefont {Brihaye}}\ and\ \bibinfo {author} {\bibfnamefont {E.}~\bibnamefont {Radu}},\ }\bibfield  {title} {\bibinfo {title} {{Black objects in the Einstein-Gauss-Bonnet theory with negative cosmological constant and the boundary counterterm method}},\ }\href {https://doi.org/10.1088/1126-6708/2008/09/006} {\bibfield  {journal} {\bibinfo  {journal} {JHEP}\ }\textbf {\bibinfo {volume} {09}},\ \bibinfo {pages} {006}},\ \Eprint {https://arxiv.org/abs/0806.1396} {arXiv:0806.1396 [gr-qc]} \BibitemShut {NoStop}%
\bibitem [{\citenamefont {Balasubramanian}\ and\ \citenamefont {Kraus}(1999)}]{Balasubramanian:1999re}%
  \BibitemOpen
  \bibfield  {author} {\bibinfo {author} {\bibfnamefont {V.}~\bibnamefont {Balasubramanian}}\ and\ \bibinfo {author} {\bibfnamefont {P.}~\bibnamefont {Kraus}},\ }\bibfield  {title} {\bibinfo {title} {{A Stress tensor for Anti-de Sitter gravity}},\ }\href {https://doi.org/10.1007/s002200050764} {\bibfield  {journal} {\bibinfo  {journal} {Commun. Math. Phys.}\ }\textbf {\bibinfo {volume} {208}},\ \bibinfo {pages} {413} (\bibinfo {year} {1999})},\ \Eprint {https://arxiv.org/abs/hep-th/9902121} {arXiv:hep-th/9902121} \BibitemShut {NoStop}%
\bibitem [{\citenamefont {Donos}\ and\ \citenamefont {Gauntlett}(2013)}]{Donos:2013cka}%
  \BibitemOpen
  \bibfield  {author} {\bibinfo {author} {\bibfnamefont {A.}~\bibnamefont {Donos}}\ and\ \bibinfo {author} {\bibfnamefont {J.~P.}\ \bibnamefont {Gauntlett}},\ }\bibfield  {title} {\bibinfo {title} {{On the thermodynamics of periodic AdS black branes}},\ }\href {https://doi.org/10.1007/JHEP10(2013)038} {\bibfield  {journal} {\bibinfo  {journal} {JHEP}\ }\textbf {\bibinfo {volume} {10}},\ \bibinfo {pages} {038}},\ \Eprint {https://arxiv.org/abs/1306.4937} {arXiv:1306.4937 [hep-th]} \BibitemShut {NoStop}%
\bibitem [{\citenamefont {Wald}(1990)}]{Wald:1990mme}%
  \BibitemOpen
  \bibfield  {author} {\bibinfo {author} {\bibfnamefont {R.~M.}\ \bibnamefont {Wald}},\ }\bibfield  {title} {\bibinfo {title} {{On identically closed forms locally constructed from a field}},\ }\href {https://doi.org/10.1063/1.528839} {\bibfield  {journal} {\bibinfo  {journal} {J. Math. Phys.}\ }\textbf {\bibinfo {volume} {31}},\ \bibinfo {pages} {2378} (\bibinfo {year} {1990})}\BibitemShut {NoStop}%
\bibitem [{\citenamefont {Bl{\'a}zquez-Salcedo}\ \emph {et~al.}(2018)\citenamefont {Bl{\'a}zquez-Salcedo}, \citenamefont {Kunz}, \citenamefont {Navarro-L{\'e}rida},\ and\ \citenamefont {Radu}}]{Blazquez-Salcedo:2017ghg}%
  \BibitemOpen
  \bibfield  {author} {\bibinfo {author} {\bibfnamefont {J.~L.}\ \bibnamefont {Bl{\'a}zquez-Salcedo}}, \bibinfo {author} {\bibfnamefont {J.}~\bibnamefont {Kunz}}, \bibinfo {author} {\bibfnamefont {F.}~\bibnamefont {Navarro-L{\'e}rida}},\ and\ \bibinfo {author} {\bibfnamefont {E.}~\bibnamefont {Radu}},\ }\bibfield  {title} {\bibinfo {title} {{Squashed, magnetized black holes in $D=5$ minimal gauged supergravity}},\ }\href {https://doi.org/10.1007/JHEP02(2018)061} {\bibfield  {journal} {\bibinfo  {journal} {JHEP}\ }\textbf {\bibinfo {volume} {02}},\ \bibinfo {pages} {061}},\ \Eprint {https://arxiv.org/abs/1711.10483} {arXiv:1711.10483 [gr-qc]} \BibitemShut {NoStop}%
\bibitem [{\citenamefont {Astefanesei}\ \emph {et~al.}(2011)\citenamefont {Astefanesei}, \citenamefont {Banerjee},\ and\ \citenamefont {Dutta}}]{Astefanesei:2010dk}%
  \BibitemOpen
  \bibfield  {author} {\bibinfo {author} {\bibfnamefont {D.}~\bibnamefont {Astefanesei}}, \bibinfo {author} {\bibfnamefont {N.}~\bibnamefont {Banerjee}},\ and\ \bibinfo {author} {\bibfnamefont {S.}~\bibnamefont {Dutta}},\ }\bibfield  {title} {\bibinfo {title} {{Moduli and electromagnetic black brane holography}},\ }\href {https://doi.org/10.1007/JHEP02(2011)021} {\bibfield  {journal} {\bibinfo  {journal} {JHEP}\ }\textbf {\bibinfo {volume} {02}},\ \bibinfo {pages} {021}},\ \Eprint {https://arxiv.org/abs/1008.3852} {arXiv:1008.3852 [hep-th]} \BibitemShut {NoStop}%
\end{thebibliography}%


\begin{thebibliography}{15}%
\makeatletter
\providecommand \@ifxundefined [1]{%
 \@ifx{#1\undefined}
}%
\providecommand \@ifnum [1]{%
 \ifnum #1\expandafter \@firstoftwo
 \else \expandafter \@secondoftwo
 \fi
}%
\providecommand \@ifx [1]{%
 \ifx #1\expandafter \@firstoftwo
 \else \expandafter \@secondoftwo
 \fi
}%
\providecommand \natexlab [1]{#1}%
\providecommand \enquote  [1]{``#1''}%
\providecommand \bibnamefont  [1]{#1}%
\providecommand \bibfnamefont [1]{#1}%
\providecommand \citenamefont [1]{#1}%
\providecommand \href@noop [0]{\@secondoftwo}%
\providecommand \href [0]{\begingroup \@sanitize@url \@href}%
\providecommand \@href[1]{\@@startlink{#1}\@@href}%
\providecommand \@@href[1]{\endgroup#1\@@endlink}%
\providecommand \@sanitize@url [0]{\catcode `\\12\catcode `\$12\catcode `\&12\catcode `\#12\catcode `\^12\catcode `\_12\catcode `\%12\relax}%
\providecommand \@@startlink[1]{}%
\providecommand \@@endlink[0]{}%
\providecommand \url  [0]{\begingroup\@sanitize@url \@url }%
\providecommand \@url [1]{\endgroup\@href {#1}{\urlprefix }}%
\providecommand \urlprefix  [0]{URL }%
\providecommand \Eprint [0]{\href }%
\providecommand \doibase [0]{https://doi.org/}%
\providecommand \selectlanguage [0]{\@gobble}%
\providecommand \bibinfo  [0]{\@secondoftwo}%
\providecommand \bibfield  [0]{\@secondoftwo}%
\providecommand \translation [1]{[#1]}%
\providecommand \BibitemOpen [0]{}%
\providecommand \bibitemStop [0]{}%
\providecommand \bibitemNoStop [0]{.\EOS\space}%
\providecommand \EOS [0]{\spacefactor3000\relax}%
\providecommand \BibitemShut  [1]{\csname bibitem#1\endcsname}%
\let\auto@bib@innerbib\@empty
\bibitem [{\citenamefont {Balasubramanian}\ and\ \citenamefont {Kraus}(1999)}]{Balasubramanian:1999re}%
  \BibitemOpen
  \bibfield  {author} {\bibinfo {author} {\bibfnamefont {V.}~\bibnamefont {Balasubramanian}}\ and\ \bibinfo {author} {\bibfnamefont {P.}~\bibnamefont {Kraus}},\ }\bibfield  {title} {\bibinfo {title} {{A Stress tensor for Anti-de Sitter gravity}},\ }\href {https://doi.org/10.1007/s002200050764} {\bibfield  {journal} {\bibinfo  {journal} {Commun. Math. Phys.}\ }\textbf {\bibinfo {volume} {208}},\ \bibinfo {pages} {413} (\bibinfo {year} {1999})},\ \Eprint {https://arxiv.org/abs/hep-th/9902121} {arXiv:hep-th/9902121} \BibitemShut {NoStop}%
\bibitem [{\citenamefont {Ammon}\ \emph {et~al.}(2016)\citenamefont {Ammon}, \citenamefont {Leiber},\ and\ \citenamefont {Macedo}}]{Ammon:2016szz}%
  \BibitemOpen
  \bibfield  {author} {\bibinfo {author} {\bibfnamefont {M.}~\bibnamefont {Ammon}}, \bibinfo {author} {\bibfnamefont {J.}~\bibnamefont {Leiber}},\ and\ \bibinfo {author} {\bibfnamefont {R.~P.}\ \bibnamefont {Macedo}},\ }\bibfield  {title} {\bibinfo {title} {{Phase diagram of 4D field theories with chiral anomaly from holography}},\ }\href {https://doi.org/10.1007/JHEP03(2016)164} {\bibfield  {journal} {\bibinfo  {journal} {JHEP}\ }\textbf {\bibinfo {volume} {03}},\ \bibinfo {pages} {164}},\ \Eprint {https://arxiv.org/abs/1601.02125} {arXiv:1601.02125 [hep-th]} \BibitemShut {NoStop}%
\bibitem [{\citenamefont {Donos}\ and\ \citenamefont {Gauntlett}(2013)}]{Donos:2013cka}%
  \BibitemOpen
  \bibfield  {author} {\bibinfo {author} {\bibfnamefont {A.}~\bibnamefont {Donos}}\ and\ \bibinfo {author} {\bibfnamefont {J.~P.}\ \bibnamefont {Gauntlett}},\ }\bibfield  {title} {\bibinfo {title} {{On the thermodynamics of periodic AdS black branes}},\ }\href {https://doi.org/10.1007/JHEP10(2013)038} {\bibfield  {journal} {\bibinfo  {journal} {JHEP}\ }\textbf {\bibinfo {volume} {10}},\ \bibinfo {pages} {038}},\ \Eprint {https://arxiv.org/abs/1306.4937} {arXiv:1306.4937 [hep-th]} \BibitemShut {NoStop}%
\bibitem [{\citenamefont {Iyer}\ and\ \citenamefont {Wald}(1994)}]{Iyer:1994ys}%
  \BibitemOpen
  \bibfield  {author} {\bibinfo {author} {\bibfnamefont {V.}~\bibnamefont {Iyer}}\ and\ \bibinfo {author} {\bibfnamefont {R.~M.}\ \bibnamefont {Wald}},\ }\bibfield  {title} {\bibinfo {title} {{Some properties of Noether charge and a proposal for dynamical black hole entropy}},\ }\href {https://doi.org/10.1103/PhysRevD.50.846} {\bibfield  {journal} {\bibinfo  {journal} {Phys. Rev. D}\ }\textbf {\bibinfo {volume} {50}},\ \bibinfo {pages} {846} (\bibinfo {year} {1994})},\ \Eprint {https://arxiv.org/abs/gr-qc/9403028} {arXiv:gr-qc/9403028} \BibitemShut {NoStop}%
\bibitem [{\citenamefont {Wald}(1990)}]{Wald:1990mme}%
  \BibitemOpen
  \bibfield  {author} {\bibinfo {author} {\bibfnamefont {R.~M.}\ \bibnamefont {Wald}},\ }\bibfield  {title} {\bibinfo {title} {{On identically closed forms locally constructed from a field}},\ }\href {https://doi.org/10.1063/1.528839} {\bibfield  {journal} {\bibinfo  {journal} {J. Math. Phys.}\ }\textbf {\bibinfo {volume} {31}},\ \bibinfo {pages} {2378} (\bibinfo {year} {1990})}\BibitemShut {NoStop}%
\bibitem [{\citenamefont {Ammon}\ \emph {et~al.}(2017)\citenamefont {Ammon}, \citenamefont {Kaminski}, \citenamefont {Koirala}, \citenamefont {Leiber},\ and\ \citenamefont {Wu}}]{Ammon:2017ded}%
  \BibitemOpen
  \bibfield  {author} {\bibinfo {author} {\bibfnamefont {M.}~\bibnamefont {Ammon}}, \bibinfo {author} {\bibfnamefont {M.}~\bibnamefont {Kaminski}}, \bibinfo {author} {\bibfnamefont {R.}~\bibnamefont {Koirala}}, \bibinfo {author} {\bibfnamefont {J.}~\bibnamefont {Leiber}},\ and\ \bibinfo {author} {\bibfnamefont {J.}~\bibnamefont {Wu}},\ }\bibfield  {title} {\bibinfo {title} {{Quasinormal modes of charged magnetic black branes and chiral magnetic transport}},\ }\href {https://doi.org/10.1007/JHEP04(2017)067} {\bibfield  {journal} {\bibinfo  {journal} {JHEP}\ }\textbf {\bibinfo {volume} {04}},\ \bibinfo {pages} {067}},\ \Eprint {https://arxiv.org/abs/1701.05565} {arXiv:1701.05565 [hep-th]} \BibitemShut {NoStop}%
\bibitem [{\citenamefont {Kovtun}(2016)}]{Kovtun:2016lfw}%
  \BibitemOpen
  \bibfield  {author} {\bibinfo {author} {\bibfnamefont {P.}~\bibnamefont {Kovtun}},\ }\bibfield  {title} {\bibinfo {title} {{Thermodynamics of polarized relativistic matter}},\ }\href {https://doi.org/10.1007/JHEP07(2016)028} {\bibfield  {journal} {\bibinfo  {journal} {JHEP}\ }\textbf {\bibinfo {volume} {07}},\ \bibinfo {pages} {028}},\ \Eprint {https://arxiv.org/abs/1606.01226} {arXiv:1606.01226 [hep-th]} \BibitemShut {NoStop}%
\bibitem [{\citenamefont {Jensen}\ \emph {et~al.}(2014)\citenamefont {Jensen}, \citenamefont {Loganayagam},\ and\ \citenamefont {Yarom}}]{Jensen:2013kka}%
  \BibitemOpen
  \bibfield  {author} {\bibinfo {author} {\bibfnamefont {K.}~\bibnamefont {Jensen}}, \bibinfo {author} {\bibfnamefont {R.}~\bibnamefont {Loganayagam}},\ and\ \bibinfo {author} {\bibfnamefont {A.}~\bibnamefont {Yarom}},\ }\bibfield  {title} {\bibinfo {title} {{Anomaly inflow and thermal equilibrium}},\ }\href {https://doi.org/10.1007/JHEP05(2014)134} {\bibfield  {journal} {\bibinfo  {journal} {JHEP}\ }\textbf {\bibinfo {volume} {05}},\ \bibinfo {pages} {134}},\ \Eprint {https://arxiv.org/abs/1310.7024} {arXiv:1310.7024 [hep-th]} \BibitemShut {NoStop}%
\bibitem [{\citenamefont {Donos}\ and\ \citenamefont {Gauntlett}(2012)}]{Donos:2012wi}%
  \BibitemOpen
  \bibfield  {author} {\bibinfo {author} {\bibfnamefont {A.}~\bibnamefont {Donos}}\ and\ \bibinfo {author} {\bibfnamefont {J.~P.}\ \bibnamefont {Gauntlett}},\ }\bibfield  {title} {\bibinfo {title} {{Black holes dual to helical current phases}},\ }\href {https://doi.org/10.1103/PhysRevD.86.064010} {\bibfield  {journal} {\bibinfo  {journal} {Phys. Rev. D}\ }\textbf {\bibinfo {volume} {86}},\ \bibinfo {pages} {064010} (\bibinfo {year} {2012})},\ \Eprint {https://arxiv.org/abs/1204.1734} {arXiv:1204.1734 [hep-th]} \BibitemShut {NoStop}%
\bibitem [{\citenamefont {Blazquez-Salcedo}\ \emph {et~al.}(2018)\citenamefont {Blazquez-Salcedo}, \citenamefont {Kunz}, \citenamefont {Navarro-Lerida},\ and\ \citenamefont {Radu}}]{Blazquez-Salcedo:2017ghg}%
  \BibitemOpen
  \bibfield  {author} {\bibinfo {author} {\bibfnamefont {J.~L.}\ \bibnamefont {Blazquez-Salcedo}}, \bibinfo {author} {\bibfnamefont {J.}~\bibnamefont {Kunz}}, \bibinfo {author} {\bibfnamefont {F.}~\bibnamefont {Navarro-Lerida}},\ and\ \bibinfo {author} {\bibfnamefont {E.}~\bibnamefont {Radu}},\ }\bibfield  {title} {\bibinfo {title} {{Squashed, magnetized black holes in D=5 minimal gauged supergravity}},\ }\href {https://doi.org/10.1007/JHEP02(2018)061} {\bibfield  {journal} {\bibinfo  {journal} {JHEP}\ }\textbf {\bibinfo {volume} {02}},\ \bibinfo {pages} {061}},\ \Eprint {https://arxiv.org/abs/1711.10483} {arXiv:1711.10483 [gr-qc]} \BibitemShut {NoStop}%
\bibitem [{\citenamefont {Blazquez-Salcedo}\ \emph {et~al.}(2017)\citenamefont {Blazquez-Salcedo}, \citenamefont {Kunz}, \citenamefont {Navarro-Lerida},\ and\ \citenamefont {Radu}}]{Blazquez-Salcedo:2016rkj}%
  \BibitemOpen
  \bibfield  {author} {\bibinfo {author} {\bibfnamefont {J.~L.}\ \bibnamefont {Blazquez-Salcedo}}, \bibinfo {author} {\bibfnamefont {J.}~\bibnamefont {Kunz}}, \bibinfo {author} {\bibfnamefont {F.}~\bibnamefont {Navarro-Lerida}},\ and\ \bibinfo {author} {\bibfnamefont {E.}~\bibnamefont {Radu}},\ }\bibfield  {title} {\bibinfo {title} {{Charged rotating black holes in Einstein-Maxwell-Chern-Simons theory with a negative cosmological constant}},\ }\href {https://doi.org/10.1103/PhysRevD.95.064018} {\bibfield  {journal} {\bibinfo  {journal} {Phys. Rev. D}\ }\textbf {\bibinfo {volume} {95}},\ \bibinfo {pages} {064018} (\bibinfo {year} {2017})},\ \Eprint {https://arxiv.org/abs/1610.05282} {arXiv:1610.05282 [gr-qc]} \BibitemShut {NoStop}%
\bibitem [{\citenamefont {Chong}\ \emph {et~al.}(2005)\citenamefont {Chong}, \citenamefont {Cvetic}, \citenamefont {Lu},\ and\ \citenamefont {Pope}}]{Chong:2005hr}%
  \BibitemOpen
  \bibfield  {author} {\bibinfo {author} {\bibfnamefont {Z.~W.}\ \bibnamefont {Chong}}, \bibinfo {author} {\bibfnamefont {M.}~\bibnamefont {Cvetic}}, \bibinfo {author} {\bibfnamefont {H.}~\bibnamefont {Lu}},\ and\ \bibinfo {author} {\bibfnamefont {C.~N.}\ \bibnamefont {Pope}},\ }\bibfield  {title} {\bibinfo {title} {{General nonextremal rotating black holes in minimal five-dimensional gauged supergravity}},\ }\href {https://doi.org/10.1103/PhysRevLett.95.161301} {\bibfield  {journal} {\bibinfo  {journal} {Phys. Rev. Lett.}\ }\textbf {\bibinfo {volume} {95}},\ \bibinfo {pages} {161301} (\bibinfo {year} {2005})},\ \Eprint {https://arxiv.org/abs/hep-th/0506029} {arXiv:hep-th/0506029} \BibitemShut {NoStop}%
\bibitem [{\citenamefont {Astefanesei}\ \emph {et~al.}(2011)\citenamefont {Astefanesei}, \citenamefont {Banerjee},\ and\ \citenamefont {Dutta}}]{Astefanesei:2010dk}%
  \BibitemOpen
  \bibfield  {author} {\bibinfo {author} {\bibfnamefont {D.}~\bibnamefont {Astefanesei}}, \bibinfo {author} {\bibfnamefont {N.}~\bibnamefont {Banerjee}},\ and\ \bibinfo {author} {\bibfnamefont {S.}~\bibnamefont {Dutta}},\ }\bibfield  {title} {\bibinfo {title} {{Moduli and electromagnetic black brane holography}},\ }\href {https://doi.org/10.1007/JHEP02(2011)021} {\bibfield  {journal} {\bibinfo  {journal} {JHEP}\ }\textbf {\bibinfo {volume} {02}},\ \bibinfo {pages} {021}},\ \Eprint {https://arxiv.org/abs/1008.3852} {arXiv:1008.3852 [hep-th]} \BibitemShut {NoStop}%
\bibitem [{\citenamefont {Cai}(2002)}]{Cai:2001dz}%
  \BibitemOpen
  \bibfield  {author} {\bibinfo {author} {\bibfnamefont {R.~G.}\ \bibnamefont {Cai}},\ }\bibfield  {title} {\bibinfo {title} {{Gauss-Bonnet black holes in AdS spaces}},\ }\href {https://doi.org/10.1103/PhysRevD.65.084014} {\bibfield  {journal} {\bibinfo  {journal} {Phys. Rev. D}\ }\textbf {\bibinfo {volume} {65}},\ \bibinfo {pages} {084014} (\bibinfo {year} {2002})},\ \Eprint {https://arxiv.org/abs/hep-th/0109133} {arXiv:hep-th/0109133} \BibitemShut {NoStop}%
\bibitem [{\citenamefont {Brihaye}\ and\ \citenamefont {Radu}(2008)}]{Brihaye:2008xu}%
  \BibitemOpen
  \bibfield  {author} {\bibinfo {author} {\bibfnamefont {Y.}~\bibnamefont {Brihaye}}\ and\ \bibinfo {author} {\bibfnamefont {E.}~\bibnamefont {Radu}},\ }\bibfield  {title} {\bibinfo {title} {{Black objects in the Einstein-Gauss-Bonnet theory with negative cosmological constant and the boundary counterterm method}},\ }\href {https://doi.org/10.1088/1126-6708/2008/09/006} {\bibfield  {journal} {\bibinfo  {journal} {JHEP}\ }\textbf {\bibinfo {volume} {09}},\ \bibinfo {pages} {006}},\ \Eprint {https://arxiv.org/abs/0806.1396} {arXiv:0806.1396 [gr-qc]} \BibitemShut {NoStop}%
\end{thebibliography}%

\end{document}


\preprint{APS/123-QED}
\title{Supplemental material for\texorpdfstring{\\}{}
``Transgression Completion for Chern-Simons Black Hole Thermodynamics"}

\author{\textbf{Rong-Gen Cai}$^{2,1,3}$}%
\author{\textbf{Li Li}$^{1,3,4}$}%
\author{\textbf{Jun-Kun Zhao}$^{1}$}
\vspace{1cm}

\affiliation{${}^{1}$Institute of Theoretical Physics, Chinese Academy of Sciences, Beijing 100190, China}
\affiliation{${}^{2}$Institute of Fundamental Physics and Quantum Technology, Ningbo University, Ningbo 315211, China}
\affiliation{${}^{3}$School of Fundamental Physics and Mathematical Sciences, Hangzhou Institute for Advanced Study, UCAS, Hangzhou 310024, China}
\affiliation{${}^{4}$School of Physical Sciences, University of Chinese Academy of Sciences, No.19A Yuquan Road, Beijing 100049, China}

\maketitle
\appendix 
\onecolumngrid
\renewcommand{\appendixname}{}

\begingroup
\hypersetup{linkcolor=black,urlcolor=black}  %
\tableofcontents
\endgroup
\vspace{.3cm}
\hrule
\vspace{.3cm}
In this Supplemental Material, we provide technical details supporting the results reported in the main Letter. Throughout this Supplemental Material, $w_{ QSR}=T I_E/V$ denotes the free energy density obtained from the renormalized Euclidean action containing the Chern-Simons interactions, whereas $w_{\mathcal T}$ denotes the corresponding result after the local Chern-Simons representative is replaced by its gauge-invariant transgression completion. The two are related by $w_{\mathcal T}=w_{ QSR}+kQ_{ cs}=w$, where the overall normalization by the physical boundary volume is understood and is displayed explicitly for the rotating and spherical backgrounds. In each example, we follow the same strategy: we first analyze the thermodynamics and derive the corresponding first law by two independent methods---the variation of the Euclidean action and the Iyer-Wald formalism---thereby identifying the Chern-Simons correction $Q_{ cs}$; we then explain its origin in terms of the transgression form.

In Section~\ref{SM-A}, we present a detailed analysis of the minimal dyonic black hole (BH) solution in Einstein-Maxwell-Chern-Simons (EMCS) theory. We first show that, for charged magnetic BHs in theories with a Chern-Simons interaction, the standard quantum statistical relation does not, by itself, furnish the thermodynamic potential for the ensemble specified by $(T,\mu,B)$. A direct variation of the on-shell Euclidean action identifies an additional Chern-Simons contribution, denoted by $kQ_{ cs}$. Once this term is included, the resulting free energy obeys the standard first law and the Smarr relation. We further confirm the same result using the Iyer-Wald formalism. We then show that this correction is generated automatically when the Chern-Simons form is replaced by its transgression completion, thereby explaining the origin of $Q_{ cs}$ while rendering the full action strictly gauge invariant.

In Section~\ref{SM-B}, we briefly review the hydrodynamic description of the system in the presence of a background magnetic field. As discussed in the main text, the uncorrected free energy $w_{ QSR}$ does not reproduce the magnetic susceptibility expected from hydrodynamics, whereas the corrected free energy does.

Remarkably, this mechanism appears to be highly universal. To demonstrate this, we consider several representative BH solutions, including five-dimensional helical BHs (Sec.~\ref{SM-C}), rotating BHs (Sec.~\ref{SM-D}), scalarized BHs (Sec.~\ref{SM-E}), BHs in higher-derivative theories (Sec.~\ref{SM-F}), and BHs in higher spacetime dimensions (Sec.~\ref{SM-G}). Taken together, these examples demonstrate the universality of $Q_{ cs}$: up to an overall factor, it retains the same characteristic radial-integral structure and restores thermodynamic consistency across changes in symmetry, topology, spacetime dimension, higher-derivative interactions, and matter content. The transgression completion provides a unified gauge-invariant action-level explanation for its origin.

\section{Minimal model of BHs with Chern-Simons interactions}\label{SM-A}
This section provides a detailed analysis of the dyonic BHs (3) in the EMCS theory, complementing the results reported in the main text.

\subsection*{A.1: Equations of motion and thermodynamics}

The general equations of motion from the five-dimensional EMCS theory in the main text read
\begin{equation} \begin{split} \label{SM-A-eom}
G_{ab}-\frac{6}{L^2}g_{ab} -\frac{1}{2}\big(F_{ac}F_b^{\ c} -\frac{1}{4}g_{ab}F_{cd}F^{cd} \big) &=0 \,, \\ 
\nabla_{b}F^{ba} +\frac{k}{8}\epsilon^{abcde}F_{bc}F_{de} &=0 \,.
\end{split}
\end{equation}
The explicit equations of motion for the dyonic BHs~(3) of the main text are given by
\begin{equation}
\begin{split}
\Big[ \frac{e^{\chi/2} h}{r}(A_t'+p A_z') \Big]'+kBA_z' &=0 \,, \\
\Big[ \frac{e^{\chi/2}}{r h} \Big(p h^2 A_t'- (f e^{-\chi} -h^2p^2) A_z' \Big) \Big]'-kBA_t' &=0 \,,\\
\frac{f''}{f}-\chi''+ \left( \frac{\chi'}{2}-\frac{3f'}{2f} \right)\chi' -\frac{3e^\chi \left( h^2p'^2 +r^2(A_t'+p A_z')^2\right)}{2f} +\frac{6h'}{r h} +\frac{r^2 A_z'^2}{2h^2} -\frac{3B^2r^2}{2f} +\frac{12-12f}{r^2f} &=0 \,,\\
\frac{f''}{f}-\chi'' -\left( \frac{h'}{h}+ \frac{3\chi'}{2} +\frac{3}{r} \right) \frac{f'}{f} +\left(\frac{h'}{h} +\frac{\chi'}{2} \right)\chi' -\frac{e^\chi \left( 2h^2p'^2 +r^2(A_t'+p A_z')^2\right)}{f} +\frac{6h'}{r h}  &=0 \,,\\
h''+\left(\frac{f'}{f} -\frac{\chi'}{2} -\frac{3}{r} \right)h'-\frac{B^2r^2h}{2f} +\frac{e^\chi h^3p'^2}{2f} +\frac{r^2A_z'^2}{2h}  &=0 \,, \\
p'' +\left( \frac{3h'}{h} +\frac{\chi'}{2} -\frac{3}{r} \right) p' -\frac{r^2A_z'(A_t' +p A_z')}{h^2} &=0 \,,
\end{split}
\end{equation}
where the prime denotes the derivative with respect to $r$. The above equations of motion have the following scaling symmetry
\begin{equation} \label{SM-A-scaling}
(r,\, t,\, x,\, y,\, z)\to \lambda (r,\, t,\, x,\, y,\, z) \,,\quad (A_t,\, A_z) \to \lambda^{-1}(A_t,\, A_z) \,,\quad B\to \lambda^{-2}B,\quad (f, \chi, h, p)\rightarrow(f, \chi, h, p)  \,,
\end{equation}
with $\lambda$ a positive constant.

To solve the above equations, we first need to determine the asymptotic expansions of the bulk fields. Near the AdS boundary $r=0$, we obtain the following expansions
\begin{equation}
\begin{split} \label{SM-A-uvseries}
f(r)&= 1+\frac{B^2}{2}r^4\ln{r}+f_4 r^4+\cdots \,, \quad \chi(r)= \chi_0+\frac{B^2}{4}r^4\ln{r}+\left( \frac{B^2}{12}+2h_4 \right) r^4+\cdots  \,, \\
h(r)&= 1+\frac{B^2}{8}r^4\ln{r}+h_4 r^4+\cdots  \,, \quad p(r)= p_4 r^4+\cdots    \,, \\
A_t(r)&= e^{-\chi_0/2} \left( \mu +A_{t2} r^2+\cdots  \right) \,,\qquad 
A_z(r)= A_{z2} r^2 +\cdots  \,.
\end{split}
\end{equation}
To match the Hawking temperature with the boundary field theory temperature, $\chi_0$ must ultimately be set to zero in the final expressions. However, to account for temperature variations in the Euclidean action, it is kept as a free parameter to properly incorporate temperature variations when carrying out variation of Euclidean action and will be set to zero in the final expressions presented in the main text.

Near the event horizon $r=r_h$, we impose the conditions $f(r_h)=A_t(r_h)=0$ to ensure a smooth horizon. Additionally, by applying the $\alpha\, symmetry$ transformation: $z\to z+\alpha t \,, p\to p -\alpha \,, A_t\to A_t +\alpha A_z$ with $\alpha$ being a constant, we can set $p(r_h)=0$. Near the BH horizon, the field expansions are given by
\begin{equation}
\begin{split}
f(r)&=f_1(r_h -r)+\cdots \,,\quad \chi(r)=\chi_1+\cdots \,,\qquad
h(r)= h_1+\cdots  \,, \\
p(r)&=p_1(r_h -r)+\cdots \,,\quad A_t(r)=A_{t1}(r_h -r)+\cdots \,,\quad A_z(r)=A_{z1}+\cdots  \,.
\end{split}
\end{equation}
Note that the scaling symmetry~\eqref{SM-A-scaling} can be used to fix the location of the BH horizon to $r_h=1$, which is convenient for numerical calculations.

With the background solutions determined, we can study the thermodynamic properties of the dyonic BHs. The on-shell action suffers from divergence and we need to regulate it by adding appropriate counterterms at the AdS boundary. The total renormalized action is given by
\begin{equation}\label{SM-A-sren}
S_{ren}=S+S_{bdy}\,,
\end{equation}
with the boundary action
\begin{equation} \label{SM-A-sbdy}
S_{bdy}=\frac{1}{16\pi G}\int d^4x\sqrt{-\gamma}\Big[ 2K-\frac{6}{L}-\frac{L\hat{R}}{2}+\frac{L^3}{4}\ln{ (\frac{r}{L}) } \left( \hat{R}_{\mu\nu}\hat{R}^{\mu\nu} -\frac{\hat{R}^2}{3}- \frac{\hat{F}^2}{L^2} \right) \Big]\,.
\end{equation}
Here $\gamma_{\mu\nu}$ is the induced metric at the conformal boundary, $K$ is the trace of extrinsic curvature, and $\hat{R}_{\mu\nu}$ denotes the Ricci tensor associated with $\gamma_{\mu\nu}$. Moreover, $\hat{F}$ is the field strength of the boundary $U(1)$ gauge field.

To obtain analytical expressions for the thermodynamic variables, we can utilize the properties of the equations of motion and the Ricci tensor to rewrite the on-shell action in terms of surface terms. The Einstein equation implies that its $(i,i)$ component satisfies $2R^i_{\ i} =R+\frac{12}{L^2}-\frac{1}{4}F^2 +F^i_{\ c}F_i^{\ c}\,, (i=t,x,y,z)$. Consequently, the Lagrangian density $\mathcal{L}$ can be rewritten as
\begin{equation} \label{SM-A-onshellL}
\mathcal{L} = R +\frac{12}{L^2}-\frac{1}{4}F^2 +\frac{k}{24} \epsilon^{abcde}A_a F_{bc} F_{de} = 2R^i_{\ i} -F^i_{\ c}F_i^{\ c}  +\frac{k}{24} \epsilon^{abcde}A_a F_{bc} F_{de}  \,,
\end{equation}
where $i$ represents again one of the coordinates $(t,x,y,z)$. Using the identity $R^a_{\ b}\xi^b=\nabla_b\nabla^a\xi^b$ for any Killing vector $\xi^a$, we can express $R^i_{\ i}$ as a total derivative. Specifically, one has
\begin{equation}
R^a_{\ b}\xi^b = \nabla_b\nabla^a\xi^b =-\nabla_b \nabla^b \xi^a = -\frac{1}{\sqrt{-g}} \partial_b \left( \sqrt{-g} g^{bc}\nabla_c\xi^a \right) =  -\frac{1}{\sqrt{-g}} \partial_r \left( \sqrt{-g} g^{rc}\nabla_c\xi^a \right) \,,
\end{equation}
where we have used the fact that the background fields depend only on the radial coordinate $r$. For each Killing vector $\partial_t, \partial_x$ and $\partial_z$, we obtain
\begin{equation}
\begin{split}
2\sqrt{-g}R^t_{\ t} &= \left[\frac{e^{-\chi/2}h}{r^4} \left( -r f'+r e^\chi h^2 p p' +f(2+r \chi') \right) \right]'  \,,\quad 
2\sqrt{-g}R^x_{\ x} = 2\sqrt{-g}R^y_{\ y} =  \left[\frac{2e^{-\chi/2} f h}{r^4} \right]'  \,,  \\
2\sqrt{-g}R^z_{\ z} &=  \left[\frac{e^{-\chi/2}}{r^4} \left( 2f(h-rh') -r e^{\chi} h^3 p p' \right) \right]' \,.
\end{split}
\end{equation}

Using the expression for the Lagrangian density~\eqref{SM-A-onshellL}, we can obtain an analytical expression for the bulk on-shell action. After some calculations, the on-shell action is given by
\begin{eqnarray}
S&=& \int d^4x\int dr \Big[ 2\sqrt{-g}R^t_{\ t}+\left(A_t N_t+\frac{k}{3}B A_t A_z \right)'+\frac{k}{3}B A_t A_z' \Big]\,, \label{SM-A-os1}   \\
&=& \int d^4x\int dr \Big[ 2\sqrt{-g}R^x_{\ x} -\frac{B^2e^{-\chi/2}h}{r} +\frac{k}{3} B (A_t' A_z -A_t A_z' ) \Big]\,,   \label{SM-A-os2}  \\
&=& \int d^4x\int dr \Big[ 2\sqrt{-g}R^z_{\ z}+\left( A_z N_z -\frac{k}{3}B A_t A_z \right)' -\frac{k}{3}B A_t' A_z \Big]\,,  \label{SM-A-os3}
\end{eqnarray}
where $N_t\equiv \frac{e^{\chi/2} h}{r}(A_t'+p A_z')$ and $N_z\equiv \frac{e^{\chi/2}}{r h} \Big(p h^2 A_t'- (f e^{-\chi} -h^2p^2) A_z' \Big)$. Additionally, we have $N_t'=-kB A_z'$ and $N_z'=kB A_t'$ as derived from the Maxwell equation.

The expectation values of the boundary stress tensor and current are obtained by varying the renormalized action~\eqref{SM-A-sren} with respect to the boundary metric $\gamma_{\mu\nu}$ and the boundary gauge field $a_\mu$, respectively. The general expressions for $\langle T_{\mu\nu}\rangle$ and $\langle J^\mu\rangle$ in the five-dimensional EMCS theory take the form
\begin{equation}\begin{split} \label{SM-A-TmnJm}
\langle T_{\mu\nu} \rangle &=  \lim_{r\to0} \frac{L^2}{r^2} \bigg[ -2K_{\mu\nu} +2\Big( K-\frac{3}{L} \Big)\gamma_{\mu\nu} +L\hat{G}_{\mu\nu} +L\ln{(\frac{r}{L})} \left( \hat{F}_{\mu\rho}\hat{F}_\nu^{\ \rho}-\frac{ \gamma_{\mu\nu} }{4} \hat{F}^2 - L^2 h_{\mu\nu}^{(4)} \right) \bigg] \,, \\
\langle J^{\mu} \rangle  &= \lim_{r\to0} \frac{L^4}{r^4} \Big[ n_r \Big( F^{\mu r} +\frac{k}{6}\epsilon^{r\mu \alpha\beta\gamma} A_\alpha F_{\beta\gamma} \Big) +L \ln{(\frac{r}{L})} \nabla_\alpha \hat{F}^{\alpha\mu}  \Big] \,,
\end{split}\end{equation}
where $h_{\mu\nu}^{(4)}$ is defined as $h_{\mu\nu}^{(4)}= \hat{R}_{\mu\rho\nu\lambda}\hat{R}^{\rho\lambda} -\frac{1}{6}\hat{\nabla}_\mu \hat{\nabla}_\nu \hat{R}+\frac{1}{2}\hat{\nabla}^2 \hat{R}_{\mu\nu}-\frac{1}{3}\hat{R}\hat{R}_{\mu\nu}-\frac{\gamma_{\mu\nu}}{4}(\hat{R}_{\rho\lambda}\hat{R}^{\rho\lambda}-\frac{\hat{R}^2}{3}+\frac{\hat{\nabla}^2 \hat{R} }{3})$.

By using the boundary expansions~\eqref{SM-A-uvseries}, the non-vanishing components of $\langle T_{\mu\nu}\rangle$ and $\langle J^\mu\rangle$ are found to be
\begin{equation}\label{SM-A-TJmn}
\begin{split}
&\langle T_{tt}\rangle =e^{-\chi_0}\left(-3f_4+8h_4+\frac{B^2}{4} \right) \,,\qquad  \mathcal{P}_\perp=\langle T_{xx}\rangle=\langle T_{yy}\rangle= -f_4 -\frac{B^2}{6}\,,\\ 
& \mathcal{P}_\parallel= \langle T_{zz}\rangle=-f_4+8h_4+\frac{B^2}{12} \,,\qquad\qquad  \langle T_{tz}\rangle=\langle T_{zt}\rangle=4p_4 \,,\\
&\langle J^t \rangle=-2e^{\chi_0/2} A_{t2} ,\,\qquad\qquad\qquad  \langle J^z \rangle= -2A_{z2} -\frac{kB\mu}{3}\,,
\end{split}
\end{equation}
where we have adopted the consistent current that includes the Chern–Simons contribution. Following~\cite{Balasubramanian:1999re}, the energy density $\epsilon$ and charge density $\rho$ of the BHs are defined as
\begin{equation} \label{SM-A-energycharge}
\epsilon =u^\mu u^\nu \langle T_{\mu\nu}\rangle=  -3f_4 +8h_4 +\frac{B^2}{4} \,,\quad  \rho= u_\mu \langle J^\mu\rangle =-2A_{t2} \,,
\end{equation}
where $u^\mu=-e^{\chi_0/2}\delta_t^\mu$ is the timelike normal vector. Note that using the equations of motion for gauge field together with the UV and IR asymptotic expansions, one finds $A_{z2}=-kB\mu/2$, and hence $\langle J^z\rangle=2kB\mu/3$. Furthermore, the chiral anomaly leads to $p_4=-kB\mu^2/8$, which implies the momentum density $\langle T_{tz}\rangle=-kB\mu^2/2$~\cite{Ammon:2016szz}. As verified numerically, both $\langle J^z\rangle$ and $\langle T_{tz} \rangle$ remain constant at fixed $B$ and exhibit no temperature dependence.

To obtain the on-shell Euclidean action $I_E$, we perform an analytic continuation by Wick-rotating $t$ to the Euclidean time coordinate $\tau=i\, t$. To ensure that both the metric and gauge field are real in the Euclidean signature, we also set $A_{t(E)}=-i A_t,\, p_{(E)}=-ip$. Consequently, the Euclidean metric and gauge field are given by
\begin{equation}
\begin{split}
ds^2_{(E)}&= \frac{1}{r^2}\Big[ \big(f e^{-\chi} +h^2 p_{(E)}^2\big)d\tau^2+2p_{(E)}h^2 d\tau dz + dx^2+dy^2 +h^2 dz^2+\frac{dr^2}{f}\Big]\,,  \\
A_{(E)}&= A_{t(E)} d\tau-\frac{B}{2}y dx+\frac{B}{2}x dy-A_z dz \,.
\end{split}
\end{equation}

Then, the Euclidean action is defined as $I_{E}=-i S_{ren}$ and the free energy density from QSR can be expressed in three equivalent forms
\begin{equation} 
\begin{split} \label{SM-A-wQSR}
w_{QSR} = -\frac{ T \ln Z}{V}=\frac{ T  I_{E} }{V} &= \epsilon -Ts-\mu\rho -\frac{e^{\chi_0/2} k B}{3}\int_0^{r_h} A_t\, A_z' d r \,, \\
&= -\langle T_{xx} \rangle + e^{\chi_0/2} B \int_0^{r_h} \Big[ \frac{B}{r} \left( e^{-\frac{\chi}{2}} h -e^{\chi_0/2} \right) +\frac{k}{3} A_t A_z' \Big]d r +B^2 \ln r_h  \,,  \\
&= -\langle T_{zz} \rangle +\frac{e^{\chi_0/2} kB}{3}\int_0^{r_h} A_t' A_z d r  \,, 
\end{split}
\end{equation}
where we have used the relation $\int_0^{r_h} A_tA_z'dr =-\int_0^{r_h} A_t'A_z dr$, which follows from the boundary condition $A_t(r=r_h)=A_z(r=0)=0$.

\subsection*{A.2: Variation of Euclidean action} \label{SM-A-a2}
Substituting the background ansatz into the bulk action $S$, the effective action of the system is given by
\begin{equation} \label{SM-A-Seff}
\begin{split}
S_{eff}&=\int d^5x \frac{e^{-\chi/2} h}{r^3} \bigg\{ -f''-\frac{2fh''}{h}+f\chi'' +f'\left( \frac{8}{r}-\frac{2h'}{h}+\frac{3\chi'}{2} \right) +\left( \frac{h'}{r h} -\frac{\chi'}{2 r} \right) \left( 8f +r  f\chi' \right) \\
&\quad +\frac{1}{2}e^{\chi} \left[ h^2 p'^2 +r^2(A_t'+p A_z')^2 \right] +\frac{12-20f}{r^2}-\frac{r^2}{2}\left( B^2+\frac{fA_z'^2}{h^2} \right) \bigg\} +\int d^5x \frac{kB}{3}\left( A_t' A_z -A_t A_z'\right) \,.
\end{split}
\end{equation}
Next, we consider analytic continuation to Euclidean time $\tau$ by performing a Wick rotation $t=-i \tau$. The Euclidean action is given by
\begin{equation}\label{SM-A-IE}
I_{E}=I+I_{bdy}\,,
\end{equation}
where $I=-i S=-i S_{eff}$ and $I_{bdy}$ is the Euclidean boundary action obtained from~\eqref{SM-A-sren}. Consequently, the variation of the total on-shell Euclidean action can be separated into two parts: $\delta I_E = \delta I+\delta I_{bdy}$. The variation of $I$ yields
\begin{equation}
\begin{split}
\frac{\delta I}{\Delta\tau V}=&-\int dr \partial_r \Bigg\{ \frac{e^{-\chi/2}h}{r^3} \left[ \left( \frac{5}{r}-\frac{h'}{h}+\chi' \right) \delta f - \delta f' \right] +\frac{ 2e^{-\chi/2} f }{r^3} \left(  \frac{\delta h}{r} - \delta h' \right)   \\
&\quad +\frac{ e^{-\chi/2} f h }{r^3} \left[ \left( \frac{f'}{2f}-\frac{\chi'}{2}-\frac{1}{r} \right) \delta \chi+ \delta \chi' \right]+\frac{ e^{\chi/2} h^3 p' }{r^3} \delta p +\frac{kB}{3} A_z \delta A_t    \\
&\quad +\frac{e^{\chi/2}h(A_t'+p A_z')}{r} \left(\delta A_t +p\delta A_z\right) -\left(\frac{kB}{3} A_t+\frac{e^{-\chi/2} fA_z'}{rh} \right) \delta A_z\Bigg\}   \\
&-\int dr \left[ \frac{kB}{3} \left( A_z' \delta A_t -\delta A_z A_t' \right)-\left( \frac{Be^{-\chi/2}h}{r} +\frac{k}{3}(A_t A_z' -A_t' A_z)\right) \delta B \right]\,,
\end{split}
\end{equation}
while the variation from boundary terms $I_{bdy}$ is given by
\begin{equation}
\begin{split}
\frac{\delta I_{bdy}}{\Delta \tau V} = &-\bigg\{ \frac{e^{-\chi/2} h}{r^3} \left[ \left( \frac{8}{r} -\frac{12+B^2r^4 \ln r}{4r \sqrt{f} } -\frac{2h'}{h}+\chi' \right) \delta f -\delta f' \right]  \\
&+ \frac{e^{-\chi/2} f}{r^3} \left[ \left( \frac{8}{r} -\frac{12+B^2r^4 \ln r}{2r\sqrt{f} } -\frac{f'}{f}+\chi' \right) \delta h -2\delta h' \right]  \\
&+ \frac{e^{-\chi/2} f h}{2r^3} \left[ \left( -\frac{8}{r} +\frac{12+B^2r^4 \ln r}{2r\sqrt{f} } +\frac{f'}{f}+\frac{2h'}{h}-\chi' \right) \delta \chi +2\delta \chi' \right]  \\
&-\left( Be^{-\chi/2} \sqrt{f}h \ln{r} \right) \delta B 
\bigg\} \bigg\lvert_{r\to 0} \,.
\end{split}
\end{equation}
Here, $\Delta \tau \equiv \int d\tau$ and $V\equiv \int dxdydz$.

Note that in the variation of $I$, there is a bulk integral term given by $k\int \frac{B}{3}(A_z' \delta A_t- \delta A_z A_t') dr$. This suggests a limitation in the method used for obtaining the background equations directly from the variation of the effective action~\eqref{SM-A-Seff}. Nevertheless, it is possible to convert this bulk integral into a total derivative term by observing that
\begin{equation}
A_z' \delta A_t- \delta A_z A_t' = \frac{1}{2} \delta \left(A_z' A_t -A_z A_t'\right) -\frac{1}{2} \left[ \delta A_z A_t-A_z \delta A_t\right ]'\,.
\end{equation}
Thus, we obtain
\begin{equation} \label{SM-A-dcs}
\begin{split}
&\int dr \frac{k B}{3} \left( A_z' \delta A_t-\delta A_z A_t' \right)   \\
&= k \delta \left[ \frac{B}{6}\int dr \left(A_z' A_t -A_z A_t'\right) \right]  -\left[ \frac{k}{6}\int dr \left(A_z' A_t -A_z A_t'\right) \right] \delta B -\frac{k B}{6}\big[ \delta A_z A_t-A_z \delta A_t\big]^{r_h}_0\,,   \\
&= k \delta \left[ \frac{B}{6}\int dr \left(A_z' A_t -A_z A_t'\right) \right]  -\left[ \frac{k}{6}\int dr \left(A_z' A_t -A_z A_t'\right) \right] \delta B \,,
\end{split}
\end{equation}
where we have used the boundary conditions $A_t(r=r_h)=A_z(r=0)=0$.

In the Euclidean metric, the spacetime near the black brane horizon resembles a cigar geometry, and hence the only remaining boundary is the AdS boundary. Using the asymptotic expansions, the variation of the total on-shell Euclidean action~\eqref{SM-A-IE} is given by
\begin{equation} \label{SM-A-deltaI}
\delta I_{E} = \delta I+\delta I_{bdy}  =\Delta\tau V \bigg\{ e^{-\chi_0/2} \left[ -\rho \delta\mu -\left(\epsilon -\mu\rho \right) \frac{\delta \chi_0}{2} \right] -k \delta Q_{cs} - e^{-\chi_0/2}M_B \delta B  \bigg\} \,,
\end{equation}
where $Q_{cs}$ is defined as
\begin{equation} \label{SM-A-Qcs}
Q_{cs}=\frac{B}{6}\int_0^{r_h} (A_z' A_t- A_z A_t') dr \,,
\end{equation}
and $M_B$ is given by
\begin{equation} \label{SM-A-MB}
M_B= -\Bigg( \int_0^{r_h} \bigg[ \frac{B}{r} \Big( e^{(\chi_0-\chi)/2} h - 1\Big) +\frac{e^{\chi_0/2} k}{2}(A_z' A_t - A_z A_t') \bigg] dr + B\ln{r_h} \Bigg) \,.
\end{equation}
Note that in the final expression, the constant $\chi_0$ in $M_B$ should be set to zero to ensure that the Hawking temperature equals the temperature at the boundary. The BH temperature is given by $T=\frac{e^{\chi_0/2}}{\Delta\tau}$. With $\Delta\tau$ fixed, the variation of $\chi_0$ is $\delta\chi_0 = 2\delta T/T$. With this relation, together with~\eqref{SM-A-deltaI} and the QSR, the variation of $w_{QSR}$ takes the form
\begin{equation} \label{SM-A-dwQSR}
\delta w_{QSR}= - \left( s + e^{\chi_0/2 }\frac{kQ_{cs}}{T} \right) \delta T -\rho \delta\mu - e^{\frac{\chi_0}{2} } k \delta Q_{cs} - M_B \delta B \,.
\end{equation}

After adding the Chern-Simons correction to $w_{ QSR}$, the corrected free energy is defined as $w= w_{ QSR} +e^{\chi_0/2}k Q_{ cs}$. Its variation is then given by
\begin{equation}
\delta w
=\delta w_{ QSR}
+\delta \left( e^{\chi_0/2}\right)kQ_{ cs}
+e^{\chi_0/2}k \delta Q_{ cs}
= -s\delta T-\rho\delta\mu-M_B\delta B \,.
\end{equation}
Thus, the free energy $w$ satisfies the standard first law of BH thermodynamics, in agreement with the discussion in the main text.

{\bf \emph{Alternative derivation of $\delta w_{QSR}$ via scaling arguments:}} Note that in the above analysis, the variation of temperature $\delta T$ is introduced by varying the parameter $\chi_0$. In contrast, one can adopt the approach of~\cite{Donos:2013cka} and introduce the temperature variation through a rescaling of the boundary coordinates. This approach yields exactly the same results, and it has the advantage that $\chi_0$ can be set to zero from the beginning. This not only further validates the above results but also paves the way for the Iyer-Wald formalism in the following section.

In the following, we illustrate how to derive the result~\eqref{SM-A-deltaI} with this method. The stationary BHs considered here are asymptotically AdS, with boundary metric and gauge field given by
\begin{equation}
ds_{bdy}^2 =\gamma_{\mu\nu} dx^\mu dx^\nu \,,\quad A_{bdy} = a_\mu dx^\mu + \frac{B}{2} \left( xdy -ydx\right) \,.
\end{equation}
The variation of the total renormalized on-shell action~\eqref{SM-A-sren} reads
\begin{equation}\begin{split}
\delta S_{ren}&=\delta S+\delta S_{bdy}=\int d^5x \sqrt{-g} \left( E_g^{ab}\delta g_{ab} +E_A^{a}\delta A_a +\nabla_a\theta^a \right) +\delta S_{bdy} \\
&=  \int d^5x \sqrt{-g} \left( E_g^{ab}\delta g_{ab} + E_A^{a} \delta A_a \right) + \int d^4x \sqrt{-\gamma} \left( n_a \theta^a \right) +\delta S_{bdy} \\
&=\int d^4x \sqrt{-\gamma} \left( \frac{1}{2} \langle T^{\mu\nu}\rangle\delta \gamma_{\mu\nu} + \langle J^\mu\rangle \delta a_\mu\right) +\int d^4x \sqrt{-\gamma_i} n_i \theta^i - \ln{(\epsilon^+)} B \delta B\,,
\end{split}\end{equation}
where $E_g^{ab}$ and $E_A^{a}$ are the equations of motion for the metric and the gauge field (see~\eqref{SM-A-eom} above), and $\nabla_a\theta^a$ is a total‑derivative term. To arrive at the last line, the on-shell condition has been applied. The term proportional to $\delta B$ arises from varying the boundary action $\delta S_{bdy}$ in the presence of a background magnetic field. Here, $\epsilon^+$ is an infinitesimal regulator introduced in the variation procedure, which will be absorbed in the final expressions.

Notably, the term $\sqrt{-\gamma_i} n_i \theta^i$ accounts for contributions from the spatial boundaries $(x, y, z)$, which become non-trivial in the presence of a background magnetic field. For the planar BH herein, we find $\sqrt{-\gamma_z}n_z\theta^z = 0$, while
\begin{equation}
\frac{ \sqrt{-\gamma_x} n_x \theta^x}{x} =\frac{ \sqrt{-\gamma_y} n_y \theta^y}{y} = \frac{1}{2} \left[ \frac{kB}{3} \left( A_z' \delta A_t -\delta A_z A_t' \right)-\left( \frac{Be^{-\chi/2}h}{r} +\frac{k}{3}(A_t A_z' -A_t' A_z)\right) \delta B \right] \,.
\end{equation}
Using expression~\eqref{SM-A-dcs}, the variation of $S_{ren}$ can be rewritten as
\begin{equation}\begin{split}
\delta S_{ren} &=\int d^4x \sqrt{-\gamma} \left( \frac{1}{2} \langle T^{\mu\nu}\rangle\delta \gamma_{\mu\nu} + \langle J^\mu\rangle \delta a_\mu\right) +\int dt \int d^3x \left( k \delta Q_{cs} + M_B\delta B \right) \,,
\end{split}\end{equation}
where $Q_{cs}$ and $M_B$ denote the radial integrals defined precisely in~\eqref{SM-A-Qcs} and~\eqref{SM-A-MB}.

Upon performing the Wick rotation $\tau = i t$ with period $\Delta\tau = 1/T$, the Euclidean action is defined as $I_E = -i S_{\text{ren}}$. Assuming the boundary metric $\gamma_{\mu\nu}$ and gauge field $a_\mu$ are periodic in $x^i$ with periods $L_i$, the variation of $I_E$ with respect to ${\gamma_{\mu\nu}, a_\mu, B}$ while holding $\Delta\tau$ and $L_i$ fixed takes the form
\begin{equation}
\delta I_E =- \int_0^{\Delta\tau} \int_0^{\{L_i\}} d\tau d^3x \sqrt{-\gamma} \left( \frac{1}{2} \langle T^{\mu\nu}\rangle\delta \gamma_{\mu\nu} + \langle J^\mu\rangle \delta a_\mu\right) -\int_0^{\Delta\tau} \int_0^{\{ L_i\}} d\tau d^3x  \left( k\delta Q_{cs} +M_B\delta B \right) \,.
\end{equation}

To obtain the variation with respect to the periods $\Delta\tau, L_i$, we rescale the boundary coordinates as $\tau =\Delta\tau \tilde{\tau} \,, x^i =L_i \tilde{x}^i$. Under this rescaling the Euclidean action satisfies
\begin{equation} \label{SM-A-Iss}
I_{E}(\gamma_{\mu\nu} , a_\mu, B ; \Delta\tau , L_i) =I_{E} (\tilde{\gamma}_{\mu\nu}, \tilde{a}_\mu, \tilde{B} ; 1,1) \,,
\end{equation}
with
\begin{equation}
\begin{split}
\tilde{\gamma}_{\mu\nu} d\tilde{x}^\mu d\tilde{x}^\nu &= \gamma_{\tau\tau} (\Delta\tau)^2 d\tilde{\tau}^2 +2 \gamma_{\tau x^i} \Delta\tau L_i d\tilde{\tau} d\tilde{x}^i +\gamma_{x^ix^j} L_iL_j d\tilde{x}^i d\tilde{x}^j \,, \\
\tilde{a}_\mu d\tilde{x}^\mu + \frac{1}{2}\tilde{B} \left( \tilde{x}^i d\tilde{x}^j - \tilde{x}^j d\tilde{x}^i \right) &= a_\tau \Delta\tau d\tilde{\tau} +a_{x^i} L_i d\tilde{x}^i +\frac{1}{2}B L_i L_j \left( \tilde{x}^i d\tilde{x}^j -\tilde{x}^j d\tilde{x}^i \right) \,.
\end{split}
\end{equation}
Using the chain rule and the relation $\sqrt{ \tilde{\gamma}}=\Delta\tau V\sqrt{ \gamma}$ (where $V=\prod_i L_i$), the total variation becomes
\begin{equation}\begin{split}
\delta I_E &=\delta\tilde{I}_E -\Delta\tau V \int_0^1 d^4\tilde{x} \sqrt{\gamma} \left( \frac{1}{2} \langle T^{\tau\tau} \rangle \gamma_{\tau\tau} \frac{\delta \left( (\Delta\tau)^2 \right)}{(\Delta\tau)^2 }  +\langle T^{\tau x^i} \rangle \gamma_{\tau x^i} \frac{\delta \left( \Delta\tau L_i \right)}{\Delta\tau L_i}  +\frac{1}{2} \langle T^{x^i x^j}\rangle \gamma_{x^i x^j} \frac{\delta \left( L_iL_j \right)}{L_i L_j }  \right) \\
&\qquad -\Delta\tau V \int_0^1 d^4\tilde{x}  \sqrt{\gamma} \left( \langle J^\tau \rangle a_\tau \frac{\delta \Delta\tau}{\Delta\tau}  +\langle J^{x^i} \rangle a_{x^i} \frac{\delta L_i}{ L_i }  \right) -\Delta\tau V\int_0^1 d^4\tilde{x} M_B  \tilde{B} \delta(L_iL_j) \,,
\end{split}
\end{equation}
where we have used identities such as $\langle\tilde{T}^{\tau\tau}\rangle = \langle T^{\tau\tau}\rangle / (\Delta\tau)^2$. For the first law of BH thermodynamics, the temperature variation is tied to the change in the Euclidean time period $\delta(\Delta\tau)$, while variations of the spatial periods $\delta L_i$ do not contribute and are therefore set to zero. Noting that $\delta \Delta\tau/\Delta\tau = -\delta T/T$, the above expression simplifies to
\begin{equation}\begin{split}
\delta I_E  &= \delta \tilde{I}_E +\frac{\delta T}{T} \int_0^{\Delta\tau} \int_0^{\{L_i\}} d\tau d^3x \sqrt{\gamma} \left( \langle T^{\tau\tau} \rangle \gamma_{\tau\tau}   +\langle T^{\tau x^i} \rangle \gamma_{\tau x^i} +\langle J^\tau\rangle a_\tau \right) \,.
\end{split}\end{equation}

Therefore, the variation of free energy density $w_{QSR}=TI_E/V$ obtained from the QSR is
\begin{equation} \begin{split} \label{SM-A-dwss}
\delta w_{QSR} &= -V^{-1} \int_0^{\{ L_i\} } d^3x \sqrt{-\gamma} \left( \frac{1}{2} \langle T^{\mu\nu} \rangle \delta \gamma_{\mu\nu} + \langle J^\mu\rangle \delta a_\mu \right) -V^{-1} \int_0^{\{ L_i\} } d^3x \sqrt{-\gamma} \left( k \delta Q_{cs} + M_B \delta B \right)  \\
& \qquad - \frac{1}{T}\left[ w_{QSR} + V^{-1} \int_0^{\{ L_i\} } d^3x \sqrt{-\gamma} \left( T^{tt} \gamma_{tt} +T^{t x^i} \gamma_{tx^i} +J^t a_t \right) \right] \delta T \,,
\end{split} \end{equation}
where the integrands are now expressed in Lorentzian signature.

For the planar BH studied here, the boundary metric is flat, as $\delta \gamma_{\mu\nu}=0$. Substituting the explicit expressions for the free energy~\eqref{SM-A-wQSR}, the energy-momentum tensor, and the current~\eqref{SM-A-TJmn}, we finally obtain
\begin{equation}
\delta w_{QSR} = - \left( s +\frac{kQ_{cs}}{T} \right) \delta T -\rho \delta\mu - k \delta Q_{cs} - M_B \delta B \,,
\end{equation}
which reproduces exactly~\eqref{SM-A-dwQSR}. To resolve this inconsistency, it is necessary to incorporate the contribution of $Q_{ cs}$ into the free energy. Consequently, for the planar BH~(3), the free energy $W$ becomes
\begin{equation}
W=T \left( I_E + \int_0^{\Delta\tau}\int_0^{\{L_i\}} d\tau d^3x kQ_{cs}\right) \,,
\end{equation}
so that the corresponding free energy density is $w=W/V=w_{QSR} +kQ_{cs}$. Following the procedure outlined in~(\ref{SM-A-Iss}-\ref{SM-A-dwss}), the variation of free energy density $\delta w$ yields
\begin{equation}
\delta w =-s\delta T -\rho\delta\mu -M_B\delta B \,.
\end{equation}
%

\subsection*{A.3: Resolution from Iyer-Wald formalism} \label{SM-A-a3}
The Iyer-Wald formalism provides a powerful framework for studying BH thermodynamics in any diffeomorphism invariant gravitational theory~\cite{Iyer:1994ys}. We begin by summarizing its key elements and then discuss its application to the thermodynamics of the particular BHs considered in this work.

{\bf\emph{Brief introduction to the Iyer-Wald formalism:}} The variation of Lagrangian $n$-form $\mathbf{L}=\mathcal{L} \epsilon$ is given by
\begin{equation} \label{SM-A-varyL}
\delta\mathbf{L} = \mathbf{E}\delta \phi +d\mathbf{\Theta} (\phi,\delta\phi) \,,
\end{equation}
where $\mathbf{E}^\phi$ represents the equations of motion for the dynamical fields $\phi=(g_{ab}, \psi)$ and $\mathbf{\Theta}(\phi,\delta\phi)$ is the symplectic potential form. The symplectic current $(n-1)$-form can be defined from $\mathbf{\Theta}$ as
\begin{equation}  \label{SM-A-sc}
\boldsymbol{\omega}(\phi,\delta_1\phi,\delta_2\phi)=\delta_1\mathbf{\Theta}(\phi,\delta_2\phi) -\delta_2\mathbf{\Theta}(\phi,\delta_1\phi) \,, 
\end{equation}
where $\delta_1\phi$ and $\delta_2\phi$ are variations of the dynamical fields.

Considering an infinitesimal diffeomorphism $\delta_\xi x^a=\xi^a(x)$, one can associate with it a Noether current $(n-1)$-form, defined by
\begin{equation} \label{SM-A-nc}
\mathbf{J}\equiv \mathbf{\Theta}(\phi,\mathcal{L}_\xi \phi) - \xi \cdot \mathbf{L} \,,
\end{equation}
where $\xi\cdot$ denotes the contraction of $\xi^a$ with the first index of $\mathbf{L}$. Under the infinitesimal diffeomorphism transformations $\delta_\xi x^a=\xi^a$, the variation of the dynamical fields equals their Lie derivatives: $\delta_\xi g_{ab}=\mathcal{L}_\xi g_{ab}, \delta_\xi \psi=\mathcal{L}_\xi \psi$. A standard calculation shows that
\begin{equation}
d\mathbf{J}=-\mathbf{E} \mathcal{L}_\xi \phi  \,,
\end{equation}
which implies that $\mathbf{J}$ is closed for all $\xi^a$ once the equations of motion are satisfied. Thus, there exists a Noether charge $(n-2)$-form $\mathbf{Q}$, locally constructed from $\phi$ and $\xi^a$, such that $\mathbf{J}=d\mathbf{Q}$ provided $\phi$ solve the equations of motion. Particularly, using the algorithm of~\cite{Wald:1990mme}, one can construct $\mathbf{Q}$ from $\mathbf{J}$.

With all the necessary ingredients in place, we can now obtain the fundamental identity following \cite{Iyer:1994ys}. To begin, considering the variation of~\eqref{SM-A-nc} and using~\eqref{SM-A-varyL} and~\eqref{SM-A-sc}, we have
\begin{equation}
\begin{split}
\delta \mathbf{J} &=\delta \mathbf{\Theta}(\phi,\mathcal{L}_\xi \phi) -\xi\cdot \delta\mathbf{L}  = \delta \mathbf{\Theta}(\phi,\mathcal{L}_\xi \phi) -\xi\cdot ( \mathbf{E}\delta\phi +d\mathbf{\Theta} ) \,, \\
&= \delta\mathbf{\Theta}(\phi,\mathcal{L}_\xi \phi) -\mathcal{L}_\xi \mathbf{\Theta}(\phi,\delta\phi) +d(\xi\cdot\mathbf{\Theta}) -\xi\cdot \mathbf{E}\delta\phi \,,  \\
&= \boldsymbol{\omega}(\phi,\delta\phi,\mathcal{L}_\xi \phi) +d(\xi\cdot\mathbf{\Theta}) -\xi\cdot \mathbf{E}\delta\phi \,,
\end{split}
\end{equation}
where we have used Cartan's magic formula $\mathcal{L}_\xi \boldsymbol{\mu}=\xi\cdot d\boldsymbol{\mu} +d(\xi\cdot\boldsymbol{\mu})$ for any differential form $\boldsymbol{\mu}$ and vector field $\xi^a$ in the second line. When $\xi^a$ corresponds to a symmetry of the dynamical fields (\emph{i.e.} $\mathcal{L}_\xi\phi=0$), it follows that $\boldsymbol{\omega}(\phi,\delta\phi,\mathcal{L}_\xi\phi)=0$. If $\phi$ satisfies the equations of motion, then $\mathbf{E}=0$ and $\mathbf{J}=d\mathbf{Q}$. Moreover, if $\delta\phi$ satisfies the linearized equations of motion, we find that $\delta\mathbf{J}=d\delta\mathbf{Q}$. Consequently, we obtain the famous fundamental identity
\begin{equation}\label{SM-A-fundamentalID}
\boldsymbol{\omega}(\phi,\delta\phi, \mathcal{L}_\xi\phi) = \delta \mathbf{J} -d(\xi \cdot \boldsymbol{\Theta} ) =d \left( \delta \mathbf{Q} -\xi \cdot \boldsymbol{\Theta} \right) =0 \,.
\end{equation}
On the other hand, the conditions that the dynamical fields $\phi=(g_{ab},\psi)$ solve the background equations of motion and $\xi$ is a Killing vector (\emph{i.e.} $\mathcal{L}_\xi \phi=0$) imply
\begin{equation} \label{SM-A-identity}
d\mathbf{Q} =\mathbf{J}=\mathbf{\Theta}(\phi,\mathcal{L}_\xi\phi)-\xi\cdot \mathbf{L} =-\xi \cdot \mathbf{L} \,.
\end{equation}

{\bf \emph{First law from Iyer-Wald formalism:}} According to the Iyer-Wald formalism, calculating the Noether charge and symplectic potential is essential for deriving the thermodynamic first law of a given gravitational theory. For the five-dimensional ($n=5$) EMCS theory, the Noether charge $3$-form $\mathbf{Q}$ and symplectic potential $4$-form $\mathbf{\Theta}$ are expressed as
\begin{equation}
\begin{split} \label{SM-A-QTheta}
\mathbf{Q} &=-\boldsymbol{\epsilon}_{abfgh} \left[ \nabla^a\xi^b + \frac{1}{2} \left( F^{ab}+\frac{k}{6}\epsilon^{aijkb} A_i F_{jk}\right)  A_c \xi^c \right] \,,\\
\mathbf{\Theta} &= \boldsymbol{\epsilon}_{a a_1a_2a_3a_4} \bigg[  \left( g^{ac}g^{bd}-g^{ab}g^{cd} \right) \nabla_b \delta g_{cd} -\left( F^{ae}+\frac{k}{6}\epsilon^{abcde} A_b F_{cd}\right) \delta A_e \bigg] \,.
\end{split}
\end{equation}

Let us choose $\xi^a$ to be the time-like Killing vector $\xi^a=(\partial_t)^a=\delta^a_t$, which becomes null at the event horizon $r=r_h$. Let $\Sigma$ be a $t=\mathrm{const.}$ space-like hypersurface with the horizon $r=r_h$ serving as its interior boundary. Then, the boundary of this hypersurface, $\partial\Sigma$, including the boundaries of the non-compact $x,\, y,\, z$ directions, is expressed as
\begin{equation}
\partial\Sigma = S_{r=r_h} \cup S_{r=0} \cup S_{x=L_x/2} \cup S_{x=-L_x/2} \cup
S_{y=L_y/2} \cup S_{y=-L_y/2} \cup S_{z=L_z/2} \cup S_{z=-L_z/2} \,,
\end{equation}
where we have considered the regulation $(L_x, L_y, L_z)$ along three spatial directions. Here $S_{r=r_h}$ denotes the boundary at $r=r_h$ and $S_{r=0}$ represents the boundary at $r=0$. The remaining terms account for the boundaries at the edges of the non-compact directions in the $x,y$ and $z$ coordinates.

Denote the intrinsic metric for $\Sigma$ and $S_\mu, (\mu=x,y,z,r)$ as $h_{\mu\nu}$ and $\gamma_{(\lambda)ij}$, respectively. Assume that $n_{(t)a}\propto \partial_a t$ is the time-like unit normal vector associated with $\Sigma$ and $n_{(\lambda)\mu}\propto \partial_\mu x_\lambda$ are the spacelike unit normal vectors associated with $S_\mu$. The induced volume element and the projection tensor associated with $\Sigma$ are
\begin{equation}
\hat{\boldsymbol{\epsilon} }_{dabc} = n_{(t)}^e \boldsymbol{\epsilon}_{edabc} \,,\quad P_{ab} = g_{ab}+n_{(t)a} n_{(t)b}\,.
\end{equation}
Note that $\Sigma$ is parameterized by the equations $x^a=x^a(y^\mu)$. Consequently, the vectors $e^a_{\ \mu}=(\partial x^a) /(\partial y^\mu)$ are tangent to $\Sigma$ and can be used to project any bulk tensor field onto the hypersurface $\Sigma$. For instance, the intrinsic metric and volume element of $\Sigma$ are given by $h_{\mu\nu}=g_{ab}e^a_{\ \mu} e^b_{\ \nu}=P_{ab}e^a_{\ \mu} e^b_{\ \nu}$ and $\boldsymbol{\epsilon}_{(4)\mu\nu\rho\sigma}=e^a_{\ \mu} e^b_{\ \nu} e^c_{\ \rho} e^d_{\ \sigma} \hat{\boldsymbol{\epsilon} }_{abcd} $. Following the same procedure, the induced volume element and projection tensor for $S_\mu$ are expressed as $\hat{\boldsymbol{\epsilon} }_{(\lambda)\mu\nu\rho} =n_{(\lambda)}^\sigma \boldsymbol{\epsilon}_{(4)\sigma\mu\nu\rho}$ and $P_{(\lambda)\mu\nu} = h_{\mu\nu}-n_{(\lambda)\mu}n_{(\lambda)\nu}$, respectively. Additionally, the intrinsic metric and volume element for $S_\mu$ are $\gamma_{ij}=h_{\mu\nu}e^\mu_{\ i} e^\nu_{\ j}=P_{(\lambda)\mu\nu}e^\mu_i e^\nu_j$ and $\boldsymbol{\epsilon}_{(i)jkl}=e^\mu_{\ j} e^\nu_{\ k} e^\rho_{\ l} \hat{\boldsymbol{\epsilon}}_{\mu\nu\rho} $ where $e^\mu_{\ i}=(\partial x^\mu) /(\partial y^i)$.

Note that the restriction of a bulk $3$-form to a three dimensional hypersurface is proportional to the volume element of that hypersurface. For instance, we have $\mathbf{Q}|_i=Q_i \boldsymbol{\epsilon}_{(i)}$, where the scalar field $Q_i$ is given by $Q_i=\mathbf{Q}_{ijk}\hat{\boldsymbol{\epsilon}}_{(i)}^{ijk}/6$ and $i\in \{x,y,z,r\}$. Therefore, we can compute the restriction of $(\delta\mathbf{Q}-\xi\cdot\mathbf{\Theta})$ to each hypersurface $S_i$. Specifically, we have $(\delta\mathbf{Q}-\xi\cdot\mathbf{\Theta})|_i= (\delta Q-\xi\cdot \Theta)_i \boldsymbol{\epsilon}_{(i)}$. Along each non-compact boundary, we find
\begin{equation}
\begin{split} \label{SM-A-bxyz}
(\delta Q -\xi \cdot \Theta)_x \sqrt{\gamma_{(x)}} &= x \bigg[ \left( \frac{B e^{-\chi/2} h }{2r}+\frac{k}{6}(2A_tA_z' -A_t'A_z) \right) \delta B +\frac{kB}{6} (A_t\delta A_z)' \bigg] \,, \\
(\delta Q -\xi \cdot \Theta)_y \sqrt{\gamma_{(y)}} &= \frac{y}{x} (\delta Q -\xi \cdot \Theta)_x \sqrt{\gamma_{(x)} }  \,,  \qquad  (\delta Q -\xi \cdot \Theta)_z \sqrt{\gamma_{(z)} } =0   \,.
\end{split}
\end{equation}

By integrating the fundamental identity~\eqref{SM-A-fundamentalID} over the hypersurface $\Sigma$, we obtain
\begin{equation}
\begin{split}   \label{SM-A-deltaH}
\delta H &=0=\int_\Sigma \boldsymbol{\omega}=\int_{\Sigma} d(\delta \mathbf{Q}- \xi \cdot \boldsymbol{\Theta}) =\int_{\partial\Sigma} (\delta \mathbf{Q} -\xi \cdot \boldsymbol{\Theta})    \\
&= \int_{ S_{ r=r_{h} } } (\delta Q -\xi \cdot \Theta)_r  \sqrt{\gamma_{(r)} } dxdydz 
-\int_{ S_{r=\epsilon^+} } (\delta Q -\xi \cdot \Theta)_r  \sqrt{\gamma_{(r)} } dxdydz \\
&\quad +\int_{S_{x=\frac{L}{2} } } (\delta Q -\xi \cdot \Theta)_x  \sqrt{\gamma_{(x)} } dydzdr 
-\int_{ S_{ x=-\frac{L}{2}} } (\delta Q -\xi \cdot \Theta)_x \sqrt{\gamma_{(x)} } dydzdr \\ 
&\quad +\int_{ S_{y=\frac{L}{2} } } (\delta Q -\xi \cdot \Theta)_y \sqrt{\gamma_{(y)} } dxdzdr 
-\int_{ S_{y=-\frac{L}{2} } } (\delta Q -\xi \cdot \Theta)_y \sqrt{\gamma_{(y)} } dxdzdr \\
&= 3\delta f_4-8\delta h_4 -\frac{B}{2}\delta B +e^{-\frac{\chi_1}{2}} f_1 \left( \frac{\delta h_1}{r_h^3} -\frac{3h_1 \delta r_h}{r_h^4} \right) +\mu\delta\rho +B (\ln{\epsilon^+}) \delta B  \\
& \qquad +\int_{\epsilon^+}^{r_h} dr  \left( \frac{B e^{-\chi/2} h }{r} +\frac{k}{2} (A_z' A_t -A_z A_t') \right) \delta B \,, 
\end{split}
\end{equation}
where we have used $\int_0^{r_h} (A_t\delta A_z)'dr= (A_t\delta A_z)|_0^{r_h}=0$. Note that the variation of the entropy density is $\delta s=\delta \left( \frac{4\pi h_1}{r_h^3} \right) =4\pi\left( \frac{\delta h_1}{r_h^3} -\frac{3h_1 \delta r_h}{r_h^4} \right)$ and the variation of energy density $\epsilon$ in~\eqref{SM-A-energycharge} reads $\delta \epsilon =-3\delta f_4 +8\delta h_4 +\frac{B}{2}\delta B$. By substituting the above equations into~\eqref{SM-A-deltaH}, we find
\begin{equation}
0= -\delta \epsilon  +T\delta s +\mu\delta\rho  +\Bigg( \int_0^{r_h} dr \Big[ \frac{B }{r} \left( e^{-\chi/2} h - 1\right)
+\frac{k}{2} (A_z' A_t -A_z A_t') \Big] +B\ln{r_h} \Bigg) \delta B \,.
\end{equation}
Thus, we arrive at the first law of BH thermodynamics
\begin{equation}\label{SM-A-first}
\delta \epsilon = T\delta s +\mu\delta\rho - M_B \delta B \,.
\end{equation}

On the other hand, using the Noether current $\mathbf{J}$,  we can also determine the expected free energy $w$ of the system. Integrating the identity~\eqref{SM-A-identity} over the hypersurface $\Sigma$ and applying Stokes's theorem, we find
\begin{equation}\label{SM-A-dQ}
\int_{\partial\Sigma} \mathbf{Q} = -\int_\Sigma \xi\cdot\mathbf{L} \,.
\end{equation}
Consequently, we have
\begin{equation}
\Big[ Q_r(r_h)-Q_r(\epsilon^+) \Big] \sqrt{\gamma_{(r)}} V = 
V \int_{\epsilon^+}^{r_h} \sqrt{-g}\mathcal{L} dr -V\int_{\epsilon^+}^{r_h} \frac{k B}{3} A_t A_z' dr \,,
\end{equation}
where $V=\int dxdydz=L_xL_yL_z$. Adding the boundary term $S_{bdy}$ of~\eqref{SM-A-sren} to both sides of the above equation and taking $\epsilon^+\to0$, we obtain
\begin{equation}  \label{SM-A-free}
w=\epsilon-Ts-\mu\rho \,.
\end{equation}
Combining~\eqref{SM-A-free} and~\eqref{SM-A-first} reproduces the results
obtained from the variation of the Euclidean action in the preceding
subsection.

Finally, using the expressions~\eqref{SM-A-os2} for the bulk on-shell action, we obtain from~\eqref{SM-A-dQ} and~\eqref{SM-A-TJmn} that
\begin{equation}
  \mathcal{P}_\parallel = \mathcal{P}_\perp+M_B B \,,
\end{equation}
with the longitudinal and transverse pressures $\mathcal{P}_\parallel$ and $\mathcal{P}_\perp$ defined in~\eqref{SM-A-TJmn}.

\subsection*{A.4: Transgression completion and the origin of $Q_{cs}$}
The Euclidean-variation and Iyer-Wald analyses above independently identify $kQ_{ cs}$ as an intrinsic contribution to the thermodynamic potential. Neither calculation, however, explains why this term is absent when the quantum statistical relation is applied to the theories containing the Chern-Simons term. We now show that $Q_{ cs}$ is precisely the finite contribution generated by its gauge-invariant transgression completion. The general construction is summarized in the End Matter of the main text; here we only apply the five-dimensional transgression form $\mathcal T_5(A,\bar A)$ to the planar background.

Replacing the Chern-Simons form $\mathcal C_5(A)$ by $\mathcal T_5(A, \bar A)$ gives the renormalized action
\begin{equation}  \label{SM-A-transgression-action}
S_{ ren}^{\mathcal T} =\frac{1}{16\pi G}\left[ \int_{\mathcal M_5}d^5x\,\sqrt{-g}\left(
R+\frac{12}{L^2}-\frac14 F_{ab}F^{ab}\right)  +\frac{k}{6}\int_{\mathcal M_5}\mathcal T_5(A,\bar A) \right]+S_{ bdy} \,,
\end{equation}
where $S_{ bdy}$ is the same boundary action given in~\eqref{SM-A-sbdy}. For the planar solution, we choose
\begin{equation}  \label{SM-A-planar-reference}
\bar A=\frac{B}{2}(x dy-y dx) \,,
\end{equation}
which lies in the same magnetic-flux sector as $A$ and satisfies $\bar F\wedge\bar F=0$. Evaluating~\eqref{SM-A-transgression-action} on-shell and setting $16\pi G=L=1$, we find
\begin{equation}\label{SM-A-transgression-onshell}
I_{E}^{\mathcal T} =I_{E}+\Delta\tau\,V\,kQ_{cs} \,, \quad  Q_{ cs}  =\frac{B}{6}\int_0^{r_h}dr \big(A_z'A_t -A_zA_t'\big) \,.
\end{equation}
Fixing the normalization of the boundary time coordinate by setting $\chi_0=0$, such that $T\Delta\tau=1$, the free energy density of the theory becomes
\begin{equation}
w_{\mathcal T} \equiv\frac{T I_{E}^{\mathcal T}}{V} =w_{ QSR}+kQ_{ cs}=w \,.
\end{equation}
Thus, the term identified independently in Secs. A.2 and A.3 is generated automatically by the transgression action.

The new action is also fully consistent with both derivations above. Its Euclidean on-shell variation directly yields
\begin{equation}
\delta w_{\mathcal T} =-s\,\delta T-\rho \delta\mu-M_B \delta B \,,
\end{equation}
because the variation of the additional term in~\eqref{SM-A-transgression-onshell} precisely cancels the $Q_{ cs}$-dependent terms in $\delta w_{ QSR}$. The bulk equation for $A$ is unchanged, while the equation for the auxiliary connection, $\bar F\wedge\bar F=0$, is identically satisfied by~\eqref{SM-A-planar-reference}. Hence no additional bulk contribution appears in the Euclidean variation. The Iyer-Wald calculation is unaffected as well. The exact term in the transgression completion shifts the symplectic potential and the Noether charge by
\begin{equation} 
\boldsymbol\Theta_{\mathcal T} =\boldsymbol\Theta-\delta\boldsymbol\mu_4 \,, \quad
\boldsymbol Q^{\mathcal T}_{\xi} =\boldsymbol Q_{\xi}-\xi\cdot\boldsymbol\mu_4 \,,
\quad  \boldsymbol\mu_4\equiv\frac{k}{6}\mathcal B_4(A, \bar A) \,.
\end{equation}
For the fixed Killing vector $\xi$ used in Sec. A.3, the two extra terms cancel identically in the integrand
\begin{equation} 
\delta\boldsymbol Q^{\mathcal T}_{\xi} -\xi\cdot\boldsymbol\Theta_{\mathcal T} =\delta\boldsymbol Q_{\xi} -\xi\cdot\boldsymbol\Theta \,.
\end{equation}
The remaining $\bar A$ contributions vanish because $\mathcal C_5(\bar A)=0$ and $\bar F\wedge\bar F=0$ for the reference connection~\eqref{SM-A-planar-reference}. Therefore, the derivation of the first law requires no modification in the Iyer-Wald formalism.

The transgression action thus provides a self-consistent resolution of the thermodynamic discrepancy. First, it explains $Q_{cs}$ as the finite contribution missing from the action based on the Chern–Simons term. Second, it promotes that representative to a strictly gauge-invariant action. The resulting action reproduces the correct thermodynamic potential and first law while leaving the bulk dynamics and the Iyer-Wald construction intact.

\subsection*{A.5: The Smarr relation}
Having established that the thermodynamic potential is $w_\mathcal{T}= T I_E^\mathcal{T}/V = w_{QSR} +kQ_{ cs}$, we now derive the corresponding Smarr relation. From the scaling symmetry~\eqref{SM-A-scaling}, one can find that, under the scale transformation $r\to \lambda r= \hat{r}$ with $\lambda$ a positive constant, the relevant physical quantities scale as
\begin{equation}
s=\lambda^3 \hat{s} \,,\quad \rho=\lambda^3 \hat{\rho} \,,\quad B=\lambda^2 \hat{B} \,, 
\end{equation}
while the energy density $\epsilon$ acquires an anomalous scaling transformation
\begin{equation}
\epsilon =\lambda^4 \hat{\epsilon} -\frac{\hat{B}^2}{2} \lambda^4 \ln{\lambda} \,.
\end{equation}
Thus, when we express $\epsilon$ as a function of $s,\rho$ and $B$, we have
\begin{equation}
\epsilon(\lambda^3\hat{s}, \lambda^3\hat{\rho}, \lambda^2\hat{B}) =\lambda^4 \hat{\epsilon} (\hat{s},\hat{\rho},\hat{B}) -\frac{\hat{B}^2}{2} \lambda^4 \ln{\lambda} \,.
\end{equation}
Taking the derivative of both sides of this equation with respect to $\lambda$ and then setting $\lambda$ to $1$, we arrive at
\begin{equation}
\left( \frac{\partial \epsilon}{\partial s} \right)_{\rho,B} (3s) +\left( \frac{\partial \epsilon}{\partial \rho} \right)_{s,B} (3\rho) +\left( \frac{\partial \epsilon}{\partial B} \right)_{s,\rho} (2B) = 4\epsilon -\frac{B^2}{2} \,.
\end{equation}
Taking advantage of the first law of thermodynamics~\eqref{SM-A-first}, we can derive the (generalized) Smarr relation
\begin{equation}
-4\epsilon+3(Ts+\mu\rho)-2M_B B =-\frac{B^2}{2} \,.
\end{equation}
The right hand side, which is non-zero when $B\neq0$, indicates a deviation from the standard Smarr formula due to the conformal anomaly. According to~\eqref{SM-A-free}, this Smarr relation effectively expresses the trace anomaly of the boundary stress-energy tensor~\emph{i.e.} $T^\mu_{\ \mu}=-B^2/2$.

\section{Magnetic susceptibility from hydrodynamics} \label{SM-B}
Relativistic hydrodynamics is crucial for describing finite-temperature interacting relativistic field theories at large distances and long time scales. This framework has found applications across various fields, including nuclear physics, condensed matter physics, astrophysics, and cosmology. The hydrodynamic variables include the local temperature $T(x)$, the local chemical potential $\mu(x)$ and the local fluid velocity $u^\mu(x)$ (normalized such that $u^\mu u_\mu=-1$). The stress-energy tensor $T^{\mu\nu}$ and current $J^\mu$ are expressed through constitutive relations in terms of hydrodynamic variables and their derivatives. 

In this section, we provide a brief overview of the hydrodynamic description of the $3+1$ dimensional chiral charged fluid in the presence of a strong external magnetic field, following~\cite{Ammon:2017ded,Kovtun:2016lfw,Jensen:2013kka}. In the presence of external background field and anomalous effects, the hydrodynamic equations governing the stress-energy tensor $T^{\mu\nu}$ and the axial current $J^\mu$ are given by
\begin{equation}
\nabla_\mu T^{\mu\nu} =F^{\nu\lambda} J_\lambda \,,\quad
\nabla_\mu J^\mu =\frac{k}{8} E^\mu B_\mu\,,
\end{equation}
where $E^\mu=F^{\mu\nu} u_\nu$ and $B^\mu=\frac{1}{2}\epsilon^{\mu\nu\alpha\beta}u_\nu F_{\alpha\beta}$ represent the electric and magnetic fields in $(3+1)$ dimensions, respectively \cite{Kovtun:2016lfw}. To leading order in the derivative expansion, the constitutive relations are expressed as
\begin{equation}
\begin{split}
\langle T^{\mu\nu}_{EFT} \rangle &= \epsilon_0 u^\mu u^\nu +P_0 \Delta^{\mu\nu} +q^\mu u^\nu+q^\nu u^\mu +M^{\mu\alpha}g_{\alpha\beta}F^{\beta\nu} +u^\mu u^\alpha \big(M_{\alpha\beta}F^{\beta\nu} -F_{\alpha\beta}M^{\beta\nu} \big) +\mathcal{O}(\partial)  \,,\\
\langle J^\mu_{EFT} \rangle &= n_0 u^\mu +\xi_B B^\mu +\mathcal{O}(\partial) \,,
\end{split}
\end{equation}
where $\epsilon_0$ is the energy density,  $P_0$ is the pressure, and $\Delta^{\mu\nu}=g^{\mu\nu} + u^\mu u^\nu$. The terms $\xi_B B^\mu$ and $q^\mu=\xi_V B^\mu$ represent chiral transport coefficients arising from the chiral anomaly, while the polarization tensor $M^{\mu\nu}=\chi_B \epsilon^{\mu\nu\alpha\beta}B_\alpha u_\beta$ is defined as the variation of the generating functional with respect to the field strength~\cite{Kovtun:2016lfw}.

In thermal equilibrium, where we choose $u^\mu=(u^t, u^x, u^y, u^z)=(1,0,0,0)$ and assume $B^\mu \propto \vec{z}$, the energy-momentum tensor and the gauge current are given by
\begin{equation} \label{SM-B-teff} 
\langle T^{\mu\nu}_{EFT}\rangle = 
\left( \begin{array}{cccc}
\epsilon_0 & 0 & 0 &  \xi_V^{(0)} B\\
0 & P_0 - \chi_{B} B^2 & 0 & 0\\
0 & 0 & P_0 - \chi_{B} B^2 & 0\\
\xi_V^{(0)} B & 0 & 0 & P_0 
\end{array} \right) +\mathcal{O}(\partial) \, ,  \quad
\langle J^\mu_{EFT} \rangle =\left(n_0,\, 0, \,0, \, \xi^{(0)}_B B \right) +\mathcal{O}(\partial) \,, 
\end{equation}
where the subscript $``0"$ and superscript $``(0)"$ denote that these quantities are evaluated in thermal equilibrium. Note that the thermodynamic relation $\epsilon_0=-P_0+Ts+\mu\rho$ has been employed to derive~\eqref{SM-B-teff}, which holds under the identification $P_0=-w$.

Finally, by comparing the results for the stress-energy tensor~\eqref{SM-B-teff} with the gravitational computations~\eqref{SM-A-TJmn}, we obtain the expression for the magnetization $M_B$ and magnetic susceptibility $\chi_B$
\begin{equation}
M_B B=\chi_B^{hydro} B^2=\langle T^{zz} \rangle -\langle T^{xx}\rangle=\frac{B^2}{4}+8h_4 \,.
\end{equation}
Thus, the magnetic susceptibility from the hydrodynamic description reads
\begin{equation}
\chi_B^{hydro}=\frac{1}{4}+\frac{8h_4}{B^2} \,,
\end{equation}
which is what we have shown in Fig.~1.

Within the hydrodynamic framework, the expression for $\chi_B^{hydro}$ is valid for the BHs analyzed in Sec.~\ref{SM-A} and Sec.~\ref{SM-E}, as well as for planar BHs with higher‑derivative corrections in Sec.~\ref{SM-F}.

\section{Helical BHs with a background magnetic field} \label{SM-C}
In this section, we analyze the thermodynamics of helical BHs, which simultaneously support helical order and a constant magnetic field, as introduced in~\cite{Ammon:2016szz}. Moving beyond the original framework—where helical order arises only spontaneously—we employ a generalized setup that also allows for its explicit generation. This extension introduces an additional conjugate pair of thermodynamic variables and enriches the system’s thermodynamic structure. Specifically, we investigate electrically and magnetically charged black branes endowed with Bianchi VII$_0$ symmetry, which can be formulated in terms of the one-form $\omega_i$ denoted as
\begin{equation}
\omega_1 = \cos{(\lambda_L z)} dx -\sin{(\lambda_L z)} dy \,, \qquad
\omega_2 = \sin{(\lambda_L z)} dx +\cos{(\lambda_L z)} dy \,, \qquad
\omega_3 = dz \,,
\end{equation}
where $\lambda_L$ represents the pitch of the helical lattice.

The Lagrangian density for this helical BH is the same as that in the main text and the configuration reads
\begin{equation} \label{SM-C-ansatz}
\begin{split}
ds^2&=\frac{1}{r^2}\Big[- \big(f e^{-\chi}-h^2p^2\big)dt^2+2ph^2 dtdz +h^2 dz^2 +u^{-2}\omega_2^2 +u^{2}\left(\omega_1 +\alpha dz+\beta dt\right)^2+\frac{dr^2}{f}\Big]\,,  \\
A&=A_t dt  +\frac{B}{2}(x dy-ydx) +b\, \omega_1-A_z dz \,,
\end{split}
\end{equation}
where $f,\, \chi,\, h,\, p,\, u,\, \alpha,\, \beta,\, A_t,\, b$ and $A_z$ are functions of the radial coordinate $r$. In such coordinates, the asymptotically AdS boundary is located at $r=0$ and the BH horizon is located at $r=r_h$. The equations corresponding to the above ansatz are lengthy and are not displayed here.

\subsection*{C.1: Asymptotic expansions and thermodynamics}
At the asymptotic AdS boundary $r=0$, we have for the metric fields that
\begin{equation}
\begin{split}
f(r)&= 1+\frac{B^2}{2}r^4\ln{r}+f_4 r^4+\cdots \,,  \quad  \chi(r)= \chi_0+ \left(\frac{B^2}{4} -\frac{\mu_b^2 \lambda_L^2}{8} \right) r^4\ln{r}+\left( \frac{B^2}{12}+2h_4 -\frac{\mu_b^2 \lambda_L^2}{24} \right) r^4+\cdots  \,, \\
h(r)&= 1+ \left(\frac{B^2}{8} -\frac{\mu_b^2 \lambda_L^2}{16} \right) r^4\ln{r}+h_4 r^4+\cdots  \,, \quad  u(r)= 1+ \frac{\mu_b^2 \lambda_L^2}{16} r^4\ln{r}+ u_4 r^4+\cdots  \,, \\
p(r) &= p_4 r^4+\cdots \,, \quad \alpha(r)=\left( \frac{b_2 B}{2\lambda_L} +\frac{\mu_b A_{z2}}{2} +\frac{\mu_b\lambda_L B}{16} \right) r^4 +\frac{\mu_b \lambda_L B}{4} r^4\ln{r}+\cdots \,,  \quad 
\beta(r)= \beta_4 r^4+\cdots \,,
\end{split}
\end{equation}
and for the gauge fields
\begin{equation}
b(r) = \mu_b + b_2 r^2 +\frac{\mu_b \lambda_L^2}{2} r^2\ln{r}+\cdots    \,, \quad
A_t(r)= e^{-\chi_0/2} \Big( \mu +A_{t2} r^2 +\cdots  \Big) \,,  \quad
A_z(r)= A_{z2} r^2 +\cdots  \,.
\end{equation}
It is important to note that we have introduced a source term $\mu_b$ for the gauge field $b(r)$ of~\eqref{SM-C-ansatz}. As a consequence, the solution presented here is more general than the original setup~\cite{Ammon:2016szz}.

Near the BH horizon, imposing the regularity condition with $f(r_h)=p(r_h)=\beta(r_h)=A_t(r_h)=0$, we obtain for the metric and gauge fields
\begin{equation}
\begin{split}
f(r)&= f_1(r_h-r)+\cdots \,,\qquad \chi(r)= \chi_1+\cdots \,,\qquad
h(r)= h_1+\cdots  \,,\qquad u(r)= u_1+\cdots \,, \\
p(r)&= p_1(r_h-r)+\cdots \,,\qquad \alpha(r)= \alpha_1+\cdots \,,\qquad  \beta(r)= \beta_1(r_h-r)+\cdots \,, \\
b(r) &= b_1 +\cdots  \,,\qquad A_t(r)= A_{t1}(r_h-r)+\cdots  \,,\qquad A_z(r)= A_{z1}+\cdots \,.
\end{split}
\end{equation}
The temperature and Bekenstein-Hawking entropy density are given by 
\begin{equation}
T=-\frac{e^{\chi_0/2} } {4\pi} f'e^{-\chi/2} \Big|_{r=r_h} \,, \qquad s=\frac{4\pi h}{r^3} \Big|_{r=r_h}  \,.
\end{equation}

The general expressions for holographic energy-momentum tensor $\langle T_{\mu\nu} \rangle$ and current $\langle J^{\mu} \rangle$ are provided in~\eqref{SM-A-TmnJm}. For helical BHs, the non-vanishing components of $\langle T_{\mu\nu}\rangle$ are found to be
\begin{equation}
\begin{split}
& \langle T_{tt}\rangle= e^{-\chi_0} \left( -3f_4+8h_4+\frac{B^2}{4} -\frac{\lambda_L^2\mu_b^2}{8} \right) \,,\quad \langle T_{tz}\rangle=\langle T_{zt}\rangle=4p_4 \,, \\
& \langle T_{tx}\rangle = \langle T_{xt}\rangle= 4\cos{(\lambda_L z)} \beta_4  \,,\qquad\qquad\quad  \langle T_{ty} \rangle =\langle T_{yt} \rangle = -4\sin{(\lambda_L z)} \beta_4 \,, \\
& \langle T_{xx}\rangle = -f_4 -\frac{B^2}{6} -\frac{\lambda_L^2 \mu_b^2}{6} +\left( 8u_4 +\frac{\lambda_L^2 \mu_b^2}{8} \right) \cos{(2\lambda_L z)} \,, \\ 
& \langle T_{yy}\rangle = -f_4 -\frac{B^2}{6} -\frac{\lambda_L^2 \mu_b^2}{6} -\left( 8u_4 +\frac{\lambda_L^2 \mu_b^2}{8} \right) \cos{(2\lambda_L z)} \,, \\
& \langle T_{xy}\rangle =\langle T_{yx}\rangle = -\left( 8u_4 +\frac{\lambda_L^2 \mu_b^2}{8}  \right)\sin{(2\lambda_L z)} \,, \\
& \langle T_{xz} \rangle = \langle T_{zx} \rangle =  \left[ \frac{2 B b_2 }{\lambda_L} +\mu_b\left(\frac{\lambda_L B}{2} +2A_{z2} \right) \right] \cos{(\lambda_L z)} \,, \quad  \langle T_{yz}\rangle= \langle T_{zy}\rangle= -\tan{(\lambda_Lz)} \langle T_{xz}\rangle \,, \\
& \langle T_{zz}\rangle=-f_4+8h_4+\frac{B^2}{12} -\frac{7\lambda_L^2 \mu_b^2}{24} \,,
\end{split}
\end{equation}
and the non-zero components of $\langle J^{\mu} \rangle$ are
\begin{equation}
\begin{split} \label{SM-C-Jmu}
& \langle J^t \rangle= e^{\chi_0/2} \left(-2A_{t2} +\frac{k \lambda_L\mu_b^2}{3} \right) \,,\,   \\
& \langle J^x\rangle +i \langle J^y \rangle = \left(2 b_2 -\frac{k \lambda_L\mu\mu_b}{3}  +\frac{\lambda_L^2 \mu_b}{2} \right) e^{-i\lambda_L z}  \,,\quad   \langle J^z \rangle= -\left( \frac{ k B\mu}{3} +2A_{z2} \right)\,.
\end{split}
\end{equation}
Therefore, the energy density $\epsilon$ and charge density $\rho$ of the helical BHs are
\begin{equation}
\epsilon =u^\mu u^\nu \langle T_{\mu\nu}\rangle=  -3f_4 +8h_4 +\frac{B^2}{4} -\frac{\lambda_L^2 \mu_b^2}{8} \,,\quad  \rho= u_\mu \langle J^\mu\rangle =-2A_{t2} +\frac{k\lambda_L \mu_b^2}{3} \,,
\end{equation}
where $u^\mu=-e^{\chi_0/2}\delta_t^\mu$ is the timelike normal vector.

Utilizing the properties of equations of motion and Ricci tensor, we can obtain an analytic expression for the on-shell action. After some calculation, we have
\begin{equation}
S=\int d^4x \int dr \Big[ 2\sqrt{-g}R^t_{\ t}+\left(A_t N_t+\frac{k}{3}B A_t A_z -\frac{\lambda_L k}{3}A_t b^2 \right)'+\frac{k}{3}B A_t A_z' \Big] \,,
\end{equation}
with $N_t\equiv \frac{e^{\chi/2}h}{r} \big[ A_t' -\beta b' +p \left( A_z'+\alpha b' \right) \big]$ and $N_t'=-k B A_z' +\lambda_L k b b'$ from the Maxwell equation. The Euclidean action $I_E$ can be obtained by performing an analytic continuation of the renormalized action $S_{ren}=S+S_{bdy}$ given by~\eqref{SM-A-sren}. From the QSR, we obtain
\begin{equation}
w_{QSR}=\frac{W}{V}= -\frac{ T \ln Z}{V}=\frac{ T  I_E }{V} = \epsilon -Ts-\mu\rho -\frac{ e^{\chi_0/2} k B}{3}\int_0^{r_h} A_t\, A_z' d r \,.
\end{equation}
Here the Bekenstein-Hawking entropy density is $s =\frac{4\pi h_1}{r_h^3}$, while the temperature $T=\frac{e^{\chi_0/2} }{\Delta \tau}=e^{-\frac{\chi_1-\chi_0}{2}} \frac{f_1}{4\pi}$ is determined by removing the conical singularity of the Euclidean metric at the horizon, with $\Delta\tau$ held fixed in the subsequent variation.

\subsection*{C.2: Thermodynamic consistency: from Chern–Simons correction to transgression}
{\bf \emph{Variation of on-shell Euclidean action:}} We consider analytic continuation to Euclidean time $\tau$ by a Wick rotation $t=-i \tau$. The total Euclidean action $I_{E}$ is given by
\begin{equation}
I_{E}=I+I_{bdy} \,,
\end{equation}
where $I=-i S =-i S_{eff}$ and $I_{bdy}$ is the Euclidean boundary action, including counterterms as in~\eqref{SM-A-sren}. The effective action 
$S_{eff}$ of $S$ is obtained by substituting the ansatz~\eqref{SM-C-ansatz} into the original action, which is given by
\begin{equation}
S_{eff}= S_g+S_A+S_{CS} \,,
\end{equation}
with the gravity sector
\begin{equation}
\begin{split}
S_{g}=&\int d^5x \frac{e^{-\chi/2} h}{r^3} \bigg\{ -f''-\frac{2fh''}{h}+f\chi'' +f'\left( \frac{8}{r}-\frac{2h'}{h}+\frac{3\chi'}{2} \right)  \\
&+\left( \frac{h'}{r h} -\frac{\chi'}{2 r} \right) \left( 8f +r  f\chi' \right) +\frac{ e^\chi u^2 }{2} \left( p\alpha' -\beta' \right)^2 -\frac{2f u'^2}{u^2} -\frac{f u^2\alpha'^2}{2h^2} \\
&+ \frac{\lambda_L^2 e^\chi p^2 (u^4-1)^2 }{2f u^4} +\frac{\lambda_L^2 e^\chi \beta^2}{2f h^2 u^2} +\frac{12-20f}{r^2} -\frac{\lambda_L^2(u^4-1)^2}{2h^2u^4} +\frac{e^\chi}{2} h^2p'^2 \bigg\} \,,
\end{split}
\end{equation}
the Maxwell sector
\begin{equation}
\begin{split}
S_A=&\int d^5 x  \frac{e^{-\chi/2} h}{r} \bigg\{ -\frac{B^2}{2} -\frac{f A_z'^2 }{2h^2} +\frac{e^\chi}{2} (A_t' +p A_z')^2  -\frac{u^2(\lambda_L b+B\alpha)^2}{2h^2}  +\frac{e^\chi u^2 ( \lambda_L b p+Bp\alpha-B\beta)^2 }{2f} \\
&+\frac{b'^2}{2} \left[ e^\chi (p\alpha-\beta)^2 -\frac{f}{u^2}-\frac{f\alpha^2}{h^2} \right] +b'\left[ e^\chi(p\alpha-\beta)(A_t'+p A_z') -\frac{f\alpha A_z'}{h^2} \right]  \bigg\} \,,
\end{split}
\end{equation}
and the Chern–Simons sector
\begin{equation}
S_{CS} =\int d^5x \bigg[ \frac{k}{3}(B A_z -\lambda_L b^2)A_t' -\frac{k}{3}B A_t A_z' +\frac{k}{3} \lambda_L A_tbb' +\frac{\lambda_L k B}{6} b (y\cos{(\lambda_L z)}+x\sin{(\lambda_L z)})A_t' \bigg] \,.
\end{equation}

In the Euclidean metric, the only remaining boundary is the AdS boundary. Thus, the variation of total Euclidean action is $\delta I_E = \delta I+\delta I_{bdy}$. Note that the terms proportional to $y\cos{(\lambda_L z)}$ or $x\sin(\lambda_L z)$ in $\delta I$ become zero when integrated along the ($x,y,z$) coordinates and we have used the same method to handle the bulk term $\int dr \frac{k B}{3} \left( A_z' \delta A_t-\delta A_z A_t' \right)$. Recalling that the temperature $T=\frac{e^{\chi_0/2} }{\Delta \tau}$, thus the variation of $\chi_0$ gives $\delta \chi_0= 2\frac{\delta T}{T}$ with $\Delta\tau$ fixed in the variation of on-shell Euclidean action. Using the QSR, the variation of $w_{QSR}$ is given by
\begin{equation}\label{SM-C-varyw}
\delta w_{QSR}= -\left( s+ \frac{e^{\chi_0/2} k Q_{cs} }{T}\right)\delta T -\rho \delta\mu - \left(2 b_2 -\frac{k \lambda_L\mu\mu_b}{3}  +\frac{\lambda_L^2 \mu_b}{2} \right) \delta\mu_b -e^{\chi_0/2} k \delta Q_{cs} -M_B \delta B \,,
\end{equation}
where the quantity $Q_{cs}$ is given by
\begin{equation} \label{SM-C-Qcs}
Q_{cs}=\frac{B}{6}\int_0^{r_h} (A_z' A_t- A_z A_t') dr \,,
\end{equation}
and $M_B$ is defined as
\begin{equation}
\begin{split}\label{SM-C-MB}
M_B =& -\bigg\{ \int_0^{r_h} \bigg[ e^{ (\chi_0 -\chi)/2 } \frac{B h}{r} - \frac{B }{r} +\frac{e^{\chi_0/2}k}{2}(A_z' A_t - A_z A_t') +e^{(\chi_0-\chi)/2 } \frac{ \alpha u^2 (\lambda_L b+B \alpha)}{r h}  \\
& \quad\quad - e^{ (\chi_0+\chi)/2 } \frac{ h u^2 (p\alpha -\beta)( \lambda_L bp+ Bp\alpha -B\beta) }{r f} \bigg] dr+ B\ln{r_h} \bigg\} \,.
\end{split}
\end{equation}
Note that in the final expression, the constant $\chi_0$ in $M_B$ should be set to zero to ensure that the Hawking temperature equals the temperature at the boundary.

From~\eqref{SM-C-varyw}, we see that the variation of $w_{ QSR}$ does not by itself take the standard form of the first law of thermodynamics. Here we have regarded $\lambda_L$ as labeling a particular helical mode and therefore do not vary it~\cite{Donos:2012wi}. Nevertheless, after supplementing $w_{ QSR}$ with the universal Chern-Simons correction, $w=w_{QSR} +e^{\chi_0/2} k Q_{cs}$, we obtain the standard form of the thermodynamic relation
\begin{eqnarray}
w &=&\epsilon-Ts-\mu\rho \,, \label{SM-C-freeW}  \\
\delta w &=&-s\delta T -\rho \delta\mu -\left(2b_2 -\frac{k \lambda_L\mu\mu_b}{3} +\frac{\lambda_L^2 \mu_b}{2} \right) \delta\mu_b -M_B\delta B\,, \label{SM-C-free1st} 
\end{eqnarray}
where $M_B$ is identified as the magnetization of the system. From~\eqref{SM-C-Jmu}, we further define
\begin{equation} \label{SM-C-Jb}
\langle J^b\rangle \equiv \sqrt{\langle J^x\rangle^2+\langle J^y\rangle^2} =\left(2b_2 -\frac{k \lambda_L\mu\mu_b}{3} +\frac{\lambda_L^2 \mu_b}{2}\right) \,, 
\end{equation}
so that the first law~\eqref{SM-C-free1st} can be written in the compact form
\begin{equation} \label{SM-C-freeE}
\delta w =-s\delta T -\rho \delta\mu -\langle J^b \rangle\delta\mu_b -M_B\delta B \,,
\end{equation}
as shown in equation~(17) of the main text.

{\bf \emph{Resolution from Iyer-Wald formalism:}} In the Iyer–Wald framework, when $\phi$ satisfy the background equations of motion, $\delta\phi$ obeys the linearized equations, and $\xi^a$ is a Killing vector, one can derive the two fundamental identities~\eqref{SM-A-fundamentalID} and~\eqref{SM-A-identity}. These relations can be directly applied to the helical BH~\eqref{SM-C-ansatz}. The corresponding Noether charge and symplectic potential then take the same form as those presented in~\eqref{SM-A-QTheta}.

Let $\xi^a$ be the time-like Killing vector $\xi^a=(\partial_t)^a=\delta^a_t$, which becomes null at the event horizon $r=r_h$. Let $\Sigma$ be a $t=$ const. space-like hypersurface with the horizon $r=r_h$ serving as its interior boundary. Note that the boundary of this hypersurface $\partial\Sigma$ includes the boundaries of the non-compact $x,\, y,\, z$ directions. We can project the 3-form $(\delta\mathbf{Q}-\xi\cdot \mathbf{\Theta})$ onto each boundary of the hypersurface $\Sigma$. It should be noted that, unlike the case in~\eqref{SM-A-bxyz}, the helical BH solution studied here does not admit a simple proportionality between $(\delta Q -\xi \cdot \Theta)_y$ and $(\delta Q -\xi \cdot \Theta)_x$,~\emph{i.e.}
\begin{equation}
(\delta Q -\xi \cdot \Theta)_y \sqrt{\gamma_{(y)}} \neq \frac{y}{x} (\delta Q -\xi \cdot \Theta)_x \sqrt{\gamma_{(x)}}\,.
\end{equation}

After a direct integration of the fundamental identity~\eqref{SM-A-fundamentalID} over the hypersurface $\Sigma$, we have
\begin{equation}
\begin{split}  \label{SM-C-deltaH}
\delta H &=0=\int_\Sigma \boldsymbol{\omega}=\int_{\Sigma} d(\delta \mathbf{Q}- \xi \cdot \boldsymbol{\Theta}) =\int_{\partial\Sigma} (\delta \mathbf{Q} -\xi \cdot \boldsymbol{\Theta})    \\
&= \int_{ S_{ r=r_{h} } } (\delta Q -\xi \cdot \Theta)_r  \sqrt{\gamma_{(r)}} dxdydz 
-\int_{ S_{r=\epsilon} } (\delta Q -\xi \cdot \Theta)_r  \sqrt{\gamma_{(r)}} dxdydz \\
&\quad +\int_{S_{x=\frac{L}{2} } } (\delta Q -\xi \cdot \Theta)_x  \sqrt{\gamma_{(x)}} dydzdr 
-\int_{ S_{ x=-\frac{L}{2}} } (\delta Q -\xi \cdot \Theta)_x \sqrt{\gamma_{(x)}} dydzdr \\ 
&\quad +\int_{ S_{y=\frac{L}{2} } } (\delta Q -\xi \cdot \Theta)_y \sqrt{\gamma_{(y)}} dxdzdr 
-\int_{ S_{y=-\frac{L}{2} } } (\delta Q -\xi \cdot \Theta)_y \sqrt{\gamma_{(y)}} dxdzdr \\
&\quad +\int_{ S_{z=\frac{L}{2} } } (\delta Q -\xi \cdot \Theta)_z \sqrt{\gamma_{(z)}} dxdydr 
-\int_{ S_{z=-\frac{L}{2} } } (\delta Q -\xi \cdot \Theta)_z \sqrt{\gamma_{(z)}} dxdydr\,, \\
&=3\delta f_4 -8 \delta h_4 -\frac{B}{2}\delta B +e^{-\frac{\chi_1}{2}} f_1 \left( \frac{\delta h_1}{r_h^3} -\frac{3h_1 \delta r_h}{r_h^4} \right) -2\mu\delta A_{t2} -\left(2b_2  -k\lambda_L \mu \mu_b +\frac{\lambda_L^2 \mu_b}{2}\right) \delta\mu_b - M_B \delta B \,,  
\end{split}
\end{equation}
where $M_B$ is exactly given by~\eqref{SM-C-MB}.

Note that the variation of $\rho$ is given by $\delta\rho=-2\delta A_{t2} +\frac{2k\lambda_L}{3} \mu_b \delta\mu_b$. Substituting the variations of energy density $\delta\epsilon$, entropy density $\delta s$ and charge density $\delta\rho$ into~\eqref{SM-C-deltaH}, we arrive at the first law of BH thermodynamics
\begin{equation}\label{SM-C-first}
\delta \epsilon = T\delta s +\mu \delta\rho  -\langle J^b\rangle \delta\mu_b - M_B \delta B \,,
\end{equation}
where $\langle J^b\rangle$ is defined in~\eqref{SM-C-Jb}.

Furthermore, we can derive an integral expression for the first law by utilizing the identity~\eqref{SM-A-identity}. By integrating this identity over the space-like hypersurface $\Sigma$ and adding proper boundary terms, we have
\begin{equation}
w =\epsilon-Ts-\mu\rho \,. \label{SM-C-free}
\end{equation}
The combination of~\eqref{SM-C-first} and~\eqref{SM-C-free} gives
\begin{equation}
\delta w =-s\delta T -\rho \delta\mu -\langle J^b\rangle \delta\mu_b -M_B\delta B \,,
\end{equation}
which is the same result obtained from the variation of Euclidean action~\eqref{SM-C-freeW} and~\eqref{SM-C-free1st}. As a consistency check, by setting the magnetic field to zero, $B=0$, we recover the previous result of~\cite{Donos:2012wi}.

{\bf\emph{Transgression interpretation:}} More importantly, the two independent approaches developed above---the variation of the Euclidean action and the Iyer-Wald formalism---lead to the same thermodynamic potential. Both show that the full free energy is $w=w_{ QSR}+kQ_{ cs}$, rather than $w_{ QSR}$ alone. This agreement admits a direct action-level explanation in terms of the transgression form. For the helical background~\eqref{SM-C-ansatz}, a natural reference connection is $\bar A_{ hel}=\frac{B}{2}(x\,dy-y\,dx)$. Replacing the Chern–Simons form with $\mathcal T_5(A,\bar A_{ hel})$ and evaluating the resulting action on-shell yields
\begin{equation}
w_{\mathcal T}=w_{ QSR}+kQ_{ cs}=w \,,
\end{equation}
where $Q_{ cs}$ is given in~\eqref{SM-C-Qcs}.

Furthermore, the variation of $w_{\mathcal T}$ consequently reproduces~\eqref{SM-C-freeE}, including the thermodynamic conjugate pair associated with the helical source $\mu_b$. The Iyer-Wald result remains unchanged: the exact form in the transgression action induces shifts in the symplectic potential and the Noether charge that cancel in $\delta\boldsymbol Q_\xi-\xi\cdot\boldsymbol\Theta$. Meanwhile, the separate $\mathcal C_5(\bar A_{ hel})$ sector vanishes because $\mathcal C_5(\bar A_{ hel})=0$ and $\bar F_{ hel}^2=0$. Thus, the transgression completion provides a natural explanation for the origin of $Q_{ cs}$.

\section{Rotating BHs with spherical topology} \label{SM-D}
In this section, we provide the technical details for the rotating BH introduced in the End Matter. We focus on BHs with two equal angular momenta, $\mathcal J_1=\mathcal J_2=\mathcal J$, governed by the EMCS action~(2) of the main text. Five-dimensional rotating BHs with equal angular momenta exhibit an enhanced $U(2)\simeq SU(2)\times U(1)$ symmetry. Following the construction of~\cite{Blazquez-Salcedo:2017ghg,Blazquez-Salcedo:2016rkj}, the general ansatz for the background fields, including a magnetic field $B$ and respecting the local $SU(2)\times U(1)$ symmetry, takes the form
\begin{equation}
\begin{split} \label{SM-D-ansatz}
ds^2 &= -F_0(r) dt^2 + \frac{dr^2}{F_1(r)} +\frac{1}{4} F_2(r) \left(\sigma_1^2+\sigma_2^2\right) +\frac{1}{4} F_3(r) \big( \sigma_3 +2p(r) dt \big)^2 \,,  \\
A &= A_t(r) dt + \frac{B\cos{\theta}}{4} d\phi  - \frac{A_z(r)}{2} \sigma_3 \,,
\end{split}
\end{equation}
where
\begin{equation} \label{SM-D-sigma}
\sigma_1= \cos{\psi} d\theta+ \sin\psi \sin\theta d\phi \,,\quad \sigma_2= -\sin{\psi} d\theta+ \cos\psi \sin\theta d\phi \,,\quad 
\sigma_3  =  d \psi + \cos \theta \, d \phi \,,
\end{equation}
are left-invariant 1-forms on $SU(2)$ expressed in terms of Euler angles $(\theta,\phi,\psi)$. The ranges of the Euler angles are $0\leq \theta\leq \pi, 0\leq \phi \leq 2\pi, 0\leq \psi\leq 4\pi$. Note that one can recover the familiar unit $3$-sphere coordinates $(\tilde{\theta},\tilde{\phi},\tilde{\psi})$ by defining $\tilde{\theta} =\theta/2 \,, \tilde{\psi} -\tilde{\phi} =\phi \,, \tilde{\psi} +\tilde{\phi} =\psi$. Therefore, we have $\left( \sigma_1^2 +\sigma_2^2 +\sigma_3^2 \right)/4= d\tilde{\theta}^2 +\sin^2\tilde{\theta} d\tilde{\phi}^2 +\cos^2 \tilde{\theta} d\tilde{\psi}^2 \equiv d\Omega_3^2$. This BH~\eqref{SM-D-ansatz} has the following scaling symmetry
\begin{equation} \label{SM-D-scaling}
t\to \lambda^{-1} t \,, \quad F_0\to \lambda^2 F_0 \,,\quad (A_t, p) \to \lambda (A_t, p) \,.
\end{equation}
Note that at the supergravity value $k=2/\sqrt{3}$, the $B=0$ limit of the ansatz~\eqref{SM-D-ansatz} contains the equal-angular-momenta sector of the general Chong-Cveti\v{c}-L\"u-Pope solution~\cite{Chong:2005hr}.

We now turn to the study of the thermodynamics of the rotating BH~\eqref{SM-D-ansatz}. For our purpose, we will consider the following reparameterization of the metric functions
\begin{equation}
F_0=\frac{L^2}{r^2}f(r)e^{-\chi(r)} \,,\quad  F_1=\frac{r^2f(r)}{L^2} \,,\quad  F_2=\frac{L^4}{r^2} \,, \quad F_3= \frac{L^4}{r^2} h(r)^2 \,.
\end{equation}
It is worth noting that the minimal planar BH ansatz is contained within the rotating BHs~\eqref{SM-D-ansatz} via a specific scaling limit. By setting $\hat{x}=\cos\theta$ and applying the transformation
\begin{equation}
(t, r, \hat{x}, \phi, \psi ) \to \lambda (t, r, 2x, 2y, 2z) \,,\quad (A_t, A_z) \to \lambda^{-1}(A_t, A_z) \,,\quad  B\to \lambda^{-2} B  \,,
\end{equation}
we find that the minimal planar BH is recovered in the limit $\lambda\to0$. Thus, the phenomena observed in the minimal planar BHs are inherently encoded within this general ansatz~\eqref{SM-D-ansatz}.

\subsection*{D.1: Far-field and near-horizon expansions, and thermodynamics}
We consider a boundary at infinity with only local AdS symmetries, which is referred to as the squashed AdS boundary. In particular, as $r\to0$, the metric functions $f$ and $h$ behave as $f\sim 1, h\sim v$, where $v$ is the squashing parameter. The far-field expansion of the background fields can be expressed as
\begin{equation}
\begin{split}
f(r) &= 1 + \frac{1}{3}\left( 8-5v^2\right) \frac{r^2}{L^2} + \bigg[ \frac{B^2}{ 2L^2} -\frac{8}{3}\left( 1-v^2\right)\left(1-3v^2\right) \bigg] \frac{r^4}{L^4} \ln{(\frac{r}{L})} + f_4 \frac{r^4}{L^4}+ \cdots \,, \\
\chi(r) &= \chi_0 +\frac{2}{3}\left(1-v^2\right) \frac{r^2}{L^2}+ \bigg[ \frac{B^2}{ 4L^2} -\frac{4}{3}\left(1-v^2\right)\left( 1-4v^2\right)  \bigg] \frac{r^4}{L^4} \ln{(\frac{r}{L})} +\chi_4 \frac{r^4}{L^4} +  \cdots  \,, \\
h(r) &=  v\left( 1 +\left(1-v^2\right) \frac{r^2}{L^2} +\bigg[ \frac{ B^2}{ 8L^2} -\frac{2}{3} \left(1-v^2\right)\left(1-4v^2\right) \bigg] \frac{r^4}{L^4} \ln{(\frac{r}{L})} + h_4 \frac{r^4}{L^4} + \cdots  \right) \,, \\
p(r) &= e^{-\chi_0/2} p_\infty+ p_4 \frac{r^4}{L^4} +\cdots   \,,\\
A_t(r) &= e^{-\chi_0/2} \left( \mu  +A_{t2} \frac{r^2}{L^2} + p_\infty B\frac{r^2}{L^2} \ln{ (\frac{r}{L})} +\cdots \right) \,,\\
A_z(r) &= A_{z2}  \frac{r^2}{L^2} -B v^2 \frac{r^2}{L^2} \ln{ (\frac{r}{L})} +\cdots    \,, 
\end{split}
\end{equation}
where $\chi_4=2h_4 + B^2/(12L^2) -(1-v^2)(5-17v^2)/9$. Note that to ensure the Hawking temperature of the BH matches that of the boundary field theory, the parameter $\chi_0$ must be set to zero by utilizing the scaling symmetry~\eqref{SM-D-scaling}. However, for convenience in varying the temperature in the on-shell Euclidean action, we retain it as a free parameter and set $\chi_0=0$ only in the final expressions. We adopt a boundary non-rotating frame by simply setting $p_\infty=0$, as thermodynamic quantities must be computed in such a frame at the AdS boundary. However, when analyzing the first law of BH thermodynamics in the next subsection, we find it more convenient to use co-rotating coordinates at the horizon. 

Working in the boundary non-rotating frame where $p_\infty=0$, we find that the boundary conformal metric takes the form of the Einstein static universe with 
\begin{equation} \label{SM-D-AdSbdy}
ds^2_{bdy} = -dt^2 +d\Omega_{(v)}^2 = -dt^2 +\frac{L^2}{4} \Big[ d\theta^2 + \sin^2\theta d\phi^2 + v^2(d\psi +\cos\theta d\phi)^2 \Big] \,.
\end{equation}
Here, $d\Omega_{(v)}^2$ denotes the metric of a squashed $S^3$. The spatial volume of the boundary metric~\eqref{SM-D-AdSbdy} is given by
\begin{equation}
V_{(v)}= \int \frac{\sin\theta L^3v}{8} d\theta d\phi d\psi =2\pi^2 L^3 v \,.
\end{equation}
In the following analysis, the thermodynamic densities are defined as their extensive counterparts divided by the volume of the squashed 3-sphere $V_{(v)}$. The AdS radius $L$ and the squashing parameter $v$ are held fixed throughout.

We consider BHs at finite temperature, for which the metric function $f(r)$ vanishes at the horizon $r=r_h$. The background fields admit the following expansions near the BH horizon
\begin{equation}\begin{split}
f(r)&= f_1 (r_h -r) +\cdots  \,,\qquad 
\chi(r)= \chi_1 +\cdots  \,,\qquad
h(r)= h_1 +\cdots  \,,\\
p(r) &= p_0 +p_1(r_h -r)+\cdots  \,,\quad
A_t(r)=A_{t0} +A_{t1}(r_h -r) +\cdots \,,\quad 
A_z(r)= A_{z1} +A_{zz}(r_h -r) +\cdots \,.
\end{split}\end{equation}
Furthermore, it is necessary to work in the $U(1)$ gauge where $A_t+p A_z$ vanishes linearly at the horizon, which enforces the condition $A_{t0}=-p_0 A_{z1}$.

The horizon angular velocity $\Omega_H$ is determined via the horizon Killing vector
\begin{equation} \label{SM-D-Killing}
\xi_H = \partial_t + 2\Omega_H \partial_\psi \,,
\end{equation}
which becomes null on the horizon. Note that the Killing field normalized at the boundary is given by $(\partial_t)^a = e^{\chi_0/2} \delta^a_t$, while the axial Killing field is simply $(\partial_\psi)^a = \delta^a_\psi$. The Hawking temperature $T$, the entropy density $s$, and the horizon angular velocity $\Omega_H$ are explicitly given by
\begin{equation}
T =-\frac{e^{\chi_0/2} } {4\pi} f'(r_h)e^{-\chi(r_h)/2} \,,\quad  s=\frac{4\pi h(r_h) L^3}{vr_h^3} \,, \quad \Omega_H =- e^{\chi_0/2}p(r_h) \,.
\end{equation}
Furthermore, the electrostatic potential entering the thermodynamics is given by the difference
\begin{equation}
\mu \equiv \xi_H^bA_b \big|_{r\to\infty} -\xi_H^bA_b \big|_{r=r_h} \,,
\end{equation}
where we used the $U(1)$ gauge condition $A_t(r_h)+p(r_h)A_z(r_h)=0$ at the horizon.

Starting from the total renormalized action $S_{ren}$ in~\eqref{SM-A-sren}, we evaluate its on-shell Euclidean action $I_E$. The free energy density $w_{QSR}$ then follows from the QSR. To derive an analytical expression for the on-shell Euclidean action, we utilize the identity $R^a_{\ b}\xi^b=\nabla_b\nabla^a\xi^b$, which holds for any Killing vector $\xi^a$. We can use this identity to recast $R^a_{\ b}$ as a total derivative. Applying this to the present BH gives
\begin{equation}
R^a_{\ b}\xi^b =\nabla_b\nabla^a\xi^b =-\nabla_b \nabla^b \xi^a = -\frac{ \partial_b \left( \sqrt{-g} g^{bc}\nabla_c\xi^a \right) }{\sqrt{-g}} = -\frac{1}{\sqrt{-g}} \Big[ \partial_r \left( \sqrt{-g} g^{rc}\nabla_c\xi^a \right) -\partial_\theta \left( \sqrt{-g} g^{\theta c}\nabla_c\xi^a \right)  \Big]\,,
\end{equation}
where we have used the fact that the metric fields depend only on~$r,\theta$. For each Killing vector $\partial_t, \partial_\psi$, we have
\begin{equation}
2\sqrt{-g}R^t_{\ t} = \left( \frac{e^{-\chi/2} L^6 h\sin\theta}{ 8 r^4 } \Big[ -r f'+r e^\chi L^2 h^2 p p' +f \left(2+r \chi' \right) \Big] \right)'  \,, \quad \sqrt{-g}R^t_{\ \psi} =  \left( \frac{e^{\chi/2} L^8 \sin\theta h^3 p' }{32r^3} \right)'  \,.
\end{equation}
Utilizing the properties of background equations and applying the identity $R^a_{\ b}\xi^b=\nabla_b\nabla^a\xi^b$ on a timelike Killing vector $\xi^a=(\partial_t)^a$, the bulk on-shell action takes
\begin{equation}
S=\int d^4x\int dr \Big[ 2\sqrt{-g}R^t_{\ t}+ \left( A_tN_t +\frac{k}{24}A_tA_z (B -2A_z) \sin{\theta} \right)' +\frac{k}{24} BA_tA_z'\sin{\theta} \Big] \,, 
\end{equation}
where $N_t\equiv \frac{ e^{\chi/2} L^4 h \sin{\theta} }{8r} \left( A_t' +p A_z'\right)$. Using the general expressions for the expectation value of energy-momentum tensor and current given in~\eqref{SM-A-TmnJm}, the non-vanishing components of $\langle T_{\mu\nu}\rangle$ for the rotating BH~\eqref{SM-D-ansatz} read
\begin{equation}\begin{split}
\langle T_{tt}\rangle &= \frac{ e^{-\chi_0}}{L} \left( -3f_4 +8h_4 +\frac{B^2}{4L^2} +\frac{64-32v^2 -23v^4}{12} \right) \,, \\
\langle T_{\theta\theta}\rangle  &= L \left( - \frac{f_4}{4} -\frac{B^2}{24L^2} +\frac{ 32-88v^2 +59v^4 }{48} \right) \,, \\
\langle T_{\phi\phi}\rangle &= L \bigg[ -\Big( v^2+1 +(v^2-1) \cos(2\theta) \Big)\frac{f_4}{8} +2v^2h_4\cos^2{\theta} +\frac{ v^2-2 +(v^2+2)\cos(2\theta) }{96L^2} B^2 \\
 &\qquad +\frac{ 32-56v^2 +139v^4- 109v^6 +(-32 +120v^2 +21v^4 -109v^6) \cos(2\theta) }{96} \bigg] \,, \\
\langle T_{\psi\psi}\rangle &= L v^2 \left( -\frac{f_4}{4} +2h_4 +\frac{B^2}{48L^2 } +\frac{ 32+80v^2 -109v^4 }{48} \right) \,, \\
\langle T_{t\psi}\rangle &= \langle T_{\psi t}\rangle= 2Lv^2p_4 \,,\qquad \langle T_{t\phi}\rangle =\langle T_{\phi t}\rangle= 2Lv^2 p_4 \cos\theta \,,\qquad  \langle T_{\phi\psi}\rangle=\langle T_{\psi\phi}\rangle= \cos\theta \langle T_{\psi\psi}\rangle  \,,
\end{split}\end{equation}
and the non-zero components of $\langle J^{\mu} \rangle$ are
\begin{equation}
\langle J^t \rangle = -e^{\chi_0/2} \frac{ 2A_{t2} }{L} \,, \qquad
\langle J^\psi \rangle = \frac{1}{L^3} \left( -\frac{4}{v^2} A_{z2} - \frac{2k}{3v} \mu B +2B  \right) \,. 
\end{equation}
The energy density $\epsilon$, angular momentum density $\mathcal{J}$, and charge density $\rho$ are defined as~\cite{Blazquez-Salcedo:2017ghg,Balasubramanian:1999re}
\begin{equation}\begin{split}
\epsilon &=u^\mu u^\nu \langle T_{\mu\nu}\rangle= \frac{1}{L} \left( -3f_4 +8h_4 +\frac{B^2}{4L^2} +\frac{64-32v^2 -23v^4}{12} \right) \,,\\ 
\mathcal{J} &= u^\mu \xi_{(\psi)}^\nu \langle T_{\mu\nu}\rangle= -2e^{\chi_0/2} Lv^2p_4 \,,\qquad \rho= u_\mu \langle J^\mu\rangle =-\frac{2A_{t2}}{L} \,,
\end{split}\end{equation}
where $u^\mu=-e^{\chi_0/2}\delta_t^\mu$ is the timelike normal vector, and $\xi_{(\psi)}^\mu=(\partial_\psi)^\mu$ is the Killing vector along the angular direction.

The renormalized on-shell action reads
\begin{equation}
\begin{split}
\frac{S_{ren}}{\Delta t V_{(v)}} &=  \frac{e^{-\chi_0/2} }{L} \left[ 3f_4 -8h_4 -\frac{B^2}{4L^2} -\frac{64-32v^2 -23v^4 }{12} -\frac{ 2\mu A_{t2} }{L} \right] + \frac{e^{-\chi_1/2} f_1 h_1 L^3}{v r_h^3} -\frac{e^{\chi_1/2} h_1^3 L^5}{v r_h^3} p_0 p_1 \\
&\qquad -\frac{e^{\chi_1/2} A_{t0} h_1 L (A_{t1} +p_0 A_{zz})}{v r_h} +\frac{k A_{t0} A_{z1} (B- 2A_{z1})}{3vL^3} +\frac{1}{L^3v} \int_0^{r_h} \frac{kB}{3} A_t A_z' dr \,.
\end{split}
\end{equation}
Consequently, the free energy density obtained from the QSR can be written as
\begin{equation}
w_{QSR} = -\frac{T \ln{Z}}{V_{(v)} }=\frac{T I_E}{V_{(v)}} = \epsilon -Ts - \mu\rho -2\Omega_H \mathcal{J}  -\frac{e^{\chi_0/2} }{L^3v} \int_0^{r_h} \frac{k B}{6} \left( A_t A_z' -A_t' A_z \right) dr \,.
\end{equation}
In deriving $w_{QSR}$, we have used two radially conserved quantities, which follow respectively from the $t$-component of the Maxwell equations and the momentum constraint associated with the $t\text{--}\psi$ component of the Einstein equations.

\subsection*{D.2: Thermodynamic consistency: from Chern–Simons correction to transgression}

In this section, we analyze the thermodynamics of the rotating BH described in~\eqref{SM-D-ansatz}, focusing on the relation between $w_{ QSR}$ and the thermodynamic first law, and on the Chern-Simons correction required to obtain the complete thermodynamic potential. We first derive the corrected thermodynamic potential from the variation of the on-shell Euclidean action and then confirm the result independently using the Iyer-Wald formalism.

{\bf \emph{Variation of on-shell Euclidean action:}}
For the Euclidean action analysis, it is convenient to work in a frame co-rotating with the BH horizon. The background ansatz~\eqref{SM-D-ansatz} possesses the residual coordinate freedom
\begin{equation}
\psi\to\tilde{\psi}\equiv\psi+2\alpha t ,
\end{equation}
under which
\begin{equation}
p(r)\to p(r)-\alpha,\qquad
A_t(r)\to A_t(r)+\alpha A_z(r),
\end{equation}
while the metric and gauge potential retain the same form. Choosing $\alpha=p(r_h)=p_0$ sets the constant horizon value of $p(r)$ to zero and therefore brings the metric to a co-rotating frame at the horizon. Accordingly, in the new coordinates $(t,\tilde{\psi},\theta,\phi,r)$, the asymptotic expansions of the relevant fields are
\begin{equation}
\begin{split}
p(r) &= -p_0+p_4\frac{r^4}{L^4}+\cdots,\\
A_t(r) &= e^{-\chi_0/2}
\left[\mu+A_{t2}\frac{r^2}{L^2}\right]
+p_0\left(A_{z2}-Bv^2\ln\frac{r}{L}\right)
\frac{r^2}{L^2}+\cdots .
\end{split}
\end{equation}
Near the horizon $r=r_h$, they behave as
$p=p_1(r_h-r)+\cdots$ and
$A_t=\widetilde A_{t1}(r_h-r)+\cdots$,
where $\widetilde A_{t1}$ denotes the linear coefficient of the transformed
electric potential. In this co-rotating coordinate system, the horizon Killing vector is generated by $(\partial_t)^a$. For the remainder of the Euclidean-action calculation, all fields are understood to be expressed in these coordinates.

Consider analytic continuation to Euclidean time $\tau$ by a Wick rotation $t=-i \tau$, under which the total Euclidean action $I_{E}$ is given by
\begin{equation}
    I_{E}=I+I_{bdy} \,,
\end{equation}
where $I=-i S =-i S_{eff}$ and $I_{bdy}$ is the Euclidean boundary action, obtained from~\eqref{SM-A-sren}. $S_{eff}$ is the effective action of $S$ with the background ansatz substituted, which is given by
\begin{equation}
\begin{split}
S_{eff}&=\Delta T V \int dr \frac{e^{-\chi/2} h L^6 }{r^3} \bigg\{ -f''-\frac{2fh''}{h}+f\chi'' +f'\left( \frac{8}{r}-\frac{2h'}{h}+\frac{3\chi'}{2} \right) +\left( \frac{h'}{r h} -\frac{\chi'}{2 r} \right) \left( 8f +r  f\chi' \right) \\
&\quad +\frac{1}{2L^2 }e^{\chi} \left[ h^2 p'^2L^4 +r^2(A_t'+ p A_z')^2 \right] +\frac{12-20f}{r^2} + \frac{8 -2h^2}{L^2} -\frac{ r^2}{ 2 L^6}\left( (B -2A_z)^2+\frac{fA_z'^2 L^2}{h^2} \right) \bigg\} \\
&\quad  +\Delta T V \int dr \frac{k(B -2A_z) }{3}\left( A_t' A_z -A_t A_z' \right) \,,
\end{split}
\end{equation}
where $\Delta T \equiv \int dt$ and $V\equiv \int \frac{ \sin{\theta} }{8} d\theta d\phi d\psi = 2\pi^2$, the volume of a 3-sphere. Since the Euclidean metric has only the AdS boundary, a direct computation gives the variation of the total on-shell Euclidean action, $\delta I_E = \delta I + \delta I_{bdy}$, as 
\begin{equation}
\frac{\delta I_E}{\Delta\tau V_{(v)} } =\bigg\{ e^{-\chi_0/2} \bigg[ - \rho \delta\mu -2\mathcal{J}\delta\Omega_H  -\Big(\epsilon -\mu\rho-2\Omega_H \mathcal{J} \Big) \frac{\delta \chi_0}{2} \bigg] - \frac{k}{L^3 v} \delta Q_{cs} - e^{-\chi_0/2}M_B \delta B  \bigg\} \,,
\end{equation}
where the charge
\begin{equation}
Q_{cs}=\frac{B}{6}\int_0^{r_h} \left( A_z' A_t -A_z A_t'\right) dr \,,
\end{equation}
and
\begin{equation} \label{SM-D-MB}
M_B = - \frac{1}{L^3v }\bigg\{ \int_0^{r_h} \bigg[ \left( \frac{ e^{(\chi_0-\chi)/2} h (B -2A_z)}{r} - \frac{ vB}{r} +\frac{e^{\chi_0/2}k}{2}(A_t A_z' -A_t' A_z)\right)  \bigg] dr + v B \ln{ (\frac{r_h}{L})} \bigg\} \,.
\end{equation}
Note that in the final expression, the constant $\chi_0$ in $M_B$ should be set to zero to ensure that the Hawking temperature equals the temperature at the boundary.

Recalling that the temperature $T=\frac{e^{\chi_0/2} }{\Delta \tau}$, thus the variation of $\chi_0$ gives $\delta \chi_0= 2\frac{\delta T}{T}$ with $\Delta\tau$ fixed in the variation of on-shell Euclidean action. Using the QSR, the variation of free energy density $w_{QSR}$ reads
\begin{equation}
\delta w_{QSR} = -\left( s+ \frac{ e^{\chi_0/2} }{L^3v} \frac{ k Q_{cs}}{T } \right)\delta T - \rho \delta\mu -2\mathcal{J} \delta\Omega_H -M_{B} \delta B -\frac{ e^{\chi_0/2} }{L^3v} k \delta Q_{cs} \,.
\end{equation}
By performing the transformation $w=w_{QSR}+ \frac{1}{L^3v}e^{\chi_0/2} k Q_{cs}$, we obtain
\begin{eqnarray}
w &=& \epsilon -Ts -\mu\rho - 2\Omega_H \mathcal{J} \,, \label{SM-D-freeW} \\
\delta w &=& -s \delta T - \rho \delta\mu -2\mathcal{J} \delta\Omega_H -M_B \delta B \,, \label{SM-D-free1st}
\end{eqnarray}
which reproduces equation (19) in the main text.

{\bf \emph{Resolution from Iyer-Wald formalism:}}
Having identified the additional Chern-Simons contribution from the Euclidean action variation, we now derive the same corrected thermodynamic potential $w$ independently within the Iyer-Wald formalism~\cite{Iyer:1994ys}. The thermodynamics of the rotating BHs described by~\eqref{SM-D-ansatz} can be analyzed using the two fundamental identities~\eqref{SM-A-fundamentalID} and~\eqref{SM-A-identity}, with the explicit forms of $\mathbf Q$ and $\mathbf\Theta$ given in~\eqref{SM-A-QTheta}.

We perform this calculation directly in the original boundary non-rotating coordinates $(t,\psi,\theta,\phi,r)$. In this part of the analysis, we set $\chi_0=0$ from the outset, corresponding to the standard normalization of the boundary time coordinate. The relevant Killing vector is the horizon generator $\xi_H$ defined in~\eqref{SM-D-Killing}. As usual in the Iyer-Wald derivation, $\xi_H$ is held fixed under the phase-space variation, $\delta\xi_H=0$.
We take $\Sigma$ to be a spacelike hypersurface of constant $t$, whose inner boundary is the bifurcation surface at $r=r_h$. Owing to the compactness of the coordinates $(\theta,\phi,\psi)$, one might initially conclude that the integration of~\eqref{SM-A-fundamentalID} over $\Sigma$ receives contributions only from the boundaries at $r=0$ and the event horizon $r=r_h$. 
However, in the presence of the magnetic field, the gauge potential used in~\eqref{SM-D-ansatz} contains singularities at the north and south poles, $\theta=0$ and $\theta=\pi$, as indicated by the divergence of
\begin{equation}
A_aA^a \sim B^2 \frac{\cos{\theta}^2}{ \sin{\theta}^2 }  +\cdots \,.
\end{equation}
Consequently, the surface integral in~\eqref{SM-A-fundamentalID} must be evaluated along a contour that excises the singularities at the south and north poles, as indicated by the dashed paths at $\theta=\epsilon^+$ and $\theta=\pi-\epsilon^+$ (with $\epsilon^+$ infinitesimal) in Fig.~\ref{fig:contour}.

\begin{figure}
\begin{center}
\begin{tikzpicture}[scale=1.6]
\def\aBig{2}
\def\bBig{1.2}
\def\aSmall{1}
\def\bSmall{0.6}
\draw[thick,blue] (0,0) ellipse (\aBig cm and \bBig cm); 
\draw[thick,blue] (0,0) ellipse (\aSmall cm and \bSmall cm);
\node at (\aBig+0.3, 0) {$r=0$};
\node at (\aSmall+0.3, 0) {$r=r_h$};
\draw[thick,magenta,->](-\aBig-0.1, 0)--(-\aBig-0.4,0) node[midway,below] {$n_{(r)}^\mu$};
\draw[thick,magenta,->](-\aSmall+0.15, 0)--(-\aSmall+0.4,0) node[midway,below] {$n_{(r)}^\mu$};
\node[above] at (0.25, \bBig+0.05) {$\theta=0$};
\node[below] at (0.25, -\bBig-0.05) {$\theta=\pi$};
\draw[red, thick](0,\bBig+0.1)--(0,-\bBig-0.1);
\draw[dashed,blue,thick]
(-0.12,\bBig) .. controls (-0.11,0.98) and (-0.05,0.72) .. (-0.04,\bSmall);
\draw[dashed,blue,thick]
(0.12,\bBig) .. controls (0.11,0.98) and (0.05,0.72) .. (0.04,\bSmall);
\draw[dashed,blue,thick]
(-0.04,-\bSmall) .. controls (-0.05,-0.72) and (-0.11,-0.98) .. (-0.12,-\bBig);
\draw[dashed,blue,thick]
(0.04,-\bSmall) .. controls (0.05,-0.72) and (0.11,-0.98) .. (0.12,-\bBig);
\end{tikzpicture}
\caption{Integration contour on the $r-\theta$ plane. $n_{(r)}^\mu$ is the space-like unit normal vector for the $r=const.$ hypersurface. The coordinate singularities are depicted by the red lines.}
\label{fig:contour}
\end{center}
\end{figure}

Considering the projection of the $3$-form $(\delta\mathbf{Q}-\xi\cdot \mathbf{\Theta})$ along a fixed value of angular coordinates, we define $\delta_i \equiv (\delta \mathbf{Q} -\xi \cdot \mathbf{\Theta})|_i =(\delta Q -\xi\cdot \Theta)_i \sqrt{\gamma_{(i)}},\,  i=(\theta,\phi,\psi)$. A direct calculation shows
\begin{equation}
\delta_\theta = \frac{\cos\theta}{2} \mathcal{G} \,,\qquad \delta_\phi =\delta_\psi =0   \,,
\end{equation}
with
\begin{equation}
\mathcal{G}= \frac{1}{L^3 v} \left[ -\left( \frac{ e^{-\chi/2} h (B -2A_z) }{r}+\frac{k}{2}(A_tA_z' -A_t'A_z) \right) \delta B -\frac{k}{6} \Big((A_t -\Omega_H A_z)(A_z \delta B +2B \delta A_z) \Big)' \right] \,.
\end{equation}
Each excised polar line gives one tubular boundary component. The outward orientations of the north and south tubes are opposite, while $\cos\theta$ also changes sign between the two poles. Consequently, the two contributions add rather than cancel.

Integrating the fundamental identity~\eqref{SM-A-fundamentalID} over the hypersurface $\Sigma$, we obtain
\begin{equation}
\begin{split}
\delta H &=0=\int_\Sigma \boldsymbol{\omega}=\int_{\Sigma} d(\delta \mathbf{Q}- \xi \cdot \boldsymbol{\Theta}) =\int_{\partial\Sigma} (\delta \mathbf{Q} -\xi \cdot \boldsymbol{\Theta})    \\
&=\int (\delta Q- \xi\cdot \Theta )_r \sqrt{\gamma_{(r)}} d\theta d\phi d\psi \bigg|_{r=0}  -\int (\delta Q- \xi\cdot \Theta )_r \sqrt{\gamma_{(r)}} d\theta d\phi d\psi \bigg|_{r=r_h}   \\
&\qquad\qquad  +\int_{r_h}^0 \frac{\mathcal{G}}{2} dr \Big( \cos{ (\epsilon^+ ) } + (-1)\cos{(\pi-\epsilon^+ )} \Big)  d\phi d\psi \bigg|_{\epsilon^+\to 0} \\
&= \int (\delta Q- \xi\cdot \Theta )_r \sqrt{\gamma_{(r)}} d\theta d\phi d\psi \bigg|_{r=0}  -\int (\delta Q- \xi\cdot \Theta )_r \sqrt{\gamma_{(r)}} d\theta d\phi d\psi \bigg|_{r=r_h} + \int_{r_h}^0 \mathcal{G} dr  d\phi d\psi \,.
\end{split}
\end{equation}
Therefore, the first law of thermodynamics reads
\begin{equation}\label{SM-D-first}
\delta\epsilon = T\delta s +\mu\delta\rho +2\Omega_H \delta\mathcal{J} -M_B \delta B \,,
\end{equation}
where 
\begin{equation}
M_B = - \frac{1}{L^3v }\bigg\{ \int_0^{r_h} \bigg[ \frac{ e^{-\chi/2} h (B -2A_z)}{r} - \frac{ vB }{r} +\frac{k}{2}(A_t A_z' -A_t' A_z)  \bigg] dr + v B \ln{ (\frac{r_h}{L})} \bigg\} \,.
\end{equation}
One finds that the magnetization $M_B$ herein coincides exactly with~\eqref{SM-D-MB} by finally setting $\chi_0=0$.

Furthermore, integrating the identity~\eqref{SM-A-identity} over the hypersurface $\Sigma$ and applying Stokes's theorem yields
\begin{equation}\label{SM-D-free}
w= w_{QSR} +\frac{1}{L^3v} kQ_{cs} =\epsilon -Ts -\mu\rho -2\Omega_H\mathcal{J} \,.
\end{equation}
Combining this result with~\eqref{SM-D-first}, we recover the same expressions~\eqref{SM-D-freeW} and~\eqref{SM-D-free1st} derived from the variation of the Euclidean action.

{\bf\emph{Transgression interpretation:}} The analysis above reveals a nontrivial agreement: the Euclidean and Iyer-Wald approaches yield the same corrected thermodynamic potential, $w=w_{ QSR} +kQ_{ cs}/(L^3v)$. This agreement can be understood directly in terms of the transgression form. For the rotating BH~\eqref{SM-D-ansatz}, we choose the reference connection
\begin{equation}
\bar A_{ rot}=\frac{B}{4}\cos\theta\,d \phi \,, \quad \bar F_{ rot}=-\frac{B}{4}\sin\theta  d\theta\wedge d\phi \,,
\end{equation}
so that $\bar F_{ rot}^2=0$. The action constructed from $\mathcal T_5(A,\bar A_{ rot})$ then yields
\begin{equation}
w_{\mathcal T} =w_{ QSR}+\frac{k}{L^3v}Q_{ cs} =\epsilon-Ts-\mu\rho-2\Omega_H\mathcal J \,.
\end{equation}

More importantly, the Euclidean variation of the transgression-completed action and the corresponding Iyer-Wald calculation both reproduce the expected first law in~\eqref{SM-D-free1st}. Thus, the transgression completion naturally explains the origin of the $Q_{ cs}$ while rendering the full action strictly gauge invariant.

\section{BHs in EMCS theory with scalar hair} \label{SM-E}
The action of EMCS theory with the scalar field reads
\begin{equation} \label{SM-E-action}
S=\frac{1}{16\pi G} \int d^5x \sqrt{-g}\Big( R+\frac{12}{L^2} -\frac{(\partial\phi)^2}{2} -V(\phi)-\frac{Z(\phi)}{4}F^2+\frac{k}{24} \epsilon^{abcde}A_a F_{bc} F_{de} \Big) \,,
\end{equation}
with gauge field strength $F_{ab}=\partial_a A_b-\partial_b A_a$. Furthermore, $G$ is Newton's constant, $L$ is the AdS radius, and $k$ is the Chern-Simons coupling. Setting $16\pi G=1$, the corresponding equations of motion read
\begin{equation}\begin{split} \label{SM-E-eom}
R_{ab}-\frac{1}{2}g_{ab}\left(R+ \frac{12}{L^2} \right)-\frac{1}{2} T_{ab} &=0 \,,  \\ 
\nabla_{b}\Big( Z(\phi)F^{ba} \Big) +\frac{k}{8} \epsilon^{abcde}F_{bc}F_{de}&=0 \,,  \\
\nabla_a\nabla^a \phi -\frac{\partial_\phi Z(\phi)}{4}F^2 -\partial_\phi V(\phi) &=0 \,,
\end{split}\end{equation}
where
\begin{equation}
T_{ab}= Z(\phi) \left(F_{ac}F_b^{\ c}-\frac{1}{4}g_{ab}F_{cd}F^{cd} \right) +\nabla_a\phi\nabla_b\phi -\frac{1}{2}g_{ab} (\partial\phi)^2 -g_{ab}V(\phi) \,.
\end{equation}

We shall consider the following background ansatz for metric and gauge field
\begin{equation}\begin{split} \label{SM-E-ansatz}
ds^2&=\frac{1}{r^2}\Big[- \big(f e^{-\chi}-h^2p^2\big)dt^2+2ph^2 dtdz+ dx^2+dy^2 +h^2 dz^2+\frac{dr^2}{f}\Big]\,, \\
A&=A_t dt-\frac{B}{2}y dx+\frac{B}{2}x dy-A_z dz \,, \qquad  \phi=\phi(r) \,,
\end{split}\end{equation}
where $f,\, \chi,\, h,\, p,\, A_t, A_z$ and $\phi$ are functions of radial coordinate $r$. In these coordinates, $r=0$ corresponds to the asymptotically AdS boundary while $r=r_h$ corresponds to the location of BH horizon. 

Note that the properties of the hairy BHs studied herein (see ansatz~\eqref{SM-E-ansatz}) hold for arbitrary forms of $Z$ and $V$. In particular, our general action~\eqref{SM-E-action} encompasses the model considered in~\cite{Astefanesei:2010dk}, which corresponds to $Z=e^{-\alpha\phi}$ and $V=0$. Note that the derivation of the Chern–Simons induced correction presented below does not rely on the specific forms of $Z(\phi)$ and $V(\phi)$, and therefore applies to arbitrary choices of these functions. The detailed asymptotic expansions and thermodynamic quantities, however, depend on the scalar couplings. To present an explicit representative example, we choose
\begin{equation}
Z(\phi)=1\,, \quad V(\phi)=\frac{m^2}{2}\phi^2+\frac{\lambda}{4}\phi^4.
\end{equation}
Furthermore, we choose $m^2=-3$ so that the operator dual to the scalar field $\phi$ has conformal dimension $3$, and its source has conformal dimension $1$.

\subsection*{E.1: Asymptotic expansions and thermodynamics}
At the asymptotic AdS boundary $r=0$, we have for the metric fields
\begin{equation}
\begin{split}
f(r)&= 1+\frac{\phi_s^2}{6}r^2+\left(\frac{B^2}{2} +\frac{\phi_s^4(1+3\lambda)}{12} \right)r^4\ln{r}+f_4 r^4+\cdots \,, \\
\chi(r)&= \chi_0+\frac{\phi_s^2}{6}r^2 +\left( \frac{B^2}{4} +\frac{\phi_s^4(1+3\lambda)}{12} \right) r^4\ln{r}+ \chi_4 r^4+\cdots  \,, \\
h(r)&= 1+\frac{B^2}{8}r^4\ln{r}+h_4 r^4+\cdots  \,,\quad 
p(r) = e^{-\chi_0/2} p_0 + p_4 r^4+\cdots    \,, \\
A_t(r)&= e^{-\chi_0/2} \left( \mu +A_{t2} r^2+\cdots  \right) \,,\qquad
A_z(r) = A_{z2} r^2 +\cdots  \,,\\
\phi(r) &= \phi_s r +\frac{\phi_s^3(1+3\lambda)}{6}  r^3\ln{r} +\phi_v r^3+\cdots \,,
\end{split}
\end{equation}
where $\chi_4=2h_4 +\frac{B^2}{12} +\frac{\phi_s\phi_v}{2} +\frac{\phi_s^4(1+3\lambda)}{144}$. Near the event horizon, imposing the regularity condition with $f(r_h)=p(r_h)=A_t(r_h)=0$, we obtain for the metric and gauge fields 
\begin{equation}\begin{split}
f(r)&= f_1(r_h-r)+\cdots \,,\qquad \chi(r)= \chi_1+\cdots  \,,\qquad
h(r)= h_1+\cdots  \,,\qquad
p(r)= p_1(r_h-r)+\cdots  \,, \\ 
A_t(r)&= A_{t1}(r_h-r)+\cdots  \,,\quad A_z(r)= A_{z1}+\cdots \,,\qquad \phi=\phi_1+\cdots \,.
\end{split}\end{equation}

The temperature and Bekenstein-Hawking entropy density are given by 
\begin{equation}
T=-\frac{e^{\chi_0/2} } {4\pi} f'e^{-\chi/2} \Big|_{r=r_h} \,, \qquad s=\frac{4\pi h}{r^3} \Big|_{r=r_h}  \,.
\end{equation}

The grand-canonical potential can be obtained within the standard holographic renormalization procedure. The renormalized on-shell action reads
\begin{equation}\label{SM-E-sren}
S_{ren}=S+S_{bdy} = S+ \frac{1}{16\pi G}\int d^4x\sqrt{-\gamma}\Big[ 2K-6-\frac{\phi^2}{2} -\ln{r} \left( \frac{\hat{F}^2}{4}+\frac{(\partial_\mu \phi)^2}{2} +\frac{1+3\lambda}{12}\phi^4 \right) \Big] \,,
\end{equation}
where $\gamma_{\mu\nu}$ is the induced metric at the conformal boundary, $K$ is the trace of extrinsic curvature, and $\hat{F}_{\mu\nu}$ is the boundary field strength. Furthermore, the relevant thermodynamic quantities of the dual theory can be identified after performing the standard holographic renormalization. The expectation value of dual stress tensor reads
\begin{equation}
\langle T_{\mu\nu} \rangle= \lim_{r\to 0} \frac{1}{r^2} \Big[-2K_{\mu\nu} +\left( 2K-6-\frac{\phi^2}{2} \right) \gamma_{\mu\nu} + \ln{(r)} t_{\mu\nu} \Big] \,, 
\end{equation}
where $t_{\mu\nu}= \hat{F}_{\mu\rho} \hat{F}_\nu^{\, \rho}-\frac{ \gamma_{\mu\nu} }{4} \left( \hat{F}^2 -\frac{1}{2}(\partial_\mu\phi)^2 +\frac{1+3\lambda}{12}\phi^4 \right)$. Its non-zero components are
\begin{equation}\begin{split}
& \langle T_{tt}\rangle= e^{-\chi_0} \left(-3f_4+8h_4+\frac{B^2}{4} +\frac{\phi_s^4}{48} +\phi_s\phi_v \right) \,, \\
& \mathcal{P}_\perp = \langle T_{xx}\rangle =\langle T_{yy}\rangle = -f_4 -\frac{B^2}{6} +\frac{3+4\lambda}{48} \phi_s^4 +\phi_s\phi_v  \,, \\
& \mathcal{P}_\parallel = \langle T_{zz}\rangle=-f_4+8h_4+\frac{B^2}{12} +\frac{3+4\lambda}{48} \phi_s^4 +\phi_s\phi_v \,,\quad \langle T_{tz}\rangle=\langle T_{zt}\rangle=4p_4 \,.
\end{split}\end{equation}
Similarly, the expectation value of the dual current reads
\begin{equation}
\langle J^{\mu} \rangle =\lim_{r\to0} \frac{1}{r^4} \Big[ n_r \left( F^{\mu r} +\frac{k}{6}\epsilon^{r\mu \alpha\beta\gamma} A_\alpha F_{\beta\gamma} \right) +\nabla_\alpha \hat{F}^{\alpha\mu} \ln{r} \Big] \,.
\end{equation}
The non-zero components of $\langle J^{\mu} \rangle$ are
\begin{equation}\begin{split} 
\langle J^t \rangle= -2e^{\chi_0/2} A_{t2}  \,,\quad  \langle J^z \rangle=  -2A_{z2} -\frac{k}{3} \mu B  \,. 
\end{split}\end{equation}
With respect to the timelike unit normal $u^\mu$, the energy density $\epsilon$ and charge density $\rho$ are defined as
\begin{equation}
\epsilon = u^\mu u^\nu \langle T_{\mu\nu}\rangle = -3f_4 + 8h_4 + \frac{B^2}{4} +\frac{\phi_s^4}{48} +\phi_s\phi_v \,,\quad
\rho = u_\mu \langle J^\mu\rangle = -2A_{t2} \,.
\end{equation}

The expectation value of the dual scalar operator is
\begin{equation}
\begin{split}
\langle O \rangle= \lim_{r\to0} \frac{1}{r^3} \left( -n_r \nabla^r \phi -\phi -\frac{1+3\lambda}{3}\phi^3 \ln{r} \right) = 2\phi_v +\frac{1+2\lambda}{4} \phi_s^3 \,.
\end{split}
\end{equation}

To obtain analytical expressions for the thermodynamic variables, we can utilize the properties of equations of motion and Ricci tensor to rewrite the on-shell action into surface terms. The Einstein equation (\ref{SM-E-eom}) implies that its $(i,i)$ component satisfies
\begin{equation}
2R^i_{\ i} =R+\frac{12}{L^2}  -\frac{(\partial \phi)^2}{2}-V(\phi) -\frac{Z(\phi)}{4}F^2 +Z(\phi) F^i_{\ c}F_i^{\ c} \,,\quad (i=t,x,y,z)
\end{equation}
Therefore, the Lagrangian density $\mathcal{L}$ can be rewritten as
\begin{equation} \label{SM-E-slag}
\mathcal{L}=2R^i_{\ i} -Z(\phi)F^i_{\ c}F_i^{\ c} +\frac{k}{24}\epsilon^{abcde}A_a F_{bc} F_{de}  \,,
\end{equation}
where $i$ is one of the coordinates $(t,x,y,z)$ and there is no sum in $i$ here. Using~\eqref{SM-E-slag} and the identity $R^a_{\ b}\xi^b=\nabla_b\nabla^a\xi^b$ for Killing vector, the bulk on-shell action can be expressed as
\begin{equation} \label{SM-E-os}
S=\Delta T V\int dr \Big[ 2\sqrt{-g}R^t_{\ t}+\left(A_t N_t+\frac{k}{3}B A_t A_z \right)'+\frac{k}{3}B A_t A_z' \Big] \,,
\end{equation}
where $\Delta T=\int dt$ and $V=\int dxdydz$. Substituting~\eqref{SM-E-os} and the ultraviolet (UV) and infrared (IR) expansions into~\eqref{SM-E-sren}, the renormalized on-shell action $S_{ren}$ reads
\begin{equation}
\frac{ S_{ren} }{\Delta T V}= e^{-\chi_0/2} \left( 3f_4 -8h_4 -\frac{B^2}{4} -\frac{\phi_s^4}{48} -\phi_s\phi_v -2\mu A_{t2} \right) +e^{-\chi_1/2 } \frac{f_1 h_1}{r_h^3} +\frac{k B}{3} \int_{0}^{r_h} A_t A_z' dr \,.
\end{equation}
After the Wick rotation $\tau=i t$, one finds the Euclidean action $I_E=-i S_{ren}$. From the QSR, we obtain
\begin{equation} 
w_{QSR}= -\frac{ T\ln Z}{V}=\frac{ T I_E}{V} =\epsilon -Ts-\mu\rho- \frac{ e^{\chi_0/2} kB}{3}\int_0^{r_h} A_t\, A_z' d r \,.
\end{equation}
Here the Bekenstein-Hawking entropy density is $s =\frac{4\pi h_1}{r_h^3}$, while the temperature $T=\frac{e^{\chi_0/2} }{\Delta \tau}=e^{-\frac{\chi_1-\chi_0}{2}} \frac{f_1}{4\pi}$ is determined by removing the conical singularity of the Euclidean metric at the horizon, with $\Delta\tau$ held fixed in the subsequent variation.

\subsection*{E.2: Thermodynamic consistency: from Chern–Simons correction to transgression}
{\bf \emph{Variation of on-shell Euclidean action:}} Consider analytic continuation to Euclidean time $\tau$ by a Wick rotation $t=-i \tau$, under which the total Euclidean action $I_{E}$ is given by
\begin{equation}
I_{E}=I+I_{bdy} \,,
\end{equation}
where $I=-i S =-i S_{eff}$ and $I_{bdy}$ is the Euclidean boundary action, including counterterms. The $S_{eff}$ is the effective action of $S$ obtained by substituting the background ansatz~\eqref{SM-E-ansatz} into~\eqref{SM-E-action}, which is given by
\begin{equation}
\begin{split}
S_{eff}&=\int d^5x \frac{e^{-\chi/2} h}{r^3} \bigg\{ -f''-\frac{2fh''}{h}+f\chi'' +f'\left( \frac{8}{r}-\frac{2h'}{h}+\frac{3\chi'}{2} \right) \\
&\quad  +\left( \frac{h'}{r h} -\frac{\chi'}{2 r} \right) \left( 8f +r  f\chi' \right) +\frac{1}{2}e^{\chi} \left[ h^2 p'^2 +r^2Z(\phi)(A_t'+p A_z')^2 \right] \\
&\quad  +\frac{12-20f}{r^2}-\frac{r^2Z(\phi)}{2}\left( B^2+\frac{fA_z'^2}{h^2} \right) -\frac{f}{2}\phi'^2 -\frac{V(\phi)}{r^2} \bigg\} +\int d^5x \frac{k B}{3}\left( A_t' A_z -A_t A_z'\right) \,.
\end{split}
\end{equation}

As the boundary of Euclidean metric is the AdS boundary, the variation of $I_E$ is given by
\begin{equation}
\delta I_E = \delta I+\delta I_{bdy}  =\Delta\tau V \bigg\{ e^{-\frac{\chi_0}{2}} \bigg[-\rho\delta\mu -\langle O\rangle \delta\phi_s -\left( \epsilon -\mu \rho \right) \frac{\delta \chi_0}{2} \bigg] -k \delta Q_{cs} - e^{-\chi_0/2} M_B \delta B  \bigg\} \,,
\end{equation}
where the charge
\begin{equation} \label{SM-E-Qcs}
Q_{cs}=\frac{B}{6}\int_0^{r_h} (A_z' A_t- A_z A_t') dr \,,
\end{equation}
and
\begin{equation} \label{SM-E-mb}
M_B = -\bigg\{ \int_0^{r_h} \bigg[ \frac{B }{r} \left(  e^{ (\chi_0-\chi)/2 } h -1 \right) +\frac{e^{\chi_0/2} k}{2}(A_z' A_t - A_z A_t') \bigg] dr+ B\ln{r_h} \bigg\} \,.
\end{equation}
Note that in the final expression, the constant $\chi_0$ in $M_B$ should be set to zero to ensure that the Hawking temperature equals the temperature at the boundary.

Recalling that the temperature $T=\frac{e^{\chi_0/2} }{\Delta \tau}$, thus the variation of $\chi_0$ gives $\delta \chi_0= 2\frac{\delta T}{T}$ with $\Delta\tau$ fixed in the variation of on-shell Euclidean action. Using the QSR, we find that the variation of $w_{QSR}$ reads
\begin{equation}
\delta w_{QSR} = -\left( s+ \frac{e^{\chi_0/2} k Q_{cs}}{T} \right)\delta T -\rho \delta\mu -\langle O\rangle \delta \phi_s -e^{\chi_0/2} k \delta Q_{cs} -M_B \delta B \,.
\end{equation}
After including the Chern-Simons correction, $w=w_{ QSR}+e^{\chi_0/2} k Q_{ cs}$, we obtain
\begin{eqnarray}
w &=&\epsilon-Ts-\mu \rho \,, \label{SM-E-freeW} \\
\delta w  &=& -s\delta T -\rho \delta\mu  -\langle O\rangle \delta \phi_s -M_B \delta B \,. \label{SM-E-free1st}
\end{eqnarray}
These relations reproduce equation~(20) in the main text.

{\bf \emph{Resolution from the Iyer-Wald formalism:}} Now, we will use the Iyer-Wald formalism~\cite{Iyer:1994ys} to prove the first law of thermodynamics in (\ref{SM-E-freeW}-\ref{SM-E-free1st}). Following the Iyer-Wald procedure~\cite{Iyer:1994ys}, the variation of Lagrangian 5-form $\mathbf{L}=\mathcal{L} \mathbf{\epsilon}$ is given by
\begin{equation}
\delta\mathbf{L} = (\mathbf{E}_g)^{ab}\delta g_{ab} +(\mathbf{E}_A)^a \delta A_a +\mathbf{E}_\phi \delta \phi +d\mathbf{\Theta} \,,
\end{equation}
where $\mathbf{\Theta}$ is the symplectic potential form defined as
\begin{equation} \label{SM-E-sp}
\mathbf{\Theta}= \boldsymbol{\epsilon}_{a a_1a_2a_3a_4} \left[  \left( g^{ac}g^{bd}-g^{ab}g^{cd} \right) \nabla_b \delta g_{cd} -\left( Z(\phi) F^{ae}+\frac{k}{6}\epsilon^{abcde} A_b F_{cd} \right) \delta A_e -\nabla^a\phi \delta\phi \right]\,.
\end{equation}
One can associate with an infinitesimal diffeomorphism $\delta_\xi x^a=\xi^a(x)$ a Noether current $4$-form, defined by $\mathbf{J}\equiv \mathbf{\Theta}(\phi,\mathcal{L}_\xi \phi) - \xi \cdot \mathbf{L}$. Thus, there exists a Noether charge $3$-form $\mathbf{Q}$, such that $\mathbf{J}=d\mathbf{Q}$, provided that the background equations of motion are satisfied. The expression for Noether charge is given by
\begin{equation}
\mathbf{Q}=-\boldsymbol{\epsilon}_{abfgh} \left[ \nabla^a\xi^b + \frac{1}{2} \left( Z(\phi) F^{ab}+\frac{k}{6}\epsilon^{aijkb} A_i F_{jk}\right)  A_c \xi^c \right] \,.
\end{equation}
Note that the scalar field does not appear in $\mathbf{Q}$, although it contributes to the symplectic potential~\eqref{SM-E-sp}.

Let the dynamical fields satisfy the background equations of motion with $\xi^a$ a Killing vector. Thus, we arrive at the fundamental identity which gives rise to the first law of BH mechanics
\begin{equation} \label{SM-E-fundamentalID}
\boldsymbol{\omega} (\phi, \delta\phi, \mathcal{L}_\xi\phi) = \delta \bold{J} -d(\xi \cdot \boldsymbol{\Theta} ) =d \left( \delta \bold{Q} -\xi \cdot \boldsymbol{\Theta} \right) =0 \,.
\end{equation}

Let us choose $\xi^a$ to be the time-like Killing vector $\xi^a=(\partial_t)^a=\delta^a_t$, which becomes null at the event horizon $r=r_h$. Let $\Sigma$ be a $t=\mathrm{const.}$ space-like hypersurface with the horizon $r=r_h$ serving as its interior boundary. Then, the boundary of hypersurface
$\Sigma$,~\emph{i.e.}~$\partial\Sigma$, is expressed as
\begin{equation}
\partial\Sigma = S_{r=r_h} \cup S_{r=0} \cup S_{x=L_x/2} \cup S_{x=-L_x/2} \cup
S_{y=L_y/2} \cup S_{y=-L_y/2} \cup S_{z=L_z/2} \cup S_{z=-L_z/2} \,,
\end{equation}
where we have considered the regulation $(L_x, L_y, L_z)$ along three spatial directions. Here $S_{r=r_h}$ denotes the boundary at $r=r_h$ and $S_{r=0}$ represents the boundary at $r=0$. The remaining terms account for the boundaries at the edges of the non-compact directions in the $x,y$ and $z$ coordinates. We can compute the restriction of $(\delta\mathbf{Q}-\xi\cdot\mathbf{\Theta})$ to each hypersurface $S_i$. Specifically, we have $(\delta\mathbf{Q}-\xi\cdot\mathbf{\Theta})|_i= (\delta Q-\xi\cdot \Theta)_i \boldsymbol{\epsilon}_{(i)}$. Along each non-compact boundary, we find
\begin{equation}
\begin{split}
(\delta Q -\xi \cdot \Theta)_x &= x \bigg[ \left( \frac{B e^{-\chi/2} h }{2r}+\frac{k}{6}(2A_tA_z' -A_t'A_z) \right) \delta B +\frac{kB}{6} (A_t\delta A_z)' \bigg] \,, \\
(\delta Q -\xi \cdot \Theta)_y &= \frac{y}{x} (\delta Q -\xi \cdot \Theta)_x  \,, \quad 
(\delta Q -\xi \cdot \Theta)_z =0   \,.
\end{split}
\end{equation}

By integrating the fundamental identity~\eqref{SM-E-fundamentalID} over the hypersurface $\Sigma$, we obtain
\begin{equation}
\begin{split}
\delta H &=0=\int_\Sigma \boldsymbol{\omega}=\int_{\Sigma} d(\delta \mathbf{Q}- \xi \cdot \boldsymbol{\Theta}) =\int_{\partial\Sigma} (\delta \mathbf{Q} -\xi \cdot \boldsymbol{\Theta})    \\
&= -\delta \epsilon +T\delta s +\mu\delta\rho - \langle O\rangle \delta \phi_s  \\
&\quad +\left( \int_0^{r_h}   \left[  \frac{B }{r}\left( e^{-\chi/2} h -1 \right) +\frac{k}{3} (2A_z' A_t -A_z A_t') \right] dr + B \ln{r_h}\right) \delta B \,. 
\end{split}
\end{equation}
After using the identity $\int_0^{r_h} A_t' A_zdr= (A_t A_z)|_0^{r_h} -\int_0^{r_h} A_t A_z'dr=-\int_0^{r_h} A_t A_z'dr$, we arrive at the first law of BH thermodynamics
\begin{equation}\label{SM-E-first}
\delta \epsilon = T\delta s +\mu\delta\rho  -\langle O\rangle \delta \phi_s  -M_B \delta B\,,
\end{equation}
where $M_B$ matches exactly the result derived via the variation of Euclidean action~\eqref{SM-E-mb}. Similarly, the integration of $d\mathbf{Q}=-\xi\cdot\mathbf{L}$ over $\Sigma$ gives
\begin{equation} \label{SM-E-free}
w=\epsilon-Ts -\mu\rho \,.
\end{equation}
Combining~\eqref{SM-E-first} and~\eqref{SM-E-free} yields the same expressions for the free energy and its variation~\eqref{SM-E-freeW} and~\eqref{SM-E-free1st}, as shown from the Euclidean action variation.

{\bf\emph{Transgression interpretation:}} The Euclidean variation and the Iyer-Wald formalism independently establish that the thermodynamic potential is $w=w_{QSR}+kQ_{cs}$, even in the presence of scalar hair. The transgression form explains this novel result at the level of the action. The scalar hair does not change the magnetic sector of the $U(1)$ gauge field. We may therefore use the same reference connection as for the planar solution
\begin{equation} 
\bar A=\frac{B}{2}(x dy-y dx) \,, \quad \bar F^2=0 \,.
\end{equation}
Upon replacing the Chern-Simons term by $\mathcal T_5(A,\bar A)$, its on-shell contribution is independent of the scalar potential $V(\phi)$ and the coupling $Z(\phi)$, and the complete free energy becomes
\begin{equation} 
w_{\mathcal T}  =w_{ QSR}+kQ_{ cs} =\epsilon-Ts-\mu\rho.
\end{equation}
Thus, the same $Q_{ cs}$ found in~\eqref{SM-E-Qcs} is the finite transgression completion even in the presence of nontrivial scalar profiles.

Furthermore, the Euclidean variation of the transgression-completed action and the corresponding Iyer-Wald calculation both reproduce the expected first law~\eqref{SM-E-free1st}, with the scalar work term $-\langle O\rangle\delta\phi_s$ unchanged.

\section{BHs with higher-derivative terms} \label{SM-F}
We consider EMCS theory extended by the Gauss-Bonnet term--a higher-curvature correction to general relativity that emerges naturally in the low-energy limit of string theory~\cite{Cai:2001dz}
\begin{equation} \label{SM-F-action}
S=\frac{1}{16\pi G} \int d^5x \sqrt{-g}\Big( R + \alpha L_{GB}+\frac{12}{L^2}-\frac{1}{4}F_{ab}F^{ab}+\frac{k}{24} \epsilon^{abcde}A_a F_{bc} F_{de} \Big)  \,.
\end{equation}
The Gauss-Bonnet term $L_{GB}$ reads
\begin{equation}
L_{GB}= R^2 -4R_{ab}R^{ab} +R_{abcd}R^{abcd}  \,.
\end{equation}
Setting $16\pi G=1$, the corresponding equations of motion are given by
\begin{equation}\begin{split} \label{SM-F-eom}
R_{ab}-\frac{1}{2}g_{ab}\left(R+ \frac{12}{L^2} \right) +\alpha H_{ab}-\frac{1}{2}\left(F_{ac}F_b^{\ c}-\frac{1}{4}g_{ab}F_{cd}F^{cd} \right)&=0 \,,  \\ 
\nabla_{b}F^{ba} +\frac{k}{8} \epsilon^{abcde}F_{bc}F_{de}&=0 \,,  
\end{split}\end{equation}
where
\begin{equation}
H_{ab} = 2\left( R_{acde}R_{b}^{\ cde} -2R_{acbd}R^{cd} -2R_{ac}R^c_{\ b} +RR_{ab} \right) -\frac{1}{2}g_{ab} L_{GB} \,.
\end{equation}

We will investigate the effects of higher-derivative terms on the thermodynamics of BHs with planar and spherical topologies, respectively. We demonstrate in detail that the phenomenon revealed in the main text persists in Einstein-Gauss-Bonnet-Maxwell-Chern-Simons (EGB-MCS) theory, and that the form of $Q_{cs}$ remains unaffected by higher-derivative corrections. More importantly, the BH entropy is given by the Wald entropy, which yields strong evidence for the robustness of the form of $Q_{cs}$.

\subsection*{F.1: BHs with planar topology}
We shall consider the following background ansatz for metric and gauge field
\begin{equation}\begin{split} \label{SM-F-ansatz1}
ds^2&= \frac{L_c^2}{r^2}\Big[- \big(f e^{-\chi}-h^2p^2\big)dt^2+2ph^2 dtdz +dx^2+dy^2+h^2 dz^2 +\frac{dr^2}{f}\Big]\,, \\
A&= A_t dt +\frac{B}{2} \left( x dy -ydx \right)  -A_z dz \,,
\end{split}\end{equation}
where $f,\, \chi,\, h,\, p,\, A_t$ and $A_z$ are functions of $r$. The asymptotically AdS boundary lies at $r=0$. Here $L_c$ denotes the effective AdS radius:
\begin{equation} \label{SM-F-leff}
L_c^2 \equiv \frac{ \left(1+\sqrt{1 -8\alpha/L^2} \right) }{2 } L^2 \,,
\end{equation}
from which there is an upper bound for the Gauss-Bonnet coefficient~\emph{i.e.} $\alpha \leq L^2/8$.

\subsubsection*{F.1.1: Asymptotic expansions and thermodynamics}
At the asymptotic AdS boundary $r=0$, we have for the metric fields
\begin{equation}\begin{split}
f(r)&= 1+\frac{B^2 L_c^2}{2 (1-4\alpha/L_c^2 ) } \frac{r^4 }{L_c^4}\ln{(\frac{r}{L_c }) }+f_4 \frac{r^4}{L_c ^4 }+\cdots \,,\qquad  
\chi(r)= \chi_0 +\frac{B^2 L_c^2}{ 4(1-4\alpha/L_c^2 ) } \frac{r^4}{L_c^4}\ln{(\frac{r}{L_c}) } + \chi_4 \frac{r^4}{L_c^4 } +\cdots  \,, \\
h(r)&= 1+ \frac{B^2 L_c^2 }{ 8(1-4\alpha/L_c^2 ) } \frac{r^4}{L_c^4}\ln{(\frac{r}{L_c}) } +h_4 \frac{r^4}{L_c^4}+\cdots  \,, \qquad p(r)= p_4 \frac{r^4}{L_c^4} +\cdots    \,, \\
A_t(r)&= e^{-\chi_0/2} \left( \mu +A_{t2} \frac{r^2}{L_c^2 } +\cdots  \right) \,,  \qquad\qquad  A_z(r)= A_{z2} \frac{r^2}{L_c^2} +\cdots  \,,
\end{split}\end{equation}
where $\chi_4 = 2h_4 + \frac{ B^2 L_c^2 }{ 12(1-4\alpha/L_c^2)}$.

Near the BH horizon $r=r_h$, imposing the regularity condition with $f(r_h)=p(r_h)=A_t(r_h)=0$, we obtain for the metric and gauge fields 
\begin{equation}\begin{split}
f(r)&=f_1 (r_h -r)+\cdots  \,,\quad \chi(r)=\chi_1 +\cdots \,,\quad h(r)=h_1 +\cdots  \,,\\
p(r)&= p_1(r_h -r)+\cdots \,,\quad A_t(r)=A_{t1}(r_h-r)+\cdots\,,\quad A_z(r) = A_{z1} +\cdots \,.
\end{split}\end{equation}
The temperature and entropy density are given by 
\begin{eqnarray}
T=-\frac{e^{\chi_0/2} } {4\pi} f'e^{-\chi/2} \Big|_{r=r_h} \,, \qquad s= \frac{4\pi h L_c^3}{r^3} \Big|_{r=r_h} \,.
\end{eqnarray}

To formulate a well-defined variational principle, one has to supplement the bulk action with proper boundary terms. In particular, the conventional Gibbons-Hawking term is given by
\begin{equation} \label{SM-F-GH}
S_{b}^{E} = \frac{1}{16\pi G} \int d^4x \sqrt{-\gamma} (2K) \,,
\end{equation}
and its counterpart for Gauss-Bonnet gravity is~\cite{Brihaye:2008xu}
\begin{equation} \label{SM-F-GHgb}
S_{b}^{GB} = \frac{ 1}{16\pi G} \int d^4x \sqrt{-\gamma}  \big[ 4 \alpha (\mathcal{J} -2\hat{G}_{\mu\nu} K^{\mu\nu}) \big] \,,
\end{equation}
where $\gamma_{\mu\nu}$ is the induced metric at the AdS boundary, $K$ is the extrinsic curvature, $\hat{G}_{\mu\nu}$ is the Einstein tensor of $\gamma_{\mu\nu}$ and $\mathcal{J}$ is the trace of
\begin{equation}
\mathcal{J}_{\mu\nu} =  \frac{1}{3} \left( 2KK_{\mu\rho} K^{\rho}_{\ \nu} +K_{\rho\sigma}K^{\rho\sigma} K_{\mu\nu} -2K_{\mu\rho}K^{\rho\sigma}K_{\sigma\nu} - K^2 K_{\mu\nu} \right) \,.
\end{equation}
The explicit expression for $\mathcal{J}$ is
\begin{equation}
\mathcal{J}= \frac{1}{3} \left( 3K K_{\mu\nu}K^{\mu\nu} -2 K_\mu^{\ \nu} K_\nu^{\ \rho} K_\rho^{\ \mu} -K^3  \right) \,.
\end{equation}
The divergences of the on-shell action could be removed by adding the counterterm. The boundary counterterm $S_{ct}$ is given by
\begin{equation}\begin{split} \label{SM-F-sct}
S_{ct}  =& \frac{1}{16\pi G}\int d^4x\sqrt{-\gamma}\Bigg\{ -\frac{6}{L_c} \left( 1- \frac{4\alpha}{3 L_c^2} \right)-\frac{L_c\hat{R}}{2} \left( 1+\frac{4\alpha}{L_c^2 } \right) \\
&\qquad +\frac{L_c^3}{4}\ln{(\frac{r}{L_c})} \left[ \left( 1 -\frac{4\alpha}{L_c^2} \right) \left( \hat{R}_{\mu\nu}\hat{R}^{\mu\nu} -\frac{\hat{R}^2 }{3} \right)- \frac{\hat{F}^2}{L_c^2} \right] \Bigg\}\,.
\end{split}\end{equation}
where $\hat{R}_{\mu\nu}$ denotes the Ricci tensor associated with $\gamma_{\mu\nu}$ and $\hat{F}$ is the field strength of the boundary $U(1)$ gauge field. Note that for the planar BH~\eqref{SM-F-ansatz1}, the
conformal boundary is flat Minkowski spacetime and thus all tensors derived from the boundary Riemann tensor $\hat{R}_{\mu\nu}$ vanish.

In summary, the renormalized on-shell action is given by
\begin{equation} \label{SM-F-sren1}
S_{ren}=S +S_{bdy} \,,
\end{equation}
where the boundary terms $S_{bdy}= S_b^{E} +S_b^{GB} + S_{ct}$. We now proceed to obtain an analytical expression for the renormalized action $S_{ren}$. Recalling that the Einstein equation in~\eqref{SM-F-eom} implies that its $(t,t)$ component takes
\begin{equation}
R^t_{\ t} +\alpha H^t_{\ t} = \frac{1}{2} \left( R+ \frac{12}{L^2} -\frac{F^2}{4} +F^t_{\ c}F_t^{\ c} \right) \,.
\end{equation}
Adding $L_{GB}$ to both sides of the equation, we get
\begin{equation}\begin{split}
2R^t_{\ t} + \alpha \left( 2H^t_{\ t} + L_{GB} \right) -F^t_{\ c}F_t^{\ c} = R +\alpha L_{GB}+ \frac{12}{L^2} -\frac{F^2}{4} \,.
\end{split}\end{equation}
Therefore, the Lagrangian density $\mathcal{L}$ can be rewritten as
\begin{equation}\begin{split} \label{SM-F-slag1}
\mathcal{L} &= R +\alpha L_{GB} +\frac{12}{L^2}-\frac{1}{4}F^2 +\frac{k}{24} \epsilon^{abcde}A_a F_{bc} F_{de}  = 2R^t_{\ t} + \alpha \left( 2H^t_{\ t} + L_{GB} \right) -F^t_{\ c}F_t^{\ c}  +\frac{k}{24} \epsilon^{abcde}A_a F_{bc} F_{de}  \,.
\end{split}\end{equation}
Remarkably, it can be shown that
\begin{equation}\begin{split} \label{SM-F-Rtt}
2\sqrt{-g}R^t_{\ t} &= \left( \frac{e^{-\chi/2} L_c^3 h}{r^4} \Big[ -r f'+r e^\chi h^2 p p'  +f(2+r \chi') \Big] \right)'  \,, \\
\alpha \sqrt{-g} (2 H^t_{\ t} + L_{GB}) &= \left(   \frac{4 \alpha e^{-\chi/2 } L_c f }{r^4} \Big[ \left(3 h-2 r h'\right) \left(r f'-r f  \chi '-2 f\right)-r e^{\chi} h^3 p' \left(p +r p'\right) \Big]  \right)'  \,.
\end{split}\end{equation}
On the other hand, we have
\begin{equation} \label{SM-F-Ftt}
-\sqrt{-g}F^t_{\ c}F_t^{\ c} = \frac{e^{\chi/2} L_c h A_t'}{r} \left(A_t' +p A_z' \right) =(A_t N_t)' + k B A_t A_z' \,,
\end{equation}
where $N_t\equiv \frac{e^{\chi/2} L_c h }{r} \big[ A_t'+p A_z' \big]$ and $N_t'=-k B A_z'$ from the Maxwell equation.

By combining~\eqref{SM-F-slag1},~\eqref{SM-F-Rtt}, and~\eqref{SM-F-Ftt}, we find that the bulk on-shell action $S$ takes the form
\begin{equation}
S = \int d^4x \int dr \left[ 2 \sqrt{-g}\left( R^t_{\ t} +\alpha H^t_{\ t} + \frac{\alpha}{2} L_{GB}  \right) +\left( A_tN_t +\frac{k}{3}B A_t A_z \right)' +\frac{k}{3} B A_t A_z'   \right] \,.
\end{equation}
Therefore, the renormalized on-shell action is given by
\begin{equation}
\frac{S_{ren} }{\Delta T V} = \frac{e^{-\chi_0/2} }{L_c} \left[ 3 \left( 1-\frac{ 4\alpha}{L_c^2} \right) f_4 -8 \left( 1-\frac{ 4\alpha}{L_c^2} \right) h_4 -\frac{B^2 L_c^2}{4 } -\mu \frac{2A_{t2}}{L_c} \right]   +\frac{e^{-\chi_1/2} f_1 h_1 L_c^3 }{r_h^3}  +\frac{kB}{3} \int_0^{r_h} A_t A_z' dr \,,
\end{equation}
where $\Delta T=\int dt$ and $V=\int dxdydz$.

The expectation value of the dual stress tensor and current are
\begin{equation}
\langle T_{\mu\nu} \rangle =  \lim_{r\to0} \left( \frac{L_c^2}{r^2}  T_{\mu\nu}^{(\gamma)} \right) \,,\quad
\langle J^{\mu} \rangle  = \lim_{r\to0} \frac{L_c^4}{r^4} \Big[ n_r \Big( F^{\mu r} +\frac{k}{6}\epsilon^{r\mu \alpha\beta\gamma} A_\alpha F_{\beta\gamma} \Big) +L_c \ln{(\frac{r}{L_c})} \nabla_\alpha \hat{F}^{\alpha\mu}  \Big] \,, 
\end{equation}
where the quasi-local stress tensor is
\begin{equation}
T_{\mu\nu}^{(\gamma)} = -2(K_{\mu\nu} -K\gamma_{\mu\nu}) - 4\alpha \left(\mathcal{Q}_{\mu\nu} - \frac{1}{3}\mathcal{Q} \gamma_{\mu\nu} \right) -\frac{6}{L_c} \left( 1-\frac{4\alpha}{3 L_c^2}\right) \gamma_{\mu\nu} +L_c \ln{(\frac{r}{L_c})} \left( \hat{F}_{\mu\rho}\hat{F}_\nu^{\ \rho}-\frac{ \gamma_{\mu\nu} }{4} \hat{F}^2 \right) \,,
\end{equation}
with $\mathcal{Q}_{\mu\nu} =3\mathcal{J}_{\mu\nu} + 2K\hat{R}_{\mu\nu} +\hat{R}K_{\mu\nu} -2K^{\rho\sigma}\hat{R}_{\rho\mu\sigma\nu} -4\hat{R}_{\mu\rho}K_\nu^{~\rho}$ and $\mathcal{Q}=\mathcal{Q}_{\mu\nu} \gamma^{\mu\nu}$.

Therefore, for the planar BH considered in this subsection (see ansatz~\eqref{SM-F-ansatz1}), the non‑zero components of $\langle T^{\mu\nu}\rangle$ and $\langle J^{\mu} \rangle$ are given by
\begin{equation}\begin{split}
\langle T_{tt}\rangle &= \frac{ e^{-\chi_0}}{ L_c} \left[ -3 \left(1- \frac{4\alpha}{L_c^2}\right) f_4 +8\left( 1-\frac{4\alpha}{L_c^2}\right) h_4 +\frac{B^2 L_c^2 }{4} \right] \,, \\
\mathcal{P}_\perp  &= \langle T_{xx}\rangle =\langle T_{yy}\rangle = -\frac{1}{L_c} \left[ \left( 1-\frac{4\alpha}{L_c^2}\right)  f_4 +\frac{B^2L_c^2}{6} \right]  \,, \\
\mathcal{P}_\parallel  &= \langle T_{zz}\rangle= \frac{ 1 }{L_c } \left[ -\left( 1-\frac{4\alpha}{L_c^2 }\right) f_4 +8\left( 1-\frac{4\alpha}{L_c^2} \right) h_4 +\frac{B^2 L_c^2 }{12}  \right]  \,, \\
\langle T_{tz}\rangle  &=\langle T_{zt}\rangle= \frac{4 p_4}{L_c} \left( 1-\frac{4\alpha}{L_c^2} \right) \,,\\
\langle J^t \rangle &= -\frac{ 2e^{\chi_0/2} A_{t2}  }{ L_c }\,,\quad  \langle J^z \rangle = -\frac{ 2 A_{z2} }{L_c}  -\frac{1}{3} kB\mu \,.
\end{split}\end{equation}
Following the standard definitions, the energy density $\epsilon$ and charge density $\rho$ are given by
\begin{equation}
\epsilon = u^\mu u^\nu \langle T_{\mu\nu}\rangle = \frac{1}{ L_c} \left[ -3 \left(1- \frac{4\alpha}{L_c^2}\right) f_4 +8\left( 1-\frac{4\alpha}{L_c^2}\right) h_4 +\frac{B^2 L_c^2 }{4} \right] ,\quad
\rho = u_\mu \langle J^\mu\rangle = -\frac{2A_{t2} }{L_c} \,,
\end{equation}
where $u^\mu$ is the timelike unit normal vector.

The on-shell Euclidean action $I_E$ is obtained by analytic continuation of~\eqref{SM-F-sren1} and can be written as
\begin{equation}
I_E=I+I_{bdy}\,,
\end{equation}
with $I=-iS$ and $I_{bdy}$ the Euclidean boundary action. From the QSR, the free energy density $w_{QSR}$ is given by
\begin{equation}
w_{QSR}= -\frac{T \ln{Z}}{V} =\frac{T I_E}{V} =\epsilon -Ts  -\mu \rho - \frac{ e^{\chi_0/2} kB }{3} \int_0^{r_h} A_t A_z' dr \,.
\end{equation}
%

\subsubsection*{F.1.2: Thermodynamic consistency: from Chern–Simons correction to transgression}
{\bf\emph{Variation of on-shell Euclidean action:}} We proceed to analyze the BH thermodynamics using the method of variation of Euclidean action. Recalling that the total Euclidean action is given by 
\begin{equation}
I_{E}=I+I_{bdy} \,,    
\end{equation}
with $I=-iS =-i S_{eff}$. The effective action $S_{eff}$ of the original bulk action $S$ is given by
\begin{equation}\begin{split}
S_{eff}= S_{EMCS} +S_{GB}\,,
\end{split}\end{equation}
where the EMCS sector
\begin{equation}\begin{split}
\begin{split}
S_{EMCS}&= \int d^5x \frac{e^{-\chi/2} h L_c^3 }{r^3} \bigg\{ -f'' -\frac{2fh''}{h} +f\chi'' +f'\left( \frac{8}{r}-\frac{2h'}{h}+\frac{3\chi'}{2} \right) \\ 
&\quad +\left( \frac{h'}{r h} -\frac{\chi'}{2 r} \right) \left( 8f +r  f\chi' \right)  +\frac{1}{2}e^{\chi} \left[ h^2 p'^2 +\frac{r^2}{L_c^2} \left(A_t'+p A_z' \right)^2 \right] \\
&\quad +\frac{12L_c^2}{r^2L^2} -\frac{20f}{r^2}-\frac{r^2}{2L_c^2}\left( B^2+\frac{fA_z'^2}{h^2} \right) \bigg\} +\int d^5x \frac{kB}{3}\left( A_t' A_z -A_t A_z'\right) \,,
\end{split}
\end{split}\end{equation}
and the Gauss-Bonnet sector
\begin{equation}\begin{split}
\begin{split}
S_{GB} &= \int d^5x  \frac{\alpha e^{-\chi/2} h L_c }{r^3}  \bigg\{ 4f^2 \left( 3- \frac{2rh'}{h}\right)\left( \frac{f''}{f} -\chi'' \right) + \frac{8r f^2 h''}{h}\left( \frac{3}{r} -\frac{f'}{f} +\chi' \right)  \\
& \, +2e^\chi f h^2 p' \left(3p' -4rp'' \right) +2f^2\left( \chi' -\frac{2f'}{f}\right)\left( \frac{24}{r} -\frac{18h'}{h} -\frac{r e^\chi h^2 p'^2}{f} \right) \\
& \, +2f^2 \left( \chi'^2 +\frac{2f'^2}{f^2} -\frac{5f'\chi'}{f} \right) \left( 3- \frac{2rh'}{h}\right) -4r e^\chi fh^2p'^2 \left( \frac{2f'}{f} +\frac{3h'}{h}\right) +\frac{24f^2(5h -4rh')}{ r^2h} \bigg\} \,.
\end{split}
\end{split}\end{equation}

As the only boundary of Euclidean metric is the AdS boundary, the variation of the total on-shell Euclidean action $\delta I_E= \delta I+\delta I_{bdy}$ is given by
\begin{equation}
\frac{\delta I_E}{\Delta\tau V } = \bigg\{ e^{-\chi_0/2} \bigg[ -\rho\delta\mu  -\Big(\epsilon -\mu\rho \Big) \frac{\delta \chi_0}{2} \bigg] -k \delta Q_{cs} - e^{-\chi_0/2} M_B \delta B  \bigg\} \,,
\end{equation}
where
\begin{equation}
Q_{cs}=\frac{B}{6}\int_0^{r_h} \left( A_z' A_t -A_z A_t'\right) dr \,, \end{equation}
and
\begin{equation}
M_B = - \bigg\{ \int_0^{r_h} \bigg[ \left( \frac{ e^{(\chi_0-\chi)/2} h B L_c }{r} - \frac{ B L_c}{r} +\frac{e^{\chi_0/2}k}{2}(A_t A_z' -A_t' A_z)\right)  \bigg] dr +B L_c \ln{(\frac{r_h}{L_c})} \bigg\} \,.
\end{equation}
Note that in the final expression, the constant $\chi_0$ in $M_B$ should be set to zero to ensure that the Hawking temperature equals the temperature at the boundary.

Recalling that the temperature $T=\frac{e^{\chi_0/2} }{\Delta \tau}$, thus the variation of $\chi_0$ gives $\delta \chi_0= 2\frac{\delta T}{T}$ with $\Delta\tau$ fixed in the variation of on-shell Euclidean action. Applying the QSR, we obtain the variation of the free energy density $w_{QSR}$ as
\begin{equation} \label{SM-F-break}
\delta w_{QSR} =-\left(s+e^{\chi_0/2} \frac{ k Q_{cs}}{T } \right)\delta T -\rho \delta\mu - e^{\chi_0/2} k \delta Q_{cs} -M_{B} \delta B \,.
\end{equation}
In the presence of higher-derivative corrections, $w_{ QSR}$ still needs to be supplemented by the Chern-Simons correction in order to yield the standard thermodynamic potential. Defining the corrected free energy as $w=w_{ QSR}+e^{\chi_0/2} k Q_{ cs}$, we obtain
\begin{eqnarray}
w &=& \epsilon -Ts -\mu \rho \,, \\
\delta w  &=& -s \delta T -\rho \delta\mu -M_B \delta B \,. \label{SM-F1-first}
\end{eqnarray}
This is precisely the standard thermodynamic form of the free energy and the first law.

{\bf\emph{Resolution from Iyer-Wald formalism:}} We now show that the same result follows from the Iyer-Wald formalism. For the EGB-MCS theory described by~\eqref{SM-F-action}, the symplectic potential $\mathbf{\Theta}$ is defined as
\begin{equation}
\begin{split} \label{SM-F-sp}
\mathbf{\Theta} &= \mathbf{\epsilon}_{aa_1a_2a_3a_4} \bigg[  \left( 1+2 \alpha R \right) \left( g^{ac}g^{bd}-g^{ab}g^{cd} \right) \nabla_b \delta g_{cd}    -8\alpha \left(   R^{d[a} \nabla^{c]}\delta g_{cd} + g^{d [a} R^{c]b} \nabla_{b}\delta g_{cd} \right) \\
& \qquad +4\alpha R^{abcd} \nabla_d \delta g_{bc} -\left( F^{ae}+\frac{k}{6}\epsilon^{abcde} A_b F_{cd}\right) \delta A_e \bigg]   \,.
\end{split}
\end{equation}
When the background fields obey the equations of motion~\eqref{SM-F-eom}, the Noether current $4$-form $\mathbf{J}\equiv \mathbf{\Theta}(\phi,\mathcal{L}_\xi\phi) - \xi\cdot \mathbf{L}$ can be expressed via a Noether charge $3$-form $\mathbf{Q}$,  which takes the form
\begin{equation} \label{SM-F-Ncharge}
\mathbf{Q} =-\boldsymbol{\epsilon}_{abfgh} \bigg[ \nabla^a\xi^b +2\alpha \left( R\nabla^a\xi^b -4 R^a_{\ c} \nabla^{[c}\xi^{b]} +R^{abcd} \nabla_c\xi_d \right) + \frac{1}{2} \left( F^{ab}+\frac{k}{6}\epsilon^{aijkb} A_i F_{jk}\right)  A_c \xi^c \bigg] \,.
\end{equation}
The thermodynamics of the BH~\eqref{SM-F-ansatz1} can then be analyzed using two fundamental identities~\eqref{SM-A-fundamentalID} and~\eqref{SM-A-identity}, which are valid when the background fields and their perturbations satisfy the field equations and $\xi$ is a Killing vector.

Let $\xi^a$ be a timelike Killing vector $\xi^a = (\partial_t)^a$ that becomes null on the horizon. We consider a $t=$const. hypersurface $\Sigma$, whose inner boundary is the horizon. Integrating the fundamental identity~\eqref{SM-A-fundamentalID} over $\Sigma$, we have
\begin{equation}\begin{split}
\delta H &= 0 =\int_\Sigma \boldsymbol{\omega} = \int_\Sigma d \left( \delta\mathbf{Q} -\xi\cdot \mathbf{\Theta} \right) = \int_{\partial\Sigma} \left( \delta\mathbf{Q} -\xi\cdot \mathbf{\Theta} \right)  \\
&= 3\left( 1- \frac{4\alpha}{L_c^2} \right) \frac{\delta f_4}{L_c} -8\left( 1- \frac{4\alpha}{L_c^2} \right) \frac{\delta h_4}{L_c} -\frac{L_c B\delta B}{2} +e^{-\chi_1/2} f_1 L_c^3 \left(  \frac{\delta h_1}{r_h^3} -\frac{3h_1 \delta r_h}{ r_h^4} \right) -\frac{ 2\mu}{L_c}  \delta A_{t2} -M_B \delta B \,,
\end{split}\end{equation}
where 
\begin{equation}
M_B = - \bigg\{ \int_0^{r_h} \bigg[ \left( \frac{ e^{-\chi/2} h B L_c }{r} - \frac{ B L_c }{r} +\frac{k}{2}(A_t A_z' -A_t' A_z)\right)  \bigg] dr +B L_c \ln{ (\frac{r_h}{L_c})} \bigg\} \,.
\end{equation}
Therefore, the first law of thermodynamics from the Iyer-Wald formalism is
\begin{equation} \label{SM-F-firstlaw}
\delta \epsilon =T \delta s + \mu \delta\rho -M_B \delta B \,,
\end{equation}
where $M_B$ coincides exactly with that obtained from the Euclidean approach.

In addition, integrating the identity~\eqref{SM-A-identity} over the hypersurface $\Sigma$, we reproduce the expected free energy $w$
\begin{equation} \label{SM-F-freed}
w =w_{QSR} +kQ_{cs} =\epsilon -Ts-\mu\rho \,.
\end{equation}
Thus, we obtain from~\eqref{SM-F-firstlaw} and~\eqref{SM-F-freed} the variation of $w$ as
\begin{equation}
\delta w=-s\delta T -\rho \delta\mu -M_B\delta B\,.
\end{equation}
Therefore, we show that the same phenomenon also emerges for BHs with Gauss–Bonnet corrections~\eqref{SM-F-ansatz1}, using two independent methods.

{\bf\emph{Transgression interpretation:}} The Euclidean and Iyer-Wald analyses above lead to the same thermodynamic potential also in the presence of the Gauss-Bonnet interaction. In particular, both require the completion $w=w_{QSR}+kQ_{cs}$. This agreement is naturally encoded by the transgression action. For the planar Gauss-Bonnet BH, we choose
\begin{equation}
\bar A=\frac{B}{2}(x dy-y dx) \,, \qquad \bar F^2=0 \,.
\end{equation}
We replace only the Chern–Simons term by the transgression form $\mathcal T_5(A,\bar A)$, while leaving the Einstein-Gauss-Bonnet bulk action and all boundary terms unchanged. Evaluating the resulting action on the background solution gives
\begin{equation}
 w_{\mathcal T} =w_{ QSR}+kQ_{ cs}  =\epsilon-Ts-\mu\rho \,,
\end{equation}
showing that the higher-curvature coupling modifies the gravitational thermodynamic data but not the transgression contribution.

In particular, the Euclidean variation of the transgression-completed action and the corresponding Iyer-Wald calculation both reproduce the expected first law~\eqref{SM-F1-first}, without introducing any additional contribution.

\subsection*{F.2: Spherical BHs}
In addition to BHs with planar topology, we further examine the effect of the Gauss-Bonnet higher-derivative term on spherical BHs~\eqref{SM-D-ansatz} in the EMCS theory. The background ansatz for the metric and gauge field that incorporates a local $SU(2)\times U(1)$ symmetry is
\begin{equation}\begin{split} \label{SM-F-ansatz2}
ds^2 &= -F_0(r) dt^2 +F_1(r) dr^2 +\frac{1}{4} F_2(r) \left(\sigma_1^2+\sigma_2^2\right) +\frac{1}{4} F_3(r) \left( \sigma_3 +2p(r) dt \right)^2 \,,  \\
A &= A_t(r) dt + \frac{B\cos{\theta}}{4} d\phi -\frac{A_z(r)}{2}\sigma_3 \,, 
\end{split}\end{equation}
where $\sigma_i (i=1,2,3)$ are left-invariant 1-forms on $SU(2)$ expressed in terms of Euler angles $(\theta,\phi,\psi)$; see~\eqref{SM-D-sigma} for details. For convenience of analysis, we consider the following reparameterization of the metric functions
\begin{equation}
F_0=\frac{L_c^2}{r^2}f(r)e^{-\chi(r)} \,,\quad  F_1=\frac{L_c^2}{r^2f(r)} \,,\quad  F_2=\frac{L_c^4}{r^2} \,, \quad F_3= \frac{L_c^4}{r^2} h(r)^2 \,,
\end{equation}
where $L_c$ is the effective AdS radius given by~\eqref{SM-F-leff}.

\subsubsection*{F.2.1: Far-field and near-horizon expansions, and thermodynamics}
We consider a BH that possesses only asymptotically local AdS symmetries at infinity, referred to as a squashed AdS boundary. In particular, as $r\to0$, the metric functions behave as $f\sim 1$ and $h\sim v$, where $v$ is the squashing parameter. The far‑field expansion of the background fields can then be written as
\begin{equation}\begin{split} \label{SM-F-uvexp}
f(r) &= 1 + \frac{1}{3}\left( 8-5v^2\right) \frac{r^2}{L_c^2} + \bigg[ \frac{B^2}{ 2(L_c^2 -4\alpha)} -\frac{8}{3} \left( 1- 4v^2 +3v^4 \right) \bigg] \frac{r^4}{L_c^4} \ln{(\frac{r}{L_c})} + f_4 \frac{r^4}{L_c^4}+ \cdots \,, \\
\chi(r) &= \chi_0 +\frac{2}{3}\left(1-v^2\right) \frac{r^2}{L_c^2}+ \bigg[ \frac{B^2}{ 4(L_c^2 -4\alpha) } -\frac{4}{3}\left(1-5v^2 +4v^4\right) \bigg] \frac{r^4}{L_c^4} \ln{(\frac{r}{L_c})} +\chi_4 \frac{r^4}{L_c^4} +  \cdots  \,, \\
h(r) &=  v\left( 1 +\left(1-v^2\right) \frac{r^2}{L_c^2} +\bigg[ \frac{ B^2}{ 8(L_c^2-4\alpha)} -\frac{2}{3} \left(1-5v^2+4v^4 \right) \bigg] \frac{r^4}{L_c^4} \ln{(\frac{r}{L_c})} + h_4 \frac{r^4}{L_c^4} + \cdots  \right) \,, \\
p(r) &=   p_4 \frac{r^4}{L_c^4} +\cdots    \,, \quad A_t(r) =  e^{-\chi_0/2} \Big[ \mu +A_{t2} \frac{r^2}{L_c^2} +\cdots \Big] \,, \quad  A_z(r) = \left(A_{z2} -B v^2 \ln{ (\frac{r}{L_c}) } \right) \frac{r^2}{L_c^2} +\cdots    \,, 
\end{split}\end{equation}
where
\begin{equation}
\chi_4= 2 h_4 + \frac{B^2}{12(L_c^2 -4\alpha) } + (1-v^2) \frac{ L_c^2(17 v^2 -5) +12(3 -7 v^2)\alpha }{ 9 (L_c^2 -4\alpha) }\,.
\end{equation}
Note that the parameter $\chi_0$ should be fixed to zero. However, for the convenience in the analysis of Euclidean action, we will keep it as a free parameter and choose $\chi_0=0$ only in the final expressions.

The factor $v$ in the asymptotic expansion of $h(r)$ can be absorbed by defining $\psi=\bar{\psi}/v$, leading to a conformal boundary metric of the following form
\begin{equation}
ds^2_{bdy} = -dt^2 + L_c^2 d\Omega_{(v)}^2 \,,
\end{equation}
where 
\begin{equation}
d\Omega_{(v)}^2 =\frac{1}{4} \Big[ d\theta^2 + \sin^2\theta d\phi^2 +(d\bar{\psi} +v\cos\theta d\phi)^2 \Big] \,.
\end{equation}
Note that the $d\Omega_{(v)}^2 $ denotes the metric of a squashed $S^3$ with $v$ the control parameter and $0\leq \theta<\pi, 0\leq \phi<2\pi, 0\leq \bar{\psi} <4\pi v$. The spatial volume of the conformal boundary metric is given by
\begin{equation}
V_{(v)}= \int \frac{\sin\theta L_c^3}{8} d\theta d\phi d\bar{\psi} = \int \frac{v \sin\theta L_c^3}{8} d\theta d\phi d\psi  =2\pi^2 L_c^3 v \,.
\end{equation}

Near the BH horizon, the background fields take the following expansions
\begin{equation}\begin{split} \label{SM-F-irexp}
f(r)&= f_1 (r_h-r)+\cdots\,, \quad \chi(r)=\chi_1 +\cdots \,,\quad
h(r)= h_1 +\cdots  \,, \\
p(r) &= p_1(r_h -r) +\cdots \,, \quad A_t(r) =A_{t1}(r_h -r) +\cdots \,, \quad A_z(r) = A_{z1} +\cdots \,,  
\end{split}\end{equation}
with $f(r_h)=p(r_h)=A_t(r_h)=0$. The full expansion is fixed by the constants $\chi_1, h_1, p_1, A_{t1}, A_{z1}$ and the horizon radius $r_h$ (with $f_1>0$). For simplicity, the case with vanishing horizon angular velocity ($p(r_h)=0$) has been considered here. It is straightforward to generalize to the case with nonzero horizon angular velocity, based on the discussion presented in Sec.~\ref{SM-D}.

The temperature and entropy density of the BH are given by
\begin{equation}\begin{split}
T =-\frac{e^{\chi_0/2} } {4\pi} f(r)'e^{-\chi(r)/2} \Big|_{r=r_h} \,,\quad s= \frac{4\pi}{v} \frac{ h(r) L_c^3}{ r^3} \left( 1 + 4\alpha r^2 \frac{ 4-h(r)^2 }{L_c^4} \right) \Bigg|_{r=r_h} \,,
\end{split}\end{equation}
where the entropy density is defined relative to $V_{(v)}$, the volume of squashed 3-sphere. It is worth noting that the entropy density $s$ takes the Wald form 
\begin{equation}\begin{split}
S_{\rm Wald} = s_{\rm Wald} V_{(v)} = \frac{1}{4G} \int_{\Sigma_H} d^3x \sqrt{\tilde{h} } \left( 1+ 2\alpha \tilde{R}\right) \,,
\end{split}\end{equation}
where $\tilde{h}$ is the determinant of the induced metric on the horizon, and $\tilde{R}$ denotes the curvature of the horizon $\Sigma_H$.

To formulate a well-defined variational principle, one has to supplement the bulk action~\eqref{SM-F-action} with proper boundary terms. The total renormalized on-shell action is defined as
\begin{equation}\label{SM-F-sren2}
S_{ren}=S+S_b^E +S_b^{GB} +S_{ct} \,.
\end{equation}
The Gibbons-Hawking term $S_b^E$ and its counterpart for Gauss-Bonnet term $S_b^{GB}$ are given by~\eqref{SM-F-GH} and~\eqref{SM-F-GHgb}, respectively. Also, the boundary counterterm $S_{ct}$ is given in~\eqref{SM-F-sct}.

Using the properties of background equations as well as applying the identity $R^a_{\ b}\xi^b=\nabla_b\nabla^a\xi^b$ on a timelike Killing vector $\xi^a=(\partial_t)^a$, the bulk on-shell action takes
\begin{equation} \label{SM-F-sbulk}
S=\int d^4x\int dr \Big[ 2\sqrt{-g} \left(R^t_{\ t} +\alpha H^t_{\ t} +\frac{\alpha}{2} L_{GB} \right)+ \left( A_tN_t +\frac{k}{24}A_tA_z (B -2A_z) \sin{\theta} \right)' +\frac{k}{24} BA_tA_z'\sin{\theta} \Big] \,,
\end{equation}
where $N_t\equiv \frac{ e^{\chi/2} L_c^4 h \sin{\theta} }{8r} \left( A_t' +p A_z'\right)$. Combining the UV and IR expansions from~\eqref{SM-F-uvexp} and~\eqref{SM-F-irexp} with the bulk on-shell action in Eq.~\eqref{SM-F-sbulk}, we obtain the renormalized on-shell action
\begin{equation}\begin{split}
\frac{S_{ren}}{\Delta T V_{(v)}} =& \frac{e^{-\chi_0/2} }{L_c } \Bigg[ \left( 1 -\frac{4\alpha}{L_c^2}\right) \left( 3f_4 -8 h_4 \right) -\frac{B^2}{4 L_c^2} -\frac{64-32v^2 -23v^4 }{12} +\frac{3 \alpha v^2 (16-13 v^2)}{ L_c^2 } -\frac{ 2\mu A_{t2}}{L_c} \Bigg]  \\
&\qquad + \frac{e^{-\chi_1/2} f_1 h_1 L_c^3}{v r_h^3} \left( 1+ 4\alpha r_h^2\frac{ 4-h_1^2 }{L_c^4} \right) +\frac{1}{L^3v} \int_0^{r_h} \frac{kB}{3} A_t A_z' dr \,,
\end{split}\end{equation}
where $\Delta T=\int dt$.

According to the renormalization procedure, the expectation values for the dual stress tensor are given by 
\begin{equation}\begin{split}
\langle T_{\mu\nu} \rangle &=  \lim_{r\to0} \frac{L_c^2}{r^2} \Bigg[ -2(K_{\mu\nu} -K\gamma_{\mu\nu}) - 4\alpha \left(\mathcal{Q}_{\mu\nu} - \frac{1}{3}\mathcal{Q} \gamma_{\mu\nu} \right) -\frac{6}{L_c} \left( 1-\frac{4\alpha}{3 L_c^2}\right) \gamma_{\mu\nu} + \\
& \quad L_c\hat{G}_{\mu\nu} \left( 1+\frac{4\alpha}{L_c^2 }\right) +L_c \ln{(\frac{r}{L_c})} \left( \hat{F}_{\mu\rho}\hat{F}_\nu^{\ \rho}-\frac{ \gamma_{\mu\nu} }{4} \hat{F}^2 \right)  -L_c^3 \ln{(\frac{r}{L_c})} \left( 1-\frac{4\alpha}{L_c^2} \right) h_{\mu\nu}^{(4)}    \Bigg] \,, 
\end{split}\end{equation}
where
\begin{equation}\begin{split}
\mathcal{Q}_{\mu\nu} & =3\mathcal{J}_{\mu\nu} + 2K\hat{R}_{\mu\nu} +\hat{R}K_{\mu\nu} -2K^{\rho\sigma}\hat{R}_{\rho\mu\sigma\nu} -4\hat{R}_{\mu\rho}K_\nu^{~\rho} \,,  \\
h_{\mu\nu}^{(4)} &= \hat{R}_{\mu\rho\nu\lambda}\hat{R}^{\rho\lambda} -\frac{1}{6}\hat{\nabla}_\mu \hat{\nabla}_\nu \hat{R}+\frac{1}{2}\hat{\nabla}^2 \hat{R}_{\mu\nu}-\frac{1}{3}\hat{R}\hat{R}_{\mu\nu}-\frac{\gamma_{\mu\nu}}{4}(\hat{R}_{\rho\lambda}\hat{R}^{\rho\lambda}-\frac{\hat{R}^2}{3}+\frac{\hat{\nabla}^2 \hat{R} }{3}) \,.
\end{split}\end{equation}
In particular, for the spherical BH solution~\eqref{SM-F-ansatz2}, the non-zero components of $\langle T_{\mu\nu}\rangle$ read
\begin{equation}\begin{split}
&\langle T_{tt}\rangle= \frac{ e^{-\chi_0}}{L_c} \bigg[ \left( 1-\frac{4\alpha}{L_c^2 } \right) \left( -3f_4 +8h_4 \right) +\frac{B^2}{4L_c^2} +\frac{64-32v^2 -23v^4}{12}  -\alpha \frac{3 v^2(16-13v^2 )}{ L_c^2 } \bigg] \,, \\
& \langle T_{\theta\theta}\rangle = L_c \bigg[ -\left( 1-\frac{4\alpha}{L_c^2 } \right) \frac{f_4}{4} -\frac{B^2}{24L_c^2} +\frac{ 32-88v^2 +59v^4 }{48} + \alpha \frac{ 64- 56v^2 -35v^4 }{36L_c^2} \bigg] \,, \\
& \langle T_{\phi\phi}\rangle = L_c \bigg[ \left( 1-\frac{4\alpha}{L_c^2 } \right) \left( -\Big( v^2+1 +(v^2-1) \cos(2\theta) \Big)\frac{f_4}{8} +2v^2h_4\cos^2{\theta} \right) \\
 &\qquad  +\frac{ v^2-2 +(v^2+2)\cos(2\theta) }{96L_c^2} B^2  +\frac{ 32-56v^2 +139v^4- 109v^6 +(-32 +120v^2 +21v^4 -109v^6) \cos(2\theta) }{96} \\
 &\qquad +\alpha \frac{64-280v^2 -163v^4 +325v^6 -(1-v^2)(64+232v^2+325v^4 ) \cos{(2\theta)}}{72L_c^2} \bigg] \,, \\
& \langle T_{\psi\psi}\rangle= L_c v^2 \bigg[ \left( 1- \frac{4\alpha}{L_c^2} \right) \left( -\frac{f_4}{4} +2h_4 \right) +\frac{B^2}{48L_c^2 } +\frac{ 32+80v^2 -109v^4 }{48}  -\alpha \frac{ 224+128v^2 -325v^4 }{36L_c^2}\bigg] \,, \\
& \langle T_{t\psi}\rangle=\langle T_{\psi t}\rangle= 2L_c v^2 p_4 \left( 1- \frac{4\alpha}{L_c^2}\right) \,,\quad  \langle T_{t\phi}\rangle=\langle T_{\phi t}\rangle=\cos\theta \langle T_{t\psi} \rangle \,,\quad  \langle T_{\phi\psi}\rangle=\langle T_{\psi\phi}\rangle= \cos\theta \langle T_{\psi\psi}\rangle   \,.
\end{split}\end{equation}

On the other hand, the expectation value of the dual current is
\begin{equation}
\langle J^{\mu} \rangle = \lim_{r\to0} \frac{L_c^4}{r^4} \Big[ n_r \Big( F^{\mu r} +\frac{k}{6}\epsilon^{r\mu \alpha\beta\gamma} A_\alpha F_{\beta\gamma} \Big) +L_c \ln{(\frac{r}{L_c})} \nabla_\alpha \hat{F}^{\alpha\mu}  \Big] \,. \nonumber
\end{equation}
Thus, the non-zero components of $\langle J^{\mu} \rangle$ are
\begin{equation}
\langle J^t \rangle =- \frac{2 e^{\chi_0/2} A_{t2} }{L_c } \,, \qquad
\langle J^\psi \rangle = \frac{1}{L_c^3} \left( -\frac{4}{v^2} A_{z2} - \frac{2k\mu B}{3v} +2B  \right) \,. 
\end{equation}

Consequently, the energy density $\epsilon$ and charge density $\rho$ are given by
\begin{equation}\begin{split}
\epsilon &= u^\mu u^\nu \langle T_{\mu\nu}\rangle = \frac{1}{L_c} \bigg[ \left( 1-\frac{4\alpha}{L_c^2 } \right) \left( -3f_4 +8h_4 \right) +\frac{B^2}{4L_c^2} +\frac{64-32v^2 -23v^4}{12}  -\alpha \frac{3 v^2(16-13v^2 )}{ L_c^2 } \bigg] ,\\
\rho &= u_\mu \langle J^\mu\rangle = -\frac{2A_{t2} }{L_c} \,,
\end{split}\end{equation}
where $u^\mu$ is the timelike unit normal vector. The total Euclidean action $I_E$ is obtained by analytic continuation of the renormalized action $S_{ren}$. Therefore, using the QSR, $w_{QSR}$ is given by
\begin{equation}
w_{QSR} = -\frac{T \ln{Z}}{V_{(v)}} =\frac{T I_E}{V_{(v)}} = \epsilon -Ts_{\rm Wald} -\mu\rho - \frac{e^{\chi_0/2}}{L_c^3 v} \int_0^{r_h} \frac{k B}{3} A_t A_z' dr \,.
\end{equation}
%

\subsubsection*{F.2.2: Thermodynamic consistency: from Chern–Simons correction to transgression}
{\bf\emph{Variation of on-shell Euclidean action:}} Consider analytic continuation to Euclidean time $\tau$ by a Wick rotation $t=-i \tau$, under which the total Euclidean action $I_{E}$ is given by
\begin{equation}
I_{E}=I+I_{bdy} \,,
\end{equation}
where $I=-i S =-i S_{eff}$ and $I_{bdy}$ is the Euclidean boundary action, including counterterms. The $S_{eff}$ is the effective action of $S$ obtained by substituting the background ansatz~\eqref{SM-F-ansatz2} into~\eqref{SM-F-action}. The variation of the total on-shell Euclidean action is given by
\begin{equation}
\frac{\delta I_E}{\Delta\tau V_{(v)} } =\bigg\{ e^{-\chi_0/2} \bigg[ - \rho \delta\mu  -\Big( \epsilon - \mu\rho \Big) \frac{\delta \chi_0}{2} \bigg] - \frac{k}{L_c^3 v} \delta Q_{cs} - e^{ -\chi_0/2} M_B \delta B  \bigg\} \,, 
\end{equation}
where
\begin{equation}
Q_{cs}=\frac{B}{6}\int_0^{r_h} \left( A_z' A_t -A_z A_t'\right) dr \,, 
\end{equation}
and
\begin{equation} \label{SM-F-mb}
M_B = - \frac{1}{L_c^3 v }\bigg\{ \int_0^{r_h} \bigg[ \left( \frac{ e^{ (\chi_0-\chi)/2} h (B -2A_z)}{r} - \frac{ vB }{r} +\frac{e^{\chi_0/2}k}{2}(A_t A_z' -A_t' A_z)\right)  \bigg] dr  + v B \ln{ (\frac{r_h}{L_c})} \bigg\} \,.
\end{equation}
Note that in the final expression, the constant $\chi_0$ in $M_B$ should be set to zero to ensure that the Hawking temperature equals the temperature at the boundary.

Recalling that the temperature $T=\frac{e^{\chi_0/2} }{\Delta \tau}$, thus the variation of $\chi_0$ gives $\delta \chi_0= 2\frac{\delta T}{T}$ with $\Delta\tau$ fixed in the variation of on-shell Euclidean action. Using the QSR, we obtain
\begin{equation}
\delta w_{QSR} = -\left( s_{\rm Wald}+ \frac{ e^{\chi_0/2} }{L_c^3 v} \frac{ k Q_{cs}}{T } \right)\delta T - \rho\delta\mu -\frac{ e^{\chi_0/2} }{L_c^3 v} k \delta Q_{cs} - M_{B} \delta B \,,
\end{equation}
Therefore, to obtain a thermodynamic potential consistent with the first law, $w_{ QSR}$ must again be supplemented by the Chern-Simons correction. The full free energy is thus defined as $w=w_{ QSR}+\frac{1}{L_c^3 v}e^{\chi_0/2}kQ_{ cs}$ and satisfies
\begin{eqnarray}
w &=& \epsilon -Ts_{\rm Wald} -\mu \rho \,, \label{SM-F-free}  \\
\delta w &=& -s_{\rm Wald} \delta T -\rho\delta\mu -M_B \delta B \,.\label{SM-F-free1st}
\end{eqnarray}
This is precisely the standard form of the free energy and the first law. From~\eqref{SM-F-free1st}, we know that $M_B$ is nothing but the magnetization of the system.

{\bf\emph{Resolution from the Iyer-Wald formalism:}} Here, we demonstrate that the results presented in~(\ref{SM-F-free}-\ref{SM-F-free1st}) can also be obtained through the Iyer–Wald formalism. Following this approach, the first law of BH thermodynamics is analyzed using the two identities~\eqref{SM-A-fundamentalID} and~\eqref{SM-A-identity}, where the symplectic potential $\mathbf{\Theta}$ and Noether charge $3$-form $\mathbf{Q}$ are specified in~\eqref{SM-F-sp} and~\eqref{SM-F-Ncharge}, respectively.

Consider a timelike Killing field $\xi^a=(\partial_t)^a$ that becomes null on the event horizon $r=r_h$. Let $\Sigma$ denote a constant-$t$ spacelike slice, whose inner boundary is the horizon $r=r_h$. The integration of the fundamental identity~\eqref{SM-A-fundamentalID} over $\Sigma$ gives rise to the first law of BH thermodynamics. It is important to note that due to the coordinate singularities at $\theta=0$ and $\theta=\pi$, a specific integration contour must be chosen, as illustrated in Fig.~\ref{fig:contour}. Furthermore, projecting the $3$-form $(\delta\mathbf{Q}-\xi\cdot \mathbf{\Theta})$ along a fixed angular direction gives $\delta_i \equiv (\delta \mathbf{Q} -\xi \cdot \mathbf{\Theta})|_i =(\delta Q -\xi\cdot \Theta)_i \sqrt{\gamma_{(i)}},\,  i=(\theta,\phi,\psi)$. A direct calculation then yields
\begin{equation}
\begin{split}
\delta_\theta &=  \frac{\cos\theta}{2} \mathcal{G} = \frac{\cos\theta}{2} \frac{1}{L_c^3 v} \bigg[ -\left( \frac{ e^{-\chi/2} h (B -2A_z) }{r}+\frac{k}{3}(2A_tA_z' -A_t'A_z) \right) \delta B -\frac{kB}{3} \left(A_t \delta A_z \right)' \bigg] \,, \\
\delta_\phi &=\delta_\psi =0   \,.
\end{split}
\end{equation}

Integrating the fundamental identity~\eqref{SM-A-fundamentalID} over the hypersurface $\Sigma$, we obtain
\begin{equation}
\begin{split}
\delta H &=0=\int_\Sigma \boldsymbol{\omega}=\int_{\Sigma} d(\delta \mathbf{Q}- \xi \cdot \boldsymbol{\Theta}) =\int_{\partial\Sigma} (\delta \mathbf{Q} -\xi \cdot \boldsymbol{\Theta})    \\
&=\int (\delta Q- \xi\cdot \Theta )_r \sqrt{\gamma_{(r)}} d\theta d\phi d\psi \bigg|_{r=0}  -\int (\delta Q- \xi\cdot \Theta )_r \sqrt{\gamma_{(r)}} d\theta d\phi d\psi \bigg|_{r=r_h}   \\
&\qquad\qquad +\int_{r_h}^0 \frac{\mathcal{G}}{2} dr \Big( \cos{ (\epsilon) } + (-1)\cos{(\pi-\epsilon)} \Big)  d\phi d\psi \bigg|_{\epsilon^+\to 0} \\
&= \int (\delta Q- \xi\cdot \Theta )_r \sqrt{\gamma_{(r)}} d\theta d\phi d\psi \bigg|_{r=0}  -\int (\delta Q- \xi\cdot \Theta )_r \sqrt{\gamma_{(r)}} d\theta d\phi d\psi \bigg|_{r=r_h} + \int_{r_h}^0 \mathcal{G} dr  d\phi d\psi \\
&= 3\left( 1- \frac{4\alpha}{L_c^2} \right) \frac{\delta f_4}{L_c} -8\left( 1- \frac{4\alpha}{L_c^2} \right) \frac{\delta h_4}{L_c} -\frac{L_c B\delta B}{2}  -\frac{2\mu}{L_c}  \delta A_{t2} -M_B \delta B  \\
&\qquad +e^{-\chi_1/2} f_1  \frac{L_c^3}{v} \left[ \left( \frac{\delta h_1}{r_h^3}  -\frac{3h_1 \delta r_h}{ r_h^4 } \right) +\frac{\alpha}{ L_c^4 } \left( \frac{4(4-3 h_1^2) \delta h_1}{r_h }  +\frac{4h_1(h_1^2 -4) \delta r_h}{ r_h^2 }\right)  \right] \,.
\end{split}
\end{equation}
Recalling that the variation of entropy density reads
\begin{equation}
T \delta s_{\rm Wald} = e^{-\chi_1/2} f_1  \frac{L_c^3}{v} \left[ \left( \frac{\delta h_1}{r_h^3}  -\frac{3h_1 \delta r_h}{ r_h^4 } \right) +\frac{\alpha}{ L_c^4 } \left( \frac{4(4-3 h_1^2) \delta h_1}{r_h }  +\frac{4h_1(h_1^2 -4) \delta r_h}{ r_h^2 }\right)  \right] \,.
\end{equation}
Therefore, the first law of thermodynamics takes the form
\begin{equation} \label{SM-F-firstlaw2}
\delta\epsilon = T\delta s_{\rm Wald} +\mu\delta\rho -M_B \delta B \,,
\end{equation}
where $M_B$ coincides precisely with that obtained from the Euclidean action variation method in~\eqref{SM-F-mb}.

Furthermore, integrating the identity~\eqref{SM-A-identity} over the hypersurface $\Sigma$, we reproduce the expected free energy $w$
\begin{equation} \label{SM-F-freed2}
    w =w_{QSR} + \frac{1}{L_c^3 v} kQ_{cs} =\epsilon -Ts_{\rm Wald}-\mu\rho \,.
\end{equation}
Thus, we obtain from~\eqref{SM-F-firstlaw2} and~\eqref{SM-F-freed2} the variation of $w$ as
\begin{equation}
\delta w=-s_{\rm Wald}\delta T -\rho \delta\mu -M_B\delta B\,.
\end{equation}
We thus conclude that the same Chern-Simons correction persists for the spherical BHs with Gauss-Bonnet corrections described by~\eqref{SM-F-ansatz2}. In particular, including the $kQ_{ cs}$ yields a thermodynamically consistent free energy, providing further evidence for the universality of the Chern-Simons correction identified in this work.

{\bf\emph{Transgression interpretation:}} The preceding Euclidean and Iyer-Wald analyses independently yield the same corrected thermodynamic potential for the spherical Gauss-Bonnet BH, $w=w_{ QSR}+kQ_{ cs}/(L_c^3v)$. The transgression form provides the corresponding action-level explanation. As in the rotating BH~\eqref{SM-D-ansatz}, a convenient choice of reference connection is $\bar A =\frac{B}{4}\cos\theta d\phi$. Replacing the Chern–Simons form with $\mathcal T_5(A,\bar A)$ yields
\begin{equation}
w_{\mathcal T}
=w_{ QSR}+\frac{k}{L_c^3v}Q_{ cs} =\epsilon-Ts_{\rm Wald}-\mu\rho \,.
\end{equation}
Here, the factor $1/(L_c^3v)$ arises solely from the definition of boundary volume, while the radial integral defining $Q_{cs}$ remains unchanged.

The Euclidean variation of the transgression-completed action and the corresponding Iyer-Wald calculation both reproduce the expected first law in~\eqref{SM-F-free1st}, leaving the Wald entropy and magnetization unchanged. Thus, the transgression completion naturally explains the origin of $Q_{ cs}$ while ensuring the strict gauge invariance of the full action.

\section{BHs in other dimensions} \label{SM-G}
The phenomenon identified in the main text can be generalized to Einstein-Maxwell theory with a Chern-Simons term ``$AF^n$" in any odd dimension $D=2n+1$ ($n\ge 2$). In what follows, we focus on the seven‑dimensional case ($n=3$) as a concrete example. The action for the seven-dimensional Einstein-Maxwell theory, including an $A\wedge F\wedge F\wedge F$ Chern-Simons term, takes the form
\begin{equation} \label{SM-G-action7}
S=\frac{1}{16\pi G} \int d^7x \sqrt{-g}\Big( R+\frac{30}{L^2} -\frac{F^2}{4} +\frac{k}{4} \epsilon^{abcdemn}A_a F_{bc} F_{de} F_{mn} \Big) \,,
\end{equation}
where the gauge field strength $F_{ab}=\partial_a A_b-\partial_b A_a$. Also, $G$ is Newton's constant, $L$ is the AdS radius, and $k$ is the Chern-Simons coupling. Setting $16\pi G=L=1$, the corresponding equations of motion read
\begin{equation}\begin{split}
R_{ab}-\frac{1}{2}g_{ab}\left( R+30 \right)-\frac{1}{2} \left(F_{ac}F_b^{\ c}-\frac{1}{4}g_{ab}F_{cd}F^{cd} \right) &=0 \,,\\ 
\nabla_{b}F^{ba} +k \epsilon^{abcdemn}F_{bc}F_{de}F_{mn} &=0 \,.  
\end{split}\end{equation}

We shall consider the following background ansatz for metric and gauge fields
\begin{equation} \begin{split} \label{SM-G-ansatz}
ds^2&=\frac{1}{r^2}\Big[- \big(f e^{-\chi}-h^2p^2\big)dt^2+2ph^2 dtdz+ dx_1^2+dx_2^2 +u(dx_3^2+dx_4^2) +h^2 dz^2+\frac{dr^2}{f}\Big]\,, \\
A &= A_t dt +\frac{ B_1}{2} \left(x_1 dx_2 -x_2 dx_1\right) +\frac{ B_2}{2} \left( x_3 dx_4 -x_4 dx_3\right) - A_z dz \,,
\end{split} \end{equation}
where $f,\, \chi,\, h,\, p,\, u,\, A_t$ and $A_z$ are functions of radial coordinate $r$. The constants $B_1$ and $B_2$ represent the background magnetic fields perpendicular to the $x_1-x_2$ and $x_3-x_4$ plane, respectively. Here, the radial coordinate $r$ spans from the asymptotically AdS boundary at $r=0$
to the BH horizon at $r=r_h$.	

\subsection*{G.1: Asymptotic expansions and thermodynamics}
At the asymptotic AdS boundary $r=0$, the bulk fields take the following form 
\begin{equation}
\begin{split}
f(r)&= 1 -\frac{7B_1^2-3B_2^2}{20}r^4 +f_6 r^6+\cdots \,, \qquad  \chi(r) = \chi_0 -\frac{3(3B_1^2 -2B_2^2)}{40}r^4 +2(u_6+h_6) r^6+\cdots  \,, \\
u(r)&= 1-\frac{B_1^2 -B_2^2}{8}r^4 +u_6 r^6+\cdots  \,,\qquad  h(r)= 1-\frac{B_1^2}{16}r^4 +h_6 r^6+\cdots  \,, \\
p(r)&= p_6 r^6+\cdots    \,, \qquad
A_t(r) = e^{-\chi_0/2} \left( \mu +A_{t4} r^4+\cdots  \right) \,,\qquad  A_z(r) =-A_{z4} r^4 +\cdots  \,.
\end{split}
\end{equation}
Note that unlike in the five-dimensional case~\eqref{SM-A-uvseries}, no logarithmic term appears in the expansions above. Also, $\chi_0$ must be set to zero to ensure the Hawking temperature matches the boundary temperature. In our analysis, however, we will treat it as a free parameter and only fix $\chi_0=0$ in the final expressions. 

Near the horizon, imposing the regularity condition with $f(r_h)=p(r_h)=A_t(r_h)=0$, we obtain the following expansions
\begin{equation}\begin{split}
f(r)&= f_1(r_h-r)+\cdots \,,\quad \chi(r)= \chi_1+\cdots  \,, \quad u(r)= u_1+\cdots  \,,\\
h(r) &=h_1+\cdots \,,\qquad\qquad  p(r)= p_1(r_h-r)+\cdots   \,,\\
A_t(r)&= A_{t1}(r_h-r)+\cdots  \,, \quad A_z(r)= A_{z1}+\cdots.      
\end{split}\end{equation}
The corresponding temperature and Bekenstein-Hawking entropy density are given by 
\begin{equation}
T=-\frac{e^{\chi_0/2} } {4\pi} f'(r)e^{-\chi(r)/2} \Big|_{r=r_h} \,, \qquad s=\frac{4\pi u(r) h(r)}{r^5} \Big|_{r=r_h}  \,.
\end{equation}

To calculate the Euclidean action, we first consider the renormalized on-shell action
\begin{equation}  \label{SM-G-sren}
S_{ren}=S+S_{bdy} \,.
\end{equation}
Here, the boundary terms $S_{bdy}$ are given by
\begin{equation}  \label{SM-G-sbdy}
S_{bdy} =\frac{1}{16\pi G}\int d^6x\sqrt{-\gamma}\left(2K-10 +\frac{\hat{F}^2}{8} \right),
\end{equation}
where $\gamma_{\mu\nu}$ is the induced metric at the conformal boundary, $K$ is the trace of extrinsic curvature, $\hat{F}_{\mu\nu}$ is the boundary field strength. Utilizing the properties of the Einstein equation and the identity $R^a_{\ b}\xi^b=\nabla_b\nabla^a\xi^b$ for any Killing vector $\xi^a$, the bulk on-shell action can be expressed as
\begin{equation} \label{SM-G-sbulk}
S= \int d^6x\int dr \Big[ 2\sqrt{-g}R^t_{\ t}+\left[\frac{e^{\chi/2}A_t u h}{r^3} \left( A_t'+pA_z'\right)+12kB_1B_2 A_t A_z \right]'+ 24 k B_1B_2 A_t A_z' \Big]\,.
\end{equation}
The combination of~\eqref{SM-G-sbulk} and~\eqref{SM-G-sbdy} thus gives an analytical expression for the renormalized on-shell action~\eqref{SM-G-sren},~\emph{i.e.}
\begin{equation}
\frac{S_{ren} }{\Delta T V} = 5 f_6 - 12 u_6 - 12 h_6 -4\mu A_{t4} +Ts + 24 kB_1B_2 e^{\chi_0/2} \int_0^{r_h} A_t A_z' dr \,,
\end{equation}
where $\Delta T=\int dt$ and $V=\int dx_1dx_2dx_3dx_4dz$.

Additionally, the expectation value of the dual boundary stress tensor and current are obtained from~\eqref{SM-G-sren} using the holographic renormalization and are given by
\begin{equation}
\begin{split}
\langle T_{\mu\nu} \rangle &= \lim_{r\to 0} \frac{1}{r^2} \Big[-2K_{\mu\nu} +\left( 2K- 10\right) \gamma_{\mu\nu} -\frac{1}{2} \left( \hat{F}_{\mu\rho} \hat{F}_\nu^{\ \rho}-\frac{ \gamma_{\mu\nu} }{4} \hat{F}^2 \right) \Big] \,, \\
\langle J^{\mu} \rangle &= \lim_{r\to0} \frac{1}{r^6} \bigg[ n_r \left( F^{\mu r} +\frac{3k}{2}\epsilon^{r\mu \alpha\beta\gamma\rho\sigma} A_\alpha F_{\beta\gamma} F_{\rho\sigma}\right) -\frac{1}{2}\nabla_\alpha \hat{F}^{\alpha\mu}  \bigg] \,.
\end{split}
\end{equation}
Specifically, the non-zero components of $\langle T_{\mu\nu}\rangle$ and $\langle J^\mu\rangle$ are
\begin{equation}
\begin{split}
&\langle T_{tt}\rangle = e^{-\chi_0} \left(-5 f_6 +12u_6 +12h_6 \right) \,, \qquad
\langle T_{x_1x_1}\rangle =\langle T_{x_2x_2}\rangle = -f_6   \,,\\
&\langle T_{x_3x_3}\rangle =\langle T_{x_4x_4}\rangle = -f_6 +6 u_6 \,, \quad
\langle T_{zz}\rangle =-f_6 +12h_6\,, \quad \langle T_{tz}\rangle=\langle T_{zt}\rangle=6 p_6 \,,\\
&\langle J^t \rangle = -4e^{\chi_0/2} A_{t4} \,,\qquad  \langle J^z \rangle=  -4A_{z4} -12k\mu B_1 B_2  \,. 
\end{split}\end{equation}
Note that the energy density $\epsilon$ and charge density $\rho$ are defined as
\begin{equation}\begin{split}
\epsilon = u^\mu u^\nu \langle T_{\mu\nu}\rangle = -5 f_6 +12u_6 +12h_6 \,,\quad 
\rho = u_\mu \langle J^\mu\rangle = -4A_{t4} \,,
\end{split}\end{equation}
where $u^\mu$ is the timelike unit normal vector.

The Euclidean action follows from analytic continuation of the renormalized on-shell action~\eqref{SM-G-sren}. Using the standard QSR, the free energy density $w_{QSR}$ can be expressed as
\begin{equation}
w_{QSR} = \epsilon-Ts-\mu\rho -24 e^{\chi_0/2} kB_1B_2 \int_0^{r_h} A_t A_z' dr \,.
\end{equation}
%

\subsection*{G.2: Thermodynamic consistency: from Chern–Simons correction to transgression}
Now, we show that the discrepancy between the QSR and the first law of thermodynamics also occurs in this seven-dimensional BH~\eqref{SM-G-ansatz}, and use the same method in the main text to resolve it. 

{\bf \emph{Variation of on-shell Euclidean action:}} Consider analytic continuation to Euclidean time $\tau$ by a Wick rotation $t=-i \tau$, under which the total Euclidean action $I_{E}$ is given by
\begin{equation}
I_{E}=I+I_{bdy} \,,
\end{equation}
where $I=-iS=-iS_{eff}$ and $I_{bdy}$ represents the Euclidean boundary action derived from~\eqref{SM-G-sbdy}. The effective action $S_{eff}$ is given by
\begin{equation}
\begin{split}
S_{eff}&=\int d^7x \frac{e^{-\chi/2} u h}{r^5} \bigg\{ -f'' -\frac{2fu''}{u} -\frac{2fh''}{h}+f\chi'' +f'\left( \frac{12}{r} -\frac{2u'}{u} -\frac{2h'}{h}+\frac{3\chi'}{2} \right) +\frac{fu'}{u} \left( \frac{12}{r} +\frac{u'}{2u} \right) \\
&\quad  +\left( \frac{h'}{r h} -\frac{\chi'}{2 r} \right) \left( 12f -\frac{2rfu'}{u} +r  f\chi' \right) +\frac{1}{2}e^{\chi} \left[ h^2 p'^2 +r^2(A_t'+p A_z')^2 \right] +\frac{30-42f}{r^2} \\
&\quad -\frac{r^2}{2} \left( B_1^2 +\frac{B_2^2}{u^2} +\frac{fA_z'^2}{h^2}\right) \bigg\} +\int d^7x \left( 12k B_1B_2 \right) \left( A_t' A_z -A_t A_z'\right) \,.
\end{split}
\end{equation}
Thus, the variation of the total Euclidean action $I_E$ is given by
\begin{equation}
\delta I_E =\delta I+\delta I_{bdy} \,.
\end{equation}

To illustrate the distinct features of the seven-dimensional case, we provide the explicit expression for $\delta I$, given by
\begin{equation}
\begin{split}
\frac{\delta I}{\Delta \tau V}=&-\int dr \partial_r \Bigg\{ \frac{e^{-\chi/2} u h }{r^5} \left[ \left( \frac{7}{r} -\frac{u'}{u} -\frac{h'}{h} +\chi' \right) \delta f - \delta f' \right] 
+\frac{ e^{-\chi/2} f h }{r^5} \left[  \left( \frac{2}{r} +\frac{u'}{u} \right) \delta u - 2\delta u' \right] \\
&+\frac{ 2 e^{-\chi/2} f u }{r^5} \left(  \frac{\delta h}{r} - \delta h' \right)  +\frac{ e^{-\chi/2} f u h }{r^5} \left[ \left( \frac{f'}{2f}-\frac{\chi'}{2}-\frac{1}{r} \right) \delta \chi+ \delta \chi' \right]+\frac{ e^{\chi/2} h^3 u p' }{r^5 } \delta p  \\
& +12kB_1 B_2 A_z \delta A_t +\frac{e^{\chi/2} u h(A_t'+p A_z')}{r^3} \left(\delta A_t +p\delta A_z\right)  -\left(12 k B_1B_2 A_t + \frac{e^{-\chi/2} f u A_z'}{r^3 h} \right) \delta A_z \Bigg\}   \\
&-\int dr \Bigg[ 24kB_1B_2 \left( A_z' \delta A_t -\delta A_z A_t' \right)-\left( \frac{B_1 e^{-\chi/2} u h}{r^3} +12 kB_2 (A_t A_z' -A_t' A_z)\right) \delta B_1 \\
& \qquad -\left( \frac{B_2 e^{-\chi/2} u h}{r^3} +12 kB_1 (A_t A_z' -A_t' A_z)\right) \delta B_2 \Bigg]\,,
\end{split}
\end{equation}
where $\Delta \tau \equiv \int d\tau$ and $V=\int dx_1dx_2dx_3dx_4dz$. Using the boundary condition $A_t(r=r_h)=A_z(r=0)=0$, we can convert the bulk integration $\int 24kB_1B_2 \left(A_z'\delta A_t -\delta A_z A_t'\right) dr$ into
\begin{equation}\begin{split}
\int dr &\Big[ 24k B_1B_2 \left( A_z' \delta A_t-\delta A_z A_t' \right) \Big] = k \delta \left[ 12B_1B_2\int dr \left(A_z' A_t -A_z A_t'\right) \right]  \\
& \qquad\qquad -\left[ 12kB_2 \int dr \left(A_z' A_t -A_z A_t'\right) \right] \delta B_1  -\left[ 12kB_1 \int dr \left(A_z' A_t -A_z A_t'\right) \right] \delta B_2  \,.
\end{split}\end{equation}
As the unique boundary of the Euclidean metric is the AdS boundary, we have the variation of total Euclidean action
\begin{equation}
\begin{split}
\delta I_E = \delta I+\delta I_{bdy}  =\Delta\tau V \bigg\{ e^{-\frac{\chi_0}{2}} \bigg[ -\rho \delta\mu  -\left( \epsilon -\mu\rho \right) \frac{\delta \chi_0}{2} -  M_{B_1} \delta B_1 - M_{B_2} \delta B_2 \bigg] - k \delta Q_{cs} \bigg\} \,,
\end{split}
\end{equation}
where
\begin{equation}\
Q_{cs}=12B_1B_2\int_0^{r_h} (A_z' A_t- A_z A_t') dr\,,
\end{equation}
together with
\begin{equation}
\begin{split} \label{SM-G-mb}
M_{B_1} &= -\bigg\{ \int_0^{r_h} \bigg[ \frac{B_1 }{r^3} \left(  e^{ (\chi_0-\chi)/2 } uh -1 \right) +24e^{\chi_0/2}k B_2(A_z' A_t - A_z A_t') \bigg] dr  - \frac{B_1}{2r_h^2} \bigg\} \,,\\
M_{B_2} &= -\bigg\{ \int_0^{r_h} \bigg[ \frac{B_2 }{r^3} \left(  \frac{e^{(\chi_0-\chi)/2}h}{u} -1 \right) +24e^{\chi_0/2} k B_1(A_z' A_t - A_z A_t') \bigg] dr  - \frac{B_2}{2r_h^2} \bigg\} \,.
\end{split}
\end{equation}
Note that in the final expression, the constant $\chi_0$ in $M_{B_1}$ and $M_{B_2}$ should be set to zero to ensure that the Hawking temperature equals the temperature at the boundary.

Recalling that the temperature $T=\frac{e^{\chi_0/2} }{\Delta \tau}$, thus the variation of $\chi_0$ gives $\delta \chi_0= 2\frac{\delta T}{T}$ with $\Delta\tau$ fixed in the variation of on-shell Euclidean action. Using the QSR, we obtain the variation of free energy density
\begin{equation}
\delta w_{QSR} = -\left( s+ \frac{e^{\chi_0/2} k Q_{cs}}{T} \right)\delta T -\rho \delta\mu -e^{\chi_0/2} k \delta Q_{cs} -M_{B_1} \delta B_1  -M_{B_2} \delta B_2 \,.
\end{equation}
Remarkably, the seven dimensional case exhibits the same structure as the five dimensional one. In the presence of the Chern-Simons interaction, $w_{ QSR}$ is not, by itself, the thermodynamic free energy density satisfying the standard first law. Instead, the full thermodynamic potential is obtained by supplementing $w_{ QSR}$ with the Chern-Simons correction, $w=w_{ QSR}+e^{\chi_0/2} k Q_{ cs}$. Then we find
\begin{eqnarray}
w &=& \epsilon-Ts -\mu\rho \,,\label{SM-G-free} \\
\delta w &=&-s\delta T -\rho \delta\mu -M_{B_1}\delta B_1 -M_{B_2} \delta B_2 \,. \label{SM-G-free1st}
\end{eqnarray}
Thus, once the Chern-Simons correction is included, the standard first law is obtained in seven dimensions.

{\bf \emph{Resolution from the Iyer-Wald formalism:}} We now independently verify~\eqref{SM-G-free} and~\eqref{SM-G-free1st} using the Iyer–Wald formalism. To study the thermodynamics of these 7D BHs~\eqref{SM-G-ansatz}, we consider two key identities~\eqref{SM-A-fundamentalID} and~\eqref{SM-A-identity}, which hold when the background fields and their perturbations satisfy the equations of motion and when $\xi$ is a Killing vector. 

For the 7D EMCS theory~\eqref{SM-G-action7}, the explicit expressions of $\mathbf{Q}$ and $\mathbf{\Theta}$ are
\begin{equation}
\begin{split}
\mathbf{Q} &=-\boldsymbol{\epsilon}_{abfghij} \left[ \nabla^a\xi^b + \frac{1}{2} \left( F^{ab}+\frac{3k}{2}\epsilon^{aijklmb} A_i F_{jk}F_{lm}\right)  A_c \xi^c \right] \,,\\
\mathbf{\Theta} &= \boldsymbol{\epsilon}_{a a_1a_2a_3a_4a_5a_6} \bigg[  \left( g^{ac}g^{bd}-g^{ab}g^{cd} \right) \nabla_b \delta g_{cd} -\left( F^{ae}+\frac{3k}{2}\epsilon^{abcdfge} A_b F_{cd}F_{fg}\right) \delta A_e \bigg] \,.
\end{split}
\end{equation}

Taking $\xi^a$ as a timelike Killing vector $\xi^a=(\partial_t)^a$ that becomes null on the horizon, we consider a $t=$const. hypersurface $\Sigma$ with
the horizon as its inner boundary. Project $\delta\mathbf{Q}-\xi\cdot\mathbf{\Theta}$ to each boundary of $\Sigma$ and denote it as $(\delta\mathbf{Q}-\xi\cdot\mathbf{\Theta})|_i= (\delta Q-\xi\cdot \Theta)_i \boldsymbol{\epsilon}_{(i)}, (i=r,x_1,x_2,x_3,x_4,z)$, where $\boldsymbol{\epsilon}_{i}$ is the volume element of each boundary.

The first law follows from integrating the fundamental identity~\eqref{SM-A-fundamentalID} over $\Sigma$, leading to
\begin{equation}
\begin{split} 
\delta H &=0=\int_\Sigma \boldsymbol{\omega}=\int_{\Sigma} d(\delta \mathbf{Q}- \xi \cdot \boldsymbol{\Theta}) =\int_{\partial\Sigma} (\delta \mathbf{Q} -\xi \cdot \boldsymbol{\Theta})    \\
&= \int_{ S_{ r=r_{h} } } (\delta Q -\xi \cdot \Theta)_r  \sqrt{\gamma_{(r)}} dx_1dx_2dx_3dx_4dz 
-\int_{ S_{r=\epsilon^+} } (\delta Q -\xi \cdot \Theta)_r  \sqrt{\gamma_{(r)}} dx_1dx_2dx_3dx_4dz \\
&\quad +\int_{S_{x_1=\frac{L}{2} } } (\delta Q -\xi \cdot \Theta)_{x_1}  \sqrt{\gamma_{(x_1)}} dx_2dx_3dx_4dzdr 
-\int_{ S_{ x_1=-\frac{L}{2}} } (\delta Q -\xi \cdot \Theta)_{x_1} \sqrt{\gamma_{(x_1)}} dx_2dx_3dx_4dzdr \\ 
&\quad +\int_{ S_{x_2=\frac{L}{2} } } (\delta Q -\xi \cdot \Theta)_{x_2} \sqrt{\gamma_{(x_2)}} dx_1dx_3dx_4dzdr 
-\int_{ S_{x_2=-\frac{L}{2} } } (\delta Q -\xi \cdot \Theta)_{x_2} \sqrt{\gamma_{(x_2)}} dx_1dx_3dx_4dzdr \\
&\quad +\int_{ S_{x_3=\frac{L}{2} } } (\delta Q -\xi \cdot \Theta)_{x_3} \sqrt{\gamma_{(x_3)}} dx_1dx_2dx_4dzdr 
-\int_{ S_{x_3=-\frac{L}{2} } } (\delta Q -\xi \cdot \Theta)_{x_3} \sqrt{\gamma_{(x_3)}} dx_1dx_2dx_4dzdr\,, \\
&\quad +\int_{ S_{x_4=\frac{L}{2} } } (\delta Q -\xi \cdot \Theta)_{x_4} \sqrt{\gamma_{(x_4)}} dx_1dx_2dx_3dzdr 
-\int_{ S_{x_4=-\frac{L}{2} } } (\delta Q -\xi \cdot \Theta)_{x_4} \sqrt{\gamma_{(x_4)}} dx_1dx_2dx_3dzdr\,, \\
&= 5\delta f_6 -12\delta u_6 -12\delta h_6 +e^{-\frac{\chi_1}{2}} f_1 \left( \frac{u_1 \delta h_1}{r_h^5} +\frac{h_1\delta u_1}{r_h^5} -\frac{5 u_1 h_1 \delta r_h}{r_h^6}  \right) -4\mu\delta A_{t4} - M_{B_1}\delta B_1 -M_{B_2}\delta B_2 \,,  
\end{split}
\end{equation}
where $M_{B_1}$ and $M_{B_2}$ are exactly given by~\eqref{SM-G-mb}. Thus, we obtain the following first law of thermodynamics
\begin{equation} \label{SM-G-firstlaw}
\delta\epsilon =T\delta s +\mu\delta\rho -M_{B_1}\delta B_1 -M_{B_2}\delta B_2 \,.
\end{equation}

Moreover, we can derive the expected free energy density $w$ by integrating the identity~\eqref{SM-A-identity} over $\Sigma$. Once again, we have
\begin{equation}  \label{SM-G-freed}
w=w_{QSR}+kQ_{cs}=\epsilon-Ts-\mu\rho \,.
\end{equation}
Moreover, combining~\eqref{SM-G-firstlaw} and~\eqref{SM-G-freed}, we obtain
\begin{equation}
\delta w =-s\delta T -\rho \delta\mu -M_{B_1}\delta B_1 -M_{B_2}\delta B_2 \,,
\end{equation}
which is the same expression shown from the variation of Euclidean action~\eqref{SM-G-free1st}.

{\bf\emph{Transgression interpretation:}} Finally, the agreement between the Euclidean variation and the Iyer-Wald analysis extends to seven dimensions: in the presence of two independent magnetic fields, both yield the corrected thermodynamic potential $w=w_{ QSR}+kQ_{ cs}$. This agreement again admits a direct action-level explanation in terms of the transgression completion. In the units $16\pi G=1$, the Chern-Simons action corresponds to
\begin{equation}
S_{ CS}=2k\int_{\mathcal M_7}\mathcal C_7(A) \,.
\end{equation}
Its gauge-invariant completion is therefore obtained by the replacement
\begin{equation}
2k\int_{\mathcal M_7}\mathcal C_7(A) \longrightarrow 2k\int_{\mathcal M_7}\mathcal T_7(A,\bar A) \,, \qquad \bar A= \frac{B_1}{2}(x_1 dx_2-x_2 dx_1) +\frac{B_2}{2}(x_3 dx_4-x_4 dx_3) \,.
\end{equation}
Although $\bar F^{\,2}$ is generally nonzero when both magnetic fields are present, the reference curvature satisfies $\bar F^{\,3}=0$, which is precisely the relevant auxiliary equation in seven dimensions. The on-shell transgression action then yields
\begin{equation}
w_{\mathcal T} =w_{ QSR}+kQ_{ cs} =\epsilon-Ts-\mu\rho \,, \qquad
Q_{ cs} =12B_1B_2\int_0^{r_h}dr  \bigl(A_z'A_t-A_zA_t'\bigr) \,.
\end{equation}
The distinct coefficient of the seven-dimensional $Q_{ cs}$ thus follows directly from the normalization of $\mathcal T_7$ and the presence of two independent magnetic fluxes.

The Euclidean variation of the transgression-completed action and the corresponding Iyer-Wald calculation both reproduce the expected first law in~\eqref{SM-G-free1st}, including the two independent magnetization terms. Thus, the transgression completion naturally explains the origin of the seven-dimensional $Q_{ cs}$ while rendering the full action strictly gauge invariant.

We have therefore demonstrated analytically that the same phenomenon also occurs for seven-dimensional BHs. As emphasized in the main text, this mechanism is expected to persist in Einstein-Maxwell theories with $A\wedge F^n$ Chern-Simons interactions in arbitrary odd dimensions, underscoring the universality of our result.

\vspace{.5cm}

\bibliography{suppl}